%% file: paper.tex
\documentclass[opre,nonblindrev]{informs_bl}

\usepackage{amsmath}
\usepackage{latexsym}
\usepackage{amssymb}
\usepackage{amsfonts}
\usepackage{subfigure}
\usepackage{tabularx}
\usepackage{multirow}
\usepackage{comment}
\usepackage{bigints}
\usepackage{algorithm}
\usepackage{algpseudocode}
\usepackage{hyperref}
\usepackage{setspace}
\usepackage{xcolor}
\usepackage{wrapfig}
\hypersetup{
backref=true,
pageanchor=true,
bookmarks,
bookmarksnumbered,
colorlinks=true,
menubordercolor= [rgb]{0, 0, 1},
linkbordercolor= [rgb]{0, 1, 1},
citecolor = [rgb]{0.6, 0, 0},
linkcolor= [rgb]{0, 0, 0.7}
}

\OneAndAHalfSpacedXI 

\usepackage{endnotes}
\let\footnote=\endnote
\let\enotesize=\normalsize
\def\notesname{Endnotes}%
\def\makeenmark{$^{\theenmark}$}
\def\enoteformat{\rightskip0pt\leftskip0pt\parindent=1.75em
  \leavevmode\llap{\theenmark.\enskip}}

\newcommand{\Z}{\ensuremath{\mathbb Z}}
\newcommand{\N}{\ensuremath{\mathbb N}}

\newcommand{\prob}[1]{\ensuremath{\text{{\bf Pr}$\left[#1\right]$}}}

\newcommand{\expct}[1]{\ensuremath{\text{{\bf E}$\left[#1\right]$}}}

\newcommand{\ceil}[1]{\ensuremath{\left\lceil#1\right\rceil}}
\newcommand{\floor}[1]{\ensuremath{\left\lfloor#1\right\rfloor}}

\newcommand{\polylog}{\operatorname{polylog}}
\newcommand{\supp}{\textsf{supp}}

\newcommand{\OPT}{\textsf{OPT}}
\newcommand{\ONOPT}{\textsf{ON-OPT}}
\renewcommand{\SS}{\textsf{SS}}
\newcommand{\PDexp}{\textsf{PD-exp}}
\newcommand{\PDexpcont}{\textsf{PD-exp-cont}}
\newcommand{\PDtquad}{\textsf{PD-tquad}}
\renewcommand{\PD}{\textsf{PD}}

\newcommand{\junk}[1]{}

\newcommand{\ignore}[1]{}

\newcommand{\inn}[2]{\langle{#1},{#2}\rangle} 

\def\bN{\mathbf{N}}
\def\bU{\mathbf{U}}
\def\bC{\mathbf{C}}
\def\bF{\mathbf{F}}
\def\bY{\mathbf{Y}}
\def\bM{\mathbf{M}}
\def\bQ{\mathbf{Q}}
\def\bz{\mathbf{z}}
\def\bx{\mathbf{x}}

\def\bn{\mathbf{n}}
\def\be{\mathbf{e}}
\def\balpha{\boldsymbol{\alpha}}
\def\mU{\mathcal{U}}
\def\mO{\mathcal{O}}
\def\mL{\mathcal{L}}
\def\mC{\mathcal{C}}
\def\mZ{\mathcal{Z}}
\def\mF{\mathcal{F}}

\def\mM{\mathcal{M}}
\def\mN{\mathcal{N}}
\def\mJ{\mathcal{J}}

\newcommand{\expctsub}[2]{\ensuremath{\text{\bf E}_{#1}\left[#2\right]}}
\def\grad{\nabla}

\usepackage[normalem]{ulem}

\usepackage{natbib}
 \bibpunct[, ]{(}{)}{,}{a}{}{,}%
 \def\bibfont{\small}%
 \def\bibsep{\smallskipamount}%
 \def\bibhang{24pt}%

\TheoremsNumberedThrough     
\ECRepeatTheorems

\EquationsNumberedThrough    

\graphicspath{{./figures/}}

\begin{document}


\RUNAUTHOR{Gupta and Radovanovi\'{c}}

\RUNTITLE{Online Stochastic Bin Packing}

\TITLE{Interior-point Based Online Stochastic Bin Packing}

\ARTICLEAUTHORS{%
\AUTHOR{Varun Gupta}
\AFF{Booth School of Business, University of Chicago, Chicago, IL 60637, \EMAIL{varun.gupta@chicagobooth.edu}} 
\AUTHOR{Ana Radovanovi\'{c}}
\AFF{Google Research, 1245 Charleston Rd., Mountain View, CA 94043, \EMAIL{anaradovanovic@google.com}}
} 

\ABSTRACT{ 
\input{tex/abstract}
}%



\maketitle

%


\input{tex/introduction}

\input{tex/model}

\input{tex/prior}
\input{tex/PDbasic}
\input{tex/nonstationary}
\input{tex/conclusions}



\bibliographystyle{ormsv080} 
\bibliography{references} 

%
\newpage


\begin{APPENDICES}
\input{tex/appendix}

\input{tex/alternate}

\input{tex/appendix2}
\end{APPENDICES}

%
%







\end{document}

%% file: tex/abstract.tex
Bin packing is an algorithmic problem that arises in diverse applications such as remnant inventory systems, shipping logistics, and appointment scheduling.  
In its simplest variant, a sequence of $T$ items (e.g., orders for raw material, packages for delivery) is revealed one at a time, and each item must be packed on arrival in an available bin (e.g., remnant pieces of raw material in inventory, shipping containers). 
The sizes of items are \emph{i.i.d.} samples from an unknown distribution, but the sizes are known when the items arrive. The goal is to minimize the number of non-empty bins (equivalently waste, defined to be the total unused space in non-empty bins).
 This problem has been extensively studied in the Operations Research and Theoretical Computer Science communities, yet all existing heuristics either rely on learning the distribution or exhibit $o(T)$ additive suboptimality compared to the optimal offline algorithm only for certain classes of distributions (those with sublinear optimal expected waste). 
In this paper, we propose a family of algorithms which are the first truly distribution-oblivious algorithms for stochastic bin packing, and achieve $\mO(\sqrt{T})$ additive suboptimality for all item size distributions. Our algorithms are inspired by approximate interior-point algorithms for convex optimization. 
In addition to regret guarantees for discrete \emph{i.i.d.} sequences, we extend our results to continuous item size distribution with bounded density, and prove a family of novel regret bounds for \emph{non-i.i.d.} input sequences. 
 To the best of our knowledge these are the first such results for non-i.i.d. and non-random-permutation input sequences for online stochastic packing. 

%% file: tex/introduction.tex
\section{Introduction} \label{sec:introduction}

Bin packing is one of the oldest resource allocation problems and has received considerable attention due to its practical relevance. 
In the classical offline version of static bin packing, a list of scalar item sizes $\{s_1,s_2,\ldots,s_T\}$ has to be partitioned into the fewest number of partitions each summing to at most $B$ (the bin size).
In the still more challenging online version, the list of item sizes is revealed one at a time, and the items must be irrevocably assigned to a bin on arrival. Online bin packing appears as a motif in many operations research problems, of which we give a small sampling below:
\begin{enumerate}
\item Remnant Scheduling/Cutting Stock: In \cite{AdelmanNemhauser1999remnant}, the authors cite the example of a fiber optic cable manufacturer which produces cables of fixed lengths. Orders for customer-specified lengths arrive and must be served online from the available inventory. The goal is to serve the demand while minimizing the rate of production of cables, or equivalently the scrap rate of remnant inventory. In this context, bins correspond to the fixed length cables, and items correspond to the customer orders which must be packed (i.e., cut) online and irrevocably.
\item Appointment scheduling: Requests for appointments arrive online and must be scheduled in a future day with available slots. Here appointments correspond to items, and the office hours during a working day correspond to bins. In addition to the remnant scheduling example, one novel feature in appointment scheduling problem is that bins depart periodically, and the goal is to minimize the time until appointment. However, at its core is a bin packing problem.
\item Transportation logistics: Items to be shipped arrive and are queued at a transshipment center. Bins correspond to shipping containers (which arrive either exogenously, or endogenously on demand), and must be packed using queued items. The goal is to minimize some combination of shipping costs (the number of bins used) and holding costs (number of items waiting to be packed).
\end{enumerate}

The common thread in all the three examples above is that items are packed irrevocably, and do not depart from the bin once packed (the bin may depart with the items). This is known as the \emph{static bin packing} problem. In this paper we will develop algorithms for the first variant of static online bin packing: an unbounded number of bins are available, items are packed irrevocably on arrival into a feasible bin, and  the goal is to minimize the number of bins used.

{\bf Adversarial bin packing:} For adversarially generated instances (that is, worst-case analysis), offline bin packing problem is NP-hard, but good approximation algorithms exist for one-dimensional packing (\cite{KarmarkarKarp82, rothvoss2013approximating}). For online bin packing, simple algorithms such as Best Fit and First Fit have an asymptotic competitive ratio of at most $\frac{17}{10}$ (\cite{johnson1974worst}); that is, the number of bins used is at most 1.7 times the number of bins used in the optimal offline packing plus an additive sublinear function of the optimal number of bins. Most recently, \cite{balogh2017new} have proposed an online packing algorithm, Advanced Harmonic (AH), with an asymptotic competitive ratio of 1.57829. For absolute competitive ratio (that is, without any additive sublinear terms), \cite{balogh2015optimal} propose an algorithm with absolute competitive ratio $5/3$ which is the best possible for online algorithms. 

{\bf Online Stochastic bin packing:} Driven by practical considerations and the lower bounds described above for adversarial instances, a common approach has been to make stochastic assumptions on the problem instance -- the number of items to be packed, $T$, is much larger than the number of items that can fit in a bin, and the item sizes are assumed to be an \emph{i.i.d.} sequence from some distribution $F$. Further, usually the item sizes and bin sizes are assumed to be integers. The performance of an online packing heuristic is measured by the expected difference between the number of bins used by the online heuristic, and the optimal-in-hindsight algorithm (ideally, we desire the suboptimality gap, or regret, that grows as $o(T)$). 

{\bf Distribution-oblivious stochastic bin packing:} 
In this paper we adopt the convention that an online algorithm that tracks the empirical frequencies of the items revealed online is called {\it learning-based}.
Under the assumption that item sizes are {\it i.i.d.}, a natural learning-based approach is to use the empirical distribution to solve an offline optimization problem (and re-solve as more items are seen), and use the solution of this optimization problem to guide the online packing algorithm. The currently best known algorithms for minimizing regret in online stochastic bin packing (\cite{RheeTalagrand93, SumOfSquares_JACM_CsirikJKOSW06}) use this approach.
While offering the smallest known regret, such learning-based algorithms are brittle to non-stationarity in input, and it is unclear how the learning should be carried out in a non-stationary setting. 
At the other end of the spectrum are what we call {\it distribution-oblivious} or {\it blind} online algorithms whose actions are purely a function of the size of the arriving item and the levels of the non-full bins in the current packing. The distinction between learning-based and distribution-oblivious algorithms is perhaps subjective, but the above sufficient conditions seem to be quite weak -- any weaker and even the Best Fit heuristic would fail to be distribution-oblivious. 
There has been an extensive study of common blind heuristics such as Best Fit, and identifying item size distributions for which these can be optimal, and for which they are provably suboptimal.
Our main driving questions in the paper are: \emph{Are there simple distribution-oblivious online packing algorithms that nonetheless achieve $o(T)$ regret for \emph{i.i.d.} inputs for all item size distributions? What robustness guarantees can we prove for such algorithms on non-stationary input sequences?} 
We answer the first question affirmatively by presenting a simple algorithm
inpired by an approximate Interior-Point (Primal-Dual) solution of the bin packing Linear Program. Our algorithms have $\mO(T^{1/2})$ regret for all item size distributions with integer support, and $\mO(T^{2/3})$ regret for all item size distributions which are mixtures of atoms on integers and a distribution with bounded density. The former guarantee matches the best known regret in \cite{SumOfSquares_JACM_CsirikJKOSW06}, except for a special class of distributions called Bounded Waste distributions for which they achieve $\mO(1)$ regret. Without the assumption of integer item sizes, but where the number of items $T$ is known to the algorithm, the best known regret is $\mO(T^{1/2} \log^{3/4}T)$ from \cite{RheeTalagrand93} . While we do not always match the best known regret for {\it i.i.d.} items we still achieve $o(T)$ regret, and in return we are able to prove a novel family of regret guarantees for non-\emph{i.i.d.} input sequences answering the second question above, further illustrating the benefit of our distribution-oblivious approach. 

%% file: tex/model.tex
\section{Model Notation and Definitions}
\label{sec:model}




\begin{figure}[t!]
\centering
\includegraphics[height=2.2in]{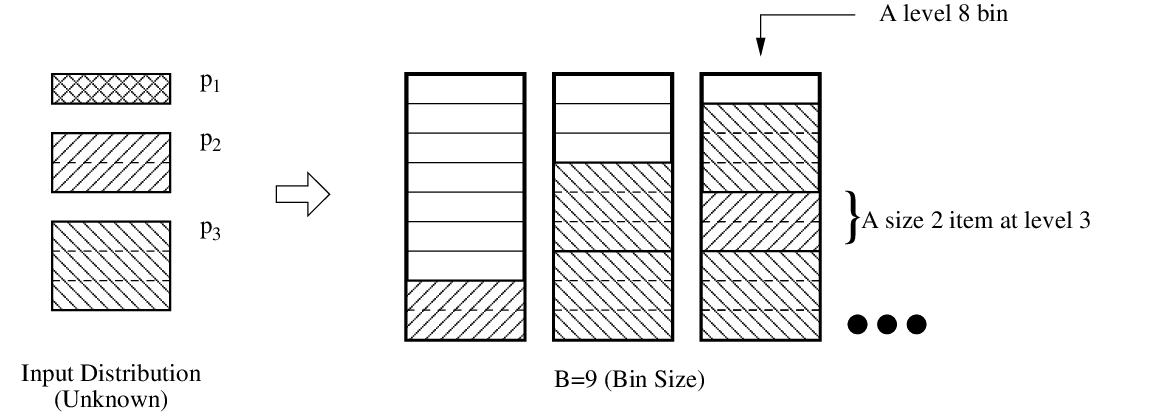}
\caption{A bin packing instance and nomenclature}
\label{fig:model}
\end{figure}

We are given an infinite collection of empty bins of capacity $B$, a scalar integer, into which a sequence $Y= \{Y_1, Y_2, \ldots, Y_T\}$ of $T$ items is to be packed online with the goal of minimizing the number of bins used. The sizes $Y_t$ are $i.i.d.$ random variables taking values in $[J]=\{1,2,\ldots, J\}$. We will sometimes refer to the size of an item as its type. We will denote the probability of type/size $j$ items by $p_j$, abbreviate the item size distribution by $F := (p_1, \ldots, p_J )$, and by $\supp(F)$ the set of item types with strictly positive probability under $F$ (i.e., the support of $F$). Unless otherwise stated, we assume $J=B-1$. A bin is called \emph{level} $h$ bin if the sizes of the items packed in the bin sum to $h$. See Figure~\ref{fig:model} for an illustration of the definitions so far.

The feasible \emph{configurations} of a bin are denoted by the set $\mC$ where a configuration $c \in \mC$ is represented as a vector $\bx_c = ( x_{c}(1), x_{c}(2), \ldots, x_{c}(J) ) \in \N^J$. The $j$th components of $\bx_c$ is the number of size $j$ items in the configuration $c$. For example, for $B=9, J=3$, $x = ( 0,\ 0,\ 2 )$ represents the configuration of the middle bin in Figure~\ref{fig:model} with two items of size $3$ and no other items. For a given collection of items, a \emph{packing} $P$ is a vector $\balpha \in \N^{\mC}$ with $\alpha_c$ denoting the number of bins in configuration $c$. We denote by $N^P(h)$ the number of level $h$ bins in packing $P$, and use $\mathbf{N}^P = ( N^P(1) , N^P(2),\ldots , N^P(B)) )$ to summarize the packing $P$ by only keeping the number of bins of each level and ignoring their configurations. 

\paragraph{Action space and algorithms:} Let $\be_h \in \{0,1\}^{B}$ for $h\in \{ 0,\ldots,B-1 \}$ denote the vector whose $(h+1)$st component is 1, and the rest are 0 (so $\be_0=(1,0,0,\ldots,0)$).
We will represent the action of packing an item in a bin of level $h$ as $\bU = \be_h$. Therefore, the set of feasible actions for an item of type $Y$ that has to be packed in state $\bN$ is given by:
\[ \mU(\bN, Y) :=  \left\{ \be_h \left| \  0 \leq h \leq B-Y ,  N(h) \geq 1  \right.  \right\}. \]
 We will use $U(i)$ to index the $(i+1)$st element of $\bU$ so that $\bU = ( U(0), U(1),\ldots, U(B-1) )$. 

Unless otherwise stated, in this work an online packing algorithm $A$ is specified by a collection of packing rules $\pi^A_t(\bN_{t-1}, Y_t) : \Z_{\geq 0}^{B} \times [B-1] \to \{0,1\}^{B}$ where $t$ denotes the time index of the item being packed, $\mathbf{N}_{t-1}$ summarizes the current packing into which the item type $Y_t$ (also the item size) is being packed. For a deterministic packing algorithm, $\pi^A_t$ is a deterministic function; for a randomized algorithm $\pi^A_t$ is a random function in which case
\[ u_t := \expct{\bU_t} := \expct{ \pi^A_t(\bN_{t-1}, Y_t) } \]
denotes the probability measure of the randomized policy $A$ (the expectation is over the randomization of policy $A$; the item type $Y_t$ is conditioned on and hence $u_t$ is a function of $\bN_{t-1},Y_t$). Here $u_t \in \widehat{\mU}(\bN_{t-1},Y_t)$:
\begin{align}
\label{eqn:feasible_simplex}
\widehat{\mU}(\bN, Y) :=  \left\{ \bU \in [0,1]^B \left| \begin{array}{l} \sum_{h=0}^{B-Y} U(h) = 1; \\  U(h)=0 , \ h \geq B-Y+1 \\  U(h)=0, \ N(h) = 0 \end{array} \right.  \right\}. 
\end{align}

To express the evolution of $\bN_t$, we next define \emph{matrices} $\bC_{y} \in \{-1,0,1\}^{B \times B}$ for $y \in [B-1]$. The $(h+1)$st column of $\bC_y$, denoted $\bC_{y,h+1}$ ($h+1 \in [B]$), represents the change in components of $\bN$ when an item of type $y$ is packed in a level $h$ bin:
\begin{align*}
\mbox{for } y \in \{1,\ldots, B-1\}, h \in \{0,\ldots, B-1\}, i \in \{1, \cdots, B\}: \quad  \bC_{y,h+1}(i) &= \begin{cases}
-1 & i = h, \\
+1 & i = h+y, \\
0 & \mbox{otherwise}.
\end{cases}
\end{align*}
We can thus write the stochastic process $\{\bN_t \}$ under item size sequence $\{Y_1,Y_2, \ldots \}$ as:
\begin{align*}
\bN_0 & = \mathbf{0}, \\
\bN_{t} &= \bN_{t-1} + \bC_{Y_t} \cdot \bU_t \ ,  \qquad \mbox{where } \bU_t \in \mU(\bN_{t-1}, Y_t);
\end{align*}
interpreting $\bN_t$ and $\bU_t$ as column vectors.

When an algorithm $A$ is used for packing items, we will denote the packing at time $t$ by $P^A_t$ and the level summary by $\bN^A_t := \bN^{P^A_t}$. The action will be denoted by $\bU^A_{t} = \pi^A_t(\bN_{t-1}, Y_t)$, so that $\bN^A_t = \bN^A_{t-1} + \bC_{Y_t}\cdot \bU^A_t$. If the $t$ items $\{Y_1,\ldots. Y_t\}$ are i.i.d. samples from distribution $F$, we denote by $P^A_{F,t}$ the random packing and by $\bN^{A}_{F,t}$ the level summary of the random packing $P^A_{F,t}$ (and thus $\bN^{A}_{F,t}$ is a random variable as well).

\paragraph{Open and Closed bins:} By open bins, we mean the bins in a packing $P$ which can receive new items. Often it is desirable to have only a bounded number of open bins. For example, in remnant scheduling, one would only like a few active pieces of remnant inventory at hand. In the case of transportation logistics, a partially filled container can not be kept undispatched indefinitely. We model such scenarios by marking some bins of $P$ as irrevocably \emph{closed} to receiving new items. We denote the number of open bins by $\widetilde{\bN} \leq \bN$. The feasible range of action in this case when packing an item of size $Y$ in state $(\bN, \widetilde{\bN})$ is $\mU(\widetilde{\bN},Y)$. In this case we let the policy depend on $\bN_t$ as well as $\widetilde{\bN}_t$. The notion of open and closed bins will be relevant in Section~\ref{sec:bounded_inventory}.

{\bf Metric:} A natural metric to minimize is the total number of bins in the packing $P$: 
\[ N^P := \sum_{h=1}^B N^P(h) . \]
(We use boldface $\bN$ for the level summary, $N$ for the total number of bins, and $N(h)$ for the number of bins of level $h$.) 
There is an alternate but equivalent performance metric that is more commonly used for one-dimensional packing -- the \emph{waste} of packing $P$:
\[ W^P := \frac{1}{B} \sum_{h=1}^{B-1} (B-h)\cdot N^P(h) . \]
That is, the waste of packing $P$ is the unused space in the bins used in $P$ (normalized so that the waste of an empty bin is 1). The expected waste of algorithm $A$ on distribution $F$ at time $T$ is given by:
\begin{align}
\label{eqn:1dwaste}
\expct{W^A_{F,T}} := \expct{W^{  P^A_{F,T} }} .
\end{align}
Similarly, we define the expected number of bins used by algorithm $A$ on distribution $F$:
\begin{align}
\label{eqn:1dcost}
\expct{N^A_{F, T}} := \expct{N^{P^A_{F,T}}}.
\end{align}
By $\OPT$ we denote an \emph{offline} algorithm that packs the item sequence $\{Y_1, \ldots, Y_T\}$ in hindsight to minimize the total number of bins used. $\expct{W_{F,T}^{\OPT}}$ and $\expct{N_{F,T}^{\OPT}}$ denote the expected waste and expected number of bins used in this optimal offline packing.

{\bf A classification of item-size distributions:} 
To describe the performance of the current state-of-the-art algorithm for stochastic online bin packing, we will need the following classification result of \cite{courcoubetis1986necessary}.

\begin{theorem}[\cite{courcoubetis1986necessary}]
Any discrete item-size distribution $F$ falls in one of three categories based on the asymptotic growth rate of waste of the optimal offline algorithm $\expct{W_{F,T}^{\OPT}}$ as a function of $T$:
\begin{enumerate}
\item Linear Waste (LW) : $\expct{W_{F,T}^{\OPT} } =  \Theta(T)$,  e.g., $B = 9, F = (p_2=0.8, p_3=0.2 )$,
\item Perfectly Packable (PP) :  $\expct{W_{F,T}^{\OPT}} =  \Theta(\sqrt{T})$, e.g. $B=9, F= ( p_2= \frac{3}{4}, p_3=\frac{1}{4} )$,
\item PP with Bounded Waste (BW) :  $\expct{W_{F,T}^{\OPT}} = \Theta(1)$, e.g. $B=9, F = ( p_2=0.5,p_3=0.5 )$ .
\end{enumerate}
\end{theorem}
The intuition for the above classification is the following. Let $\mathcal{X}^* \subset \mathcal{X} := \{ \bx_c : c \in \mC\}$ represent the set of configuration vectors which are perfectly packed (that is, level $B$ bins). For example, for $B=9, J=3$, the vectors $(0, 0, 3)$ and $( 0, 3, 1) $ are in $\mathcal{X}^*$. The set of vectors $\mathcal{X}^*$ generate a convex cone representing item frequency vectors which can be packed (with fractional bins allowed) with zero waste. A distribution $F$ in the interior (or relative interior if $\supp(F)$ is not $[J]$) of this cone is a Bounded Waste distribution because an empirical sample of $T$ items from $F$ remains in the interior of this set after discarding $\mO(1)$ items. A distribution that is outside the convex cone generated by $\mathcal{X}^*$ will be a Linear Waste distribution. Distributions on the boundary of the convex cone will have $\Theta(\sqrt{T})$ waste since $\Theta(\sqrt{T})$ items must be discarded from an empirical sample of size $T$ so that the remaining list of items can be packed with zero waste. In the concluding section, we add to this classification by introducing a rather broad class of distributions which we call \emph{Vertex Dual} distributions which we conjecture are more meaningful from an algorithmic point of view. Vertex Dual distributions also turn out to be universal in the sense that any non-discrete random perturbation of a given distribution is a Vertex Dual distribution almost surely.

%% file: tex/prior.tex
\section{Prior Work and Our Contributions} 
\label{sec:prior}

In this section we focus on the prior work in stochastic bin packing where item sizes are random but known at the time of packing, and leave out the discussion on adversarial models of online bin packing. The relevant prior literature can be partitioned into algorithms which are distribution-aware or actively learn the distribution, and distribution-oblivious algorithms.

\paragraph{Distribution-aware online packing:} \cite{AdelmanNemhauser1999remnant} consider the problem of minimizing scrap for remnant scheduling (also called the 1-d cutting stock problem) which, as mentioned earlier, is a rephrasing of the bin packing problem. The authors propose an algorithm that learns the item size distribution, and uses the duals of a bin packing Linear Program (LP) while making packing decisions. \cite{RheeTalagrand93} propose a packing heuristic which uses all the item sizes seen so far to form a bin packing LP relaxation and prove that when the item sizes are $i.i.d.$ from a general distribution (the support of the distribution can be continuous), their algorithm has regret $\mO(T^{1/2} \log^{3/4}T)$. Another relevant work is by \cite{iyengar2004exponential} where the authors devise a control policy for a loss network based on solving an offline LP, and then controlling the system online so as to minimize the deviation from the solution of the LP. The authors do not however explicitly describe the application of their algorithm to static bin packing.

\paragraph{Distribution-oblivious online packing:} Most of the work on analysis of distribution-oblivious algorithms for stochastic bin packing has been carried out in the theoretical computer science community, beginning with analysis of First Fit and Best Fit heuristics for which worst-case performance in non-stochastic settings were known from earlier. When the bin size is $1$ and item size distribution is Unif$[0,1]$, \cite{Shor_packing} proved that the expected waste under First Fit (pack in the oldest feasible bin) grows as $\Theta(T^{\frac{2}{3}})$. For Best Fit (pack in the fullest feasible bin), \cite{LeightonShor_matching} proved this to be $\Theta(T^{\frac{1}{2}}\log^{\frac{3}{4}}{T})$. While the results mentioned above are for Unif$[0,1]$ distribution, we mention them in the distribution-oblivious category because the algorithms do not exploit the distributional information. Finally, \cite{Shor91} proposed a scheme that achieves $\Theta(T^{\frac{1}{2}} \log^{\frac{1}{2}}T)$ expected waste, matching a lower bound of $\Omega(T^{\frac{1}{2}}\log^{\frac{1}{2}}T)$ proved by \cite{Shor_packing}. However, the algorithm in \cite{Shor91} is tailored to Unif$[0,1]$ distribution and hence is not distribution-oblivious.

 For discrete item sizes, when the item sizes are uniformly distributed over $\left\{ \frac{1}{B}, \frac{2}{B}, \ldots, \frac{J}{B}\right\}$, \cite{Coffmanetal_STOC} proved the expected waste for $J=B$ or $J=B-1$ grows as $\Theta(T B^{\frac{1}{2}})$ for First Fit, and $\Theta(T^{\frac{1}{2}}\log{B})$ for Best Fit. For $J=B-2$, bounded expected waste for Best Fit was proved by \cite{KenyonRabaniSinclair_SODA}, and for First Fit (using Random Fit as an intermediate step) by \cite{AlbersMitzenmacher_SODA}. \cite{KenyonMitzenmacher_FOCS} proved that the waste under Best Fit is linear when $J=\alpha B$, $ \frac{99}{150} < \alpha< \frac{100}{150}$, and $B$ large enough, but is conjectured to hold for all $0<\alpha<1$. (Interestingly, Best Fit has linear expected waste even for the benign case of $B=6$ and items of size $2,3$ with equal probability, but this appears to not have been a compelling reason to seek alternatives to BF.)

\textbf{Sum of Squares ($\SS$) rule 
\cite{SelfOrganizing_Csirik99,SumOfSquares_JACM_CsirikJKOSW06}:
} The SS heuristic is in some sense the state-of-the-art bin packing policy when item sizes and bin size $B$ are integral, and works as follows: 
Suppose $\bN_{t-1}$ represents the state of packing after seeing $t-1$ items. On arrival of the $t$th item $Y_t$, it is packed in a feasible bin so as to minimize the value of the following sum-of-squares potential function of the resulting packing $\bN_t$:
\[ ss(\bN) \triangleq \sum_{h=1}^{B-1} N(h)^2 . \]
That is, the action is given by,
\[  \bU^{\SS}_t \in \argmin_{ \bU \in \mU(\bN_{t-1}, Y_t) } ss\left( \bN_{t-1} + \bC_{Y_t} \cdot \bU \right) \]

\cite{SumOfSquares_JACM_CsirikJKOSW06} prove that for PP distributions, the waste under $\SS$ is indeed $\mO(\sqrt{T})$. Further, for BW distributions, the waste of $\SS$ is $\mO(\log{T})$ which can be reduced to $\mO(1)$ by learning the support of the distribution. The heuristic embodies the intuition that for PP and BW distributions, no $N_h$ corresponding to a partially filled bin should grow very large. Hence by penalizing $N_h$ quadratically the heuristic discourages accumulation of bins of `dead-end' levels (those levels from which we can not create a level $B$ bin). 
This intuition does not hold for Linear Waste distributions for which $N_h$ for some non-full level must grow as $\Theta(T)$, and for linear waste distributions $\SS$ achieves $\Theta(T)$ additive suboptimality. That is, $\SS$ is not asymptotically optimal for LW distributions. To rectify this, the authors propose to tune the policy by injecting `phantom' items of size 1 at the smallest rate which makes the new distribution,
$F^+$, perfectly packable. 
However, computing the optimal rate of injection of these phantom size 1 items requires learning the distribution $F$ and solving a Linear Program (called ``waste LP''). Therefore, it is not a distribution-oblivious packing algorithm. Our proposed heuristics obtain $\Theta(\sqrt{T})$ additive suboptimality for all distributions while being truly `blind'.

\paragraph{Online Convex Optimization (OCO) for Online Packing/Covering Problems:} A related thread of research is the literature exploiting online convex optimization tools such as Online Mirror Descent and Multiplicative Update algorithm to solve online packing and covering problems (e.g., \cite{gupta2014experts, agrawal2015fast}). In online packing and covering problems, an item $Y_t$ is associated with a set of feasible actions $A_t \subset \Re^{m}$. The algorithm must choose an action $a_t \in A_t$ for $Y_t$ without knowing the sequence of future arrivals, while obeying packing/covering constraints on $\sum_t a_t$ and maximizing a reward function. The arrival models considered in such papers are either $i.i.d.$ from an unknown distribution, or random permutation of a possibly adversarially generated sequence of items. To unify the presentation with the literature on OCO an stochastic networks, we have cast the bin packing problem in Section~\ref{sec:model} in a similar framework by associating with each item the set of action vectors $\mU(\bN, Y)$ denoting feasible placements. One crucial difference is that in our setting the set of feasible actions is state-dependent. A second crucial difference is that the entries in the ${\bC}$ matrices in bin packing are both positive and negative, while they are non-negative in the online packing/covering literature. In fact, we believe that robustness results similar to what we prove for non-\emph{i.i.d.} sequences can be carried over to Online Packing/Covering Problems handled in OCO framework.  

\paragraph{Bounded space bin packing: } An online bin packing algorithm is called $I$-bounded space if at any time it has at most $I$ active/open bins available to pack the arriving items. In the worst-case setting where the item size sequence is completely adversarial, \cite{lee1985simple} present an algorithm called $\textsf{HARMONICM}$ and prove that it has a competitive ratio of 1.692. They also prove an almost matching 1.691 lower bound for all $\mO(1)$-bounded space online algorithms. \cite{csirik2002resource} extend their results to the resource augmentation setting. Here the online algorithm is allowed to use bins of size $b\geq 1$, while the optimal offline uses bins of size 1. In the stochastic setting where item sizes are {\it i.i.d.}, \cite{naaman2008average} present results on the competitive ratio of common packing heuristics (Next Fit, Best Fit, Harmonic)  by analyzing the Markov chain induced on the state of the active bins. Most of the theoretical results are for uniform distributions, and for 1- or 2- bounded space algorithms due to tractability of the Markov chain. A related problem is the bounded space {\it bin cover} problem where bins must be filled to {\it at least} their size (called demand) of $B$ but can be potentially filled to more than the demand and the goal is to pack items to maximize the number of bins with satisfied demand. For the stochastic case, \cite{asgeirsson2009bounded} present several heuristics for the bounded space bin cover problem by casting it as a Markov Decision Process and approximating the bias of the value function (also called relative value function) through a modified transition kernel. While no formal guarantees are proved, such heuristics can be useful for the bounded space bin packing problem with {\it i.i.d.} item sizes as well.

\paragraph{Static packing models with bin departures:} Although not the object of study in the present paper, we briefly mention the literature on static bin packing where bins arrive and/or depart. 

\cite{CoffmanStolyar_packing} study a model where items arrive continuously and queue up. At discrete time instants ($t=0,1,2,\ldots$) a bin arrives, is filled using the items currently in queue using some packing scheme (e.g. Best Fit (pack the largest item, then the next largest and so on), First-Fit (try to pack the oldest item)), and then the bin departs immediately.
 The authors prove sufficient stability conditions for discrete item sizes with symmetric distributions. \cite{Gamarnik_packing} studies the stability for general item size distributions via Lyapunov functions and provides a numerical algorithm for checking stability to arbitrary precision. \cite{GamarnikSquillante05} further investigate the steady-state behavior of Best Fit via Lyapunov function analysis and matrix analytic techniques. In particular, they find that the sufficient condition for stability of symmetric distributions do not carry over to asymmetric distributions. \cite{ShahTsitsiklis_packing} study the lower and upper bounds for asymptotic order of growth rate of queue length in heavy traffic under symmetric distributions. 

\cite{Lelarge_online_packing} studies a model with an infinite collection of bins where items are packed on arrival, and the oldest bin departs at discrete time steps. The performance metric investigated is the number of partially filled bins, and the total size of items packed in the partially filled bins. The main result is that for symmetric item size distributions, both First Fit (FF) and Best Fit (BF) are stable, but the volume of queued items is asymptotically larger under FF in heavy traffic. 

\subsection{Main Ideas and Results}
The main insight behind our algorithm can be understood via the failure of Sum-of-Squares algorithm on Linear Waste distributions. The Sum-of-Squares penalty function $ ss(\bN) \triangleq \sum_{h=1}^{B-1} N(h)^2 $ attempt to encourage creating bins of all levels to enable flexibility in future actions. However, it attempts to do so by penalizing creating too many bins of a given level via the quadratic term $N(h)^2$. This has the side-effect of trying to equalize the number of bins of different levels which backfires for Linear Waste distribution where some $N(h)$ for $h<B$ must grow linearly as $\Theta(T)$, and thus all $N(h)$ end up growing as $\Theta(T)$. In our approach, {\it instead of penalizing large $N(h)$, we penalize small $N(h)$}, e.g., via a penalty of the form $e^{-N(h)}$. This discourages actions which deplete bins of levels with small $N(h)$. This alone is not sufficient, and we also have to penalize actions which open new bins which we achieve via the penalty $\sum_{h=1}^B N(h)$. This surprisingly turn out to be enough! (Please see Appendix~\ref{sec:interp} for further discussion on the comparison of Sum-of-Squares to our approach.)

A brief summary of the results in the paper follows:
\begin{enumerate}
\item \textbf{A new online bin packing algorithm:} In Section~\ref{sec:PD_nodepartures} we present our {\bf Primal-Dual ($\PD$) family} of online bin packing algorithms. Our algorithms are inspired by gradient descent for solving an Interior-point relaxation of the bin packing LP, and pack items so as to greedily minimize a penalized-Lagrangian. Choosing different barrier functions (Interior-point view), or penalty functions (penalized Lagrangian view), results in different algorithms in the family. For example, with exponential penalty function (Algorithm~\ref{alg:PDexp_level}), this entails greedily minimizing the following potential function at time $t$:
\[\mL^{\mbox{\footnotesize exp}}_t(\bN) := \sum_{ h=1 }^B N(h) + \frac{\kappa}{\epsilon_t} \sum_{h = 1}^{B-1} e^{-\epsilon_t \cdot N(h)}
\]
We prove that for the appropriate choice of constant $\kappa$ and $\epsilon_t = \Theta\left( \frac{1}{\sqrt{t}} \right)$, the algorithm achieves 
\[ \expct{N_{F,T}^{\PD}} \leq \expct{N^{\OPT}_{F,T}}+\sqrt{4BT} \]
for all discrete distributions $F$. (Theorems~\ref{thm:PDexp_fixed_epsilon}-\ref{thm:martingale})

To compare with existing results, the Sum-of-Squares algorithm \cite{SelfOrganizing_Csirik99,SumOfSquares_JACM_CsirikJKOSW06} gets an additive suboptimality of $\mO(\log T)$ for Bounded Waste distributions, $\mO(\sqrt{BT})$ for Perfectly Packable distributions, and $\mO(T)$ for Linear Waste distributions. Therefore, while $\PDexp$ gives a higher additive suboptimality for Bounded Waste distributions, this is still sublinear in $T$. In exchange, we get a significant improvement for the case of Linear Waste distributions.

 In Appendix~\ref{sec:interp}, we provide an interpretation of our algorithms as online mirror ascent to solve the dual maximization problem, where the choice of barrier/penalty functions now map to choice of distance generating functions on the space of duals of the bin packing LP. We also provide a nice interpretation of our algorithms as ``patching'' the $\SS$ algorithm to work for LW distributions.

\item {\bf Bounded inventory guarantees:} The $\PD$ algorithm of Theorem~\ref{thm:PDexp_fixed_epsilon} keeps all the bins created in its working inventory ($\Theta(T)$), which may not be desirable. We prove that if the online algorithm is only allowed to keep at most $I$ bins open at any time, then a modified algorithm $\PDtquad$ yields a packing with $\expct{N_{F,T}^{\PD}} \leq \left( 1+ B/I \right) \expct{N^{\OPT}_{F,T}} + B\cdot I/2$, and that there are distributions for which no online algorithm, even distribution-aware, can have a competitive ratio $\left( 1 + o(1/I) \right)$.
(Theorems~\ref{thm:bounded_upperbound}-\ref{thm:bounded_lowerbound})

\item {\bf Continuous item size distributions:} We extend our $\PD$ algorithm in Algorithm~\ref{alg:PDexp_continuous} to work with continuous item size distributions. We prove that if the item size distribution is a mixture of a discrete distribution with integer support and a continuous distribution with density bounded by $D$, then our modified $\PDexpcont$ algorithm packs with an additive suboptimality of $\mO((D+\sqrt{B})T^{2/3})$. (Theorem~\ref{thm:PDexp_continuous})
\item \textbf{Guarantees against non-$i.i.d.$ input:} In Section~\ref{sec:nonstationary} we turn to proving performance guarantees for the $\PD$ Algorithm~\ref{alg:PDexp_level} for input item size sequence $\{Y_1, Y_2, \ldots, Y_T \}$ which need not be $i.i.d.$ from a single distribution $F$. Specifically we consider the case where a sequence of non-identical distributions $\{F_1, F_2,\ldots, F_T\}$ is generated first (umknown to the algorithm), and the items are sampled independently from this sequence of distributions (Theorem~\ref{thm:adversarial_smoothed}).

\end{enumerate}


%% file: tex/PDbasic.tex
\section{Online Stochastic Bin Packing with \emph{i.i.d.} arrivals} \label{sec:PD_nodepartures}

In this section we present our algorithm for stochastic online bin packing that achieves additive $\mO(\sqrt{T})$ suboptimality for $i.i.d.$ item sizes for all distributions $F$. 
In Section~\ref{sec:PDlevel}, we present a Linear Program (LP) for finding the optimal fractional packing given the distribution $F$, and propose a general recipe for converting math programs into optimization algorithm via an interior-point/penalized-Lagrangian transformation. In Section~\ref{sec:analysis} we formally present the online algorithm for stochastic bin packing and the performance guarantees. To illustrate the main proof technique, in Section~\ref{sec:proofsketch} we present the proof of one of our main theorems. Remaining proofs are postponed to Appendix~\ref{sec:proofs}. 
In Section~\ref{sec:bounded_inventory} we present a result on competitive ratio of our online packing algorithm when there is a bound on the number of bins allowed to be kept open. In Section~\ref{sec:PDcontinuous} we present an extension of our algorithm to continuous item size distributions with bounded density.


\subsection{Preliminaries}
\label{sec:PDlevel}

\subsubsection{Bin-packing LP}
We begin with the Linear Program for the following offline one-dimensional bin packing problem (\cite{SumOfSquares_JACM_CsirikJKOSW06}): Given the distribution $F=(p_1,p_2,\ldots, p_J)$ and bin size $B$, what is the \emph{average number of bins used per item} in the optimal packing? We denote this optimal value by $b(F)$,
\[ b(F) := \lim_{T \to \infty} \frac{\expct{N^{\OPT}_{F,T}}}{T} .\]
 E.g., for $B=9, J=3, F = (p_2 = 1/2, p_3 = 1/2 )$, the optimal fractional packing has bins in configurations $(0,3,1)$ (three items of size 2, and one of size 3) and $(0,0,3)$ (three items of size 3) and hence $b(F) = \frac{p_2}{3} + \frac{p_3-p_2/3}{3} = 5/18$.

To solve for $b(F)$, we will write an LP with the following decision variables:
\begin{align*}
v(j,h) & := \mbox{fraction of jobs which are of size $j$ and are packed at level $h$ in their bin,} \\
 & \qquad  (j \in [B-1], h \in \{0,\ldots, B-1\}). 
\end{align*}  
With the above, we are led to the following LP:
\begin{align*}
b(F) &= \min_{ \mathbf{v}} \sum_{j=1}^{B-1} v(j, 0) & \mathbf{(P_{1d-level})}\\
\mbox{subject to } \quad &  \forall h \in [1,B-1] \ : \ \sum_{j=1}^h v(j,h-j) \geq \sum_{j=1}^{B-h} v(j,h) & \mbox{(\it no floating items)}\\
 & \forall j \in [J] \ : \ \sum_{h=0}^{B-j} v(j,h)  = p_j & \mbox{(\it mass balance)}\\
 & \forall j \in [J] , h \geq B-j+1\ : \ v(j,h)  = 0 & \\
& \forall j \in [J]; h \in [B-j] \ : \ v(j,h) \geq 0
\end{align*}
The mass balance constraint for size $j$ dictates that the overall fraction of jobs of size $j$ in the solution $\mathbf{v}$ be $p_j$. The \emph{no floating items} constraint says that the total fraction of jobs which sit at level $h$ (the expression $\sum_{j=1}^{B-j} v(j,h)$) should not be more than the total fraction of jobs that end at level $h$ (the expression $\sum_{j=1}^h v(j,h-j)$) since any item of former type must be placed just above an item of the latter type in its bin. The deficit
\[ n(h) \triangleq  \sum_{j=1}^{h} v(j,h-j) - \sum_{j=1}^{B-h} v(j,h)\]
denotes the number of level $h$ bins per item in the optimal fractional packing. 
The objective function counts the fraction of overall items which are packed at level $0$, and hence the number of bins per item: $\sum_{h=1}^{B} n(h)$. The expression:
\begin{align}
\label{eqn:wF}
w(F) &= b(F) - \frac{\sum_j j \cdot p_j}{B}
\end{align}
denotes the \emph{waste per item} under optimal packing. Thus for Linear Waste distributions $w(F) > 0$, while for Perfectly Packable (including Bounded Waste) distributions $w(F)=0$.
\begin{proposition}\label{prop:flowLP_props}
\begin{enumerate}
\item The optimal solution $b(F)$ is convex in $F$. That is, for two distributions $F = \{ p_1, \ldots, p_J \}$, $G = \{q_1, \ldots, q_J\}$, and scalar $0 \leq \lambda \leq 1$:
\[ b\left( \lambda F + (1-\lambda) G \right)  \leq \lambda \cdot b(F) + (1-\lambda) \cdot b(G).\]
\item The expected number of bins in the optimal packing of $F$ is lower bounded as:
\[ \expct{N^{\OPT}_{F,T}} \geq T \cdot  b(F).  \]
\end{enumerate}
\end{proposition}
See Appendix~\ref{sec:proofs} for proof.

Our approach is quite straightforward in hindsight: Rather than learn the distribution $F$ and then solve the bin packing LP as has been done in the past, we transform this LP into an interior-point/Lagrangian type objective function by using barrier/penalty functions for the {\it (no floating items)} constraint. We then perform stochastic gradient descent to solve this problem, (the stochasticity coming from the random arrival of items).
Our algorithms also have a Primal-Dual interpretation (hence the name PD), but unlike typical Primal-Dual algorithms we do not explicitly maintain dual variables. Instead, the duals are implicitly tracked via a map from the primal variables. We explain this in the following subsection where we begin with a general template for design of Interior point based Primal-Dual algorithms to build intuition, and then present our algorithm and guarantees formally. (See Appendix~\ref{sec:OMDinterp} for an interpretation of this algorithm as stochastic mirror ascent for solving the maximization problem dual to $\mathbf{P_{1d-level}}$.) 

\subsubsection{A template for Primal-Dual algorithms}
Consider the following convex minimization problem with a single constr,nt 
\begin{align*}
\mbox{minimize} \  & f(x) \\
\mbox{subject to} \ & g(x) \leq 0,
\end{align*}
where $f,g$ are convex and $f \geq 0$ is bounded from below. The interior point/penalty function approach to optimization is to convert the constrained optimization problem into an unconstrained optimization problem by imposing a smooth convex increasing penalty function $\Phi_\epsilon(\cdot)$ on the constraint and moving this penalty into the objective function:
\[ \mbox{minimize} \quad \mathcal{L}(x) =  f(x) + \Phi_\epsilon(g(x))\]
Given the optimal solution $x_\epsilon^*$ to the above optimization problem, an approximation to the value of the dual $\lambda_g$ for the constraint $g(x) \leq 0$ can be obtained by comparing the first order optimality condition for the unconstrained problem:
\begin{align*}
\left[ \grad{f} +  \frac{\partial \Phi_\epsilon(g)}{\partial g} \cdot \grad g \right]_{x_\epsilon^*} = 0
\end{align*}
to the KKT stationarity condition for the constrained problem:
\begin{align*}
\left[ \grad{f} +  \lambda \cdot \grad g \right]_{x^*} = 0
\end{align*}
as $\lambda \approx \lambda_\epsilon := \left. \frac{\partial \Phi_\epsilon(g)}{\partial g} \right|_{g = g(x_\epsilon^*)}$. More precisely, under an appropriate choice of $\Phi_{\epsilon}(\cdot)$ such as the ones we mention shortly, the following are true:
\begin{enumerate}
\item $\lambda_\epsilon$ is a feasible dual, and $(x^*_\epsilon, \lambda_\epsilon)$ are a primal-dual pair: 
\[ \lambda_\epsilon \geq 0, \quad x^*_\epsilon \in \argmin_{x} f(x) + \lambda_\epsilon \cdot g(x) .\]
\item
\begin{enumerate}
\item Either, $x^*_\epsilon$ is feasible, and $(x^*_\epsilon, \lambda_\epsilon)$ satisfy approximate complementary slackness: 
\[ 0 \geq  \lambda_\epsilon \cdot g(x^*_\epsilon) \geq  - \mO(\epsilon^\gamma)  \mbox{\ \  for some \ \ } \gamma \geq 0,  \]
and as a consequence $x^*_\epsilon$ is approximately optimal:
\[ f(x^*_\epsilon) \leq f(x^*) + \lambda(g(x^*) - g(x^*_\epsilon) \leq f(x^*) + \mO(\epsilon^\gamma).  \]
\item Or, $f(x^*_\epsilon) \leq f(x^*)$, and $x^*_\epsilon$ is approximately feasible: $ g(x^*_\epsilon) = \mO(\epsilon^{\nu})$ for some $\nu \geq 0$ (this uses the assumption that $f$ is bounded from below).
\end{enumerate}
\end{enumerate}

\begin{figure}
\centering
\subfigure[Quadratic: $\Phi_\epsilon(g) = \frac{1}{2\epsilon} \cdot (g^+)^2$]{ \includegraphics[width=2in]{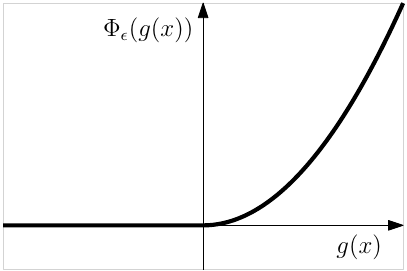}}
\hspace{0.5in} \subfigure[Translated Quadratic: $\Phi_{\epsilon,\eta}(g) = \frac{1}{2\epsilon} \cdot \left( (g+\eta)^+ \right)^2$]{ \includegraphics[width=2in]{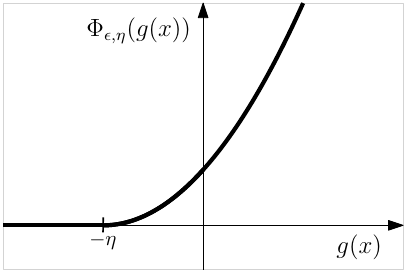}} \\
\subfigure[Exponential: $\Phi_\epsilon(g) = \epsilon \cdot e^{g/\epsilon}$]{ \includegraphics[width=2in]{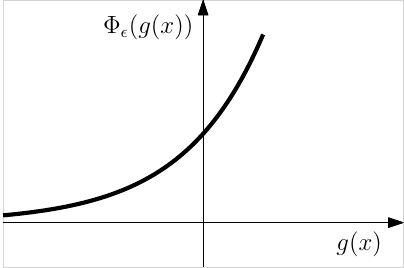}}
\hspace{0.5in} \subfigure[$\log$-barrier: $\Phi_\epsilon(g) = - \epsilon \cdot \log{(-g)}$]{ \includegraphics[width=2in]{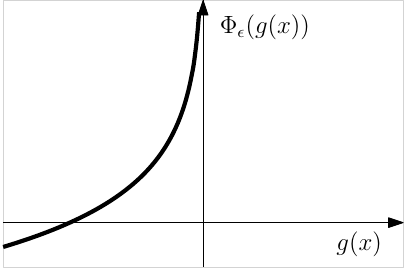}}
\caption{Illustration of four barrier/penalty functions for the constraint $g(x) \leq 0$.}
\label{fig:penalty}
\end{figure}

Depending on the choice of penalty function $\Phi(\cdot)$, there is a full menu of relaxations, each with its own mapping of primal to dual variables (see Figure~\ref{fig:penalty}):
\begin{itemize}
\item Quadratic:
\[ \mathcal{L}^{\mbox{\footnotesize quad}}(x) = f(x) + \frac{1}{2\epsilon}\left( g(x)^+ \right)^2  \hspace{1in} Dual: \lambda_\epsilon = \frac{g(x)^+}{\epsilon} \]
For the dual variables to drive the algorithm, we \emph{must violate the primal constraints}. That is, we always have a primal infeasible solution with $f(x^*_\epsilon) \leq f(x^*)$ but approximately feasible: $g(x^*_\epsilon) \leq \sqrt{2\epsilon (f(x^*) - f(x^*_\epsilon)}$. Primal-dual heuristics with quadratic penalty are common in controlling queueing systems (e.g., \cite{TassiulasEphremides92, Stolyar2005}) because queues essentially are temporary violations of capacity constraints and map to the corresponding duals.
\item Translated Quadratic:
\[ \mathcal{L}^{\mbox{\footnotesize tquad}}(x) = f(x) + \frac{1}{2\epsilon}\left( (g(x)+\eta)^+ \right)^2   \hspace{1in} Dual: \lambda_\epsilon = \frac{(g(x)+\eta)^+}{\epsilon} \]
A small variation over Quadratic penalty which allows duals to be non-zero even when the constraint is satisfied, but yields zero duals when constraints are $\eta$-far from violation. 
In Section~\ref{sec:SSrevisit}, we show that the Primal-Dual algorithms corresponding to Quadratic and Translated Quadratic penalty functions can be interpreted as two different `patches' to the Sum-of-Squares heuristic.
\item Exponential penalty:
\[\mathcal{L}^{\mbox{\footnotesize exp}}(x) = f(x) + \epsilon \cdot e^{\frac{g(x)}{\epsilon}} \hspace{1in} Dual: \lambda_\epsilon = e^{\frac{g(x)}{\epsilon}} \]
The dual variables are always non-zero even when the primal solution is feasible.
 Exponential duals are very popular for worst-case (non-stochastic) online packing and covering problems (e.g., \cite{AwerbuchKhandekar08, PlotkinShmoysTardos95, BuchbinderJainNaor07}), and in prediction with experts' advice. Our main PD heuristic is precisely the Lagrangian relaxation of $\mathbf{P_{1d-level}}$ with exponential penalty. This method is also equivalent to Online Mirror Descent algorithm for maximizing the dual of $\mathbf{P_{1d-level}}$ with entropy regularizer as we explain in Appendix~\ref{sec:OMDinterp}. 
\item $\log$-barrier:
 \[ \mathcal{L}^{\log}(x) = f(x) - \epsilon \cdot \log( - g(x))   \hspace{1in} Dual: \lambda_\epsilon = -\frac{ \epsilon}{g(x)} \]
The solution is constrained to be always primal feasible and approximately optimal: $f(x^*_\epsilon) \leq f(x^*) + \epsilon$. Since, $\log$ is a self-concordant barrier function, $\log$-barriers offer provable convergence guarantees for Newton-Raphson iteration. Therefore they are often used in interior point algorithms for convex optimization (see \cite{BoydVandenberghe}). We will not discuss $\log$-barrier based Primal-Dual heuristics in this paper as they give worse regret guarantees (See Appendix~\ref{sec:generalizedPD} for analysis of $\log$-barrier based Primal-Dual heuristic).  
\end{itemize}
In each case, as $\epsilon, \eta \to 0$, the penalty function approaches the barrier penalty, and $\epsilon, \eta$ control the violation of constraints (for quadratic penalty), or the loss in objective function (for exponential, translated quadratic, and $\log$-barrier).

\subsubsection{From Bin-packing LP to Primal-Dual}
\label{sec:LP_to_PD}

In this Section, we make the mapping of stochastic online packing to Primal-Dual algorithms via Interior Point view more formal. The final algorithm and analysis are presented in Section~\ref{sec:analysis}.

To begin with, we rewrite $\mathbf{P_{1d-level}}$ as follows:
\begin{align*}
\min_{ \mathbf{n} \in \mN} \ &\  \sum_{h=1}^{B} n(h) & \mathbf{(P'_{1d-level})}\\
\mbox{subject to } \ & \  \forall h \in [1,B-1] \ : n(h) \geq 0 
\end{align*}
where the set $\mN$ is given by the constraints:
\[ \bn \in \mN  \iff \exists \ \left\{ v(j,h) \right\}_{j\in [J], h\in [0, B-1]}  \ : \ 
\left\{ \begin{array}{l}
\forall h \in [B]: \ n(h) =  \sum_{j=1}^{h} v(j,h-j) - \sum_{j=1}^{B-h} v(j,h), \\
\forall j \in [J]: \ \sum_{h=1}^{B-j} v(j,h) = p_j, \\
\forall j \in [J], h \geq  B-j+1 : \ v(j,h) = 0, \\ 
\forall j \in [J], h \in  [B-j] : \ v(j,h) \geq  0 .
\end{array} \right. \]
That is, other than the {\it (no floating items)} constraint, we move all the other constraints of $\mathbf{P_{1d-level}}$ into the definition of the feasible set $\mN$.

Applying the interior-point framework, the variable $x$ will correspond to the vector $\bn$, the objective function $f(x)$ will correspond to the objective function $\sum_{h=1}^B n(h)$, and the constraints $g(x)\leq 0$ will correspond to the constraints $-n(h) \leq 0$ of $\mathbf{P'_{1d-level}}$. This gives the penalized Lagrangian:
\[ \mL(\bn) = \sum_{h=1}^B n(h) + \sum_{h=1}^{B-1} \Phi_\epsilon( -n(h) ). \]

The online packing algorithm will correspond to solving for the minimizer of the above penalized Lagrangian via Frank-Wolfe iterations. At iteration $t$, our approximate solution will be $\bn_t := \frac{\bN_t}{t}$. Now in principle we would like to solve for the next iterate as $\bn_{t+1} = \frac{t}{t+1}\bn_t + \frac{1}{t+1}\delta_{t+1}$ where the direction of movement $\delta_{t+1}$ solves:
\begin{align} 
\label{eqn:IP_gradient}
\delta_{t+1} \in 
\argmin_{ \delta \in \mN } \inn{\nabla \mL(\bn_t)}{\delta}
 \end{align}
where $\inn{\cdot}{\cdot}$ denotes inner product. Instead, when the online packing algorithm is presented item $Y_{t+1}$, we solve for:
\[ \bU_{t+1} \in \argmin_{\bU \in \widehat{\mU}(\bn_t, Y_{t+1})} \inn{\nabla \mL(\bn_t)}{ \bC_{Y_{t+1}} \cdot \bU}  \] 
and set $\delta_{t+1} = \bC_{Y_{t+1}} \cdot \bU_{t+1}$ giving $\bn_{t+1} = \frac{\bN_{t}+\bC_{Y_{t+1}}\cdot \bU_{t+1} }{t+1} = \frac{\bN_{t+1}}{t+1}$. Since solving for the minimizer in \eqref{eqn:IP_gradient} decomposes over each item type $j$, we have
\[ \expct{  \bC_{Y_{t+1}} \cdot \bU_{t+1} } \in   \argmin_{ \delta \in \mN } \inn{\nabla \mL(\bn_t)}{\delta}.
\]
Therefore, $\bN_{t+1} - \bN_t = \bC_{Y_{t+1}}\cdot \bU_{t+1}$ can be viewed as a stochastic subgradient of the penalized Lagrangian. The algorithm we present in the next section is only a minor variation on the foregoing discussion: instead of linearizing the penalized Lagrangian, we will instead solve for the minimizer directly:
\[ \bU_{t+1} \in \argmin_{\bU \in \widehat{\mU}(\bn_t, Y_{t+1})} \mL\left( \frac{ \bN_t + \bC_{Y_{t+1}}\cdot \bU }{t+1}\right).  \] 

\subsection{The Primal-Dual algorithm for bin packing}
\label{sec:analysis}
In this section we focus on the Exponential penalty function based interior point relaxation, and formally develop the corresponding Primal-Dual packing algorithm. The interior point relaxation with Exponential penalty function for $\mathbf{P_{1d-level}}$ is given by:
\[ \mL^{exp}(n) = \sum_{h=1}^B n_h + \kappa \epsilon \sum_{h=1}^{B-1} e^{ - n_h /\epsilon}  \]
where $\epsilon, \kappa$ are parameters which we will optimize later. Recall that $n(h)$ denotes the number of level $h$ bins per item, while the total number of bins of level $h$ at time $t$ is denoted by $N_t(h)$. Therefore, $n(h) \sim \frac{N_t(h)}{t}$. Substituting and multiplying throughout by $t$: 
\[ \mL^{exp}(\bN_t) = \sum_{h=1}^B N_t(h) + \kappa \epsilon t \sum_{h=1}^{B-1} e^{ - \frac{N_t(h) }{\epsilon t} } \]
which we write more generally as:
\begin{align}
\mL_t^{exp}(\bN_t) &= \sum_{h=1}^B N_t(h) + \frac{\kappa}{\epsilon_t} \sum_{h=1}^{B-1} e^{ - \epsilon_t N_t(h)  } .
\end{align}
The proposed Primal-Dual algorithm (Algorithm~\ref{alg:PDexp_level}) places arriving items so as to greedily minimize the above penalized-Lagrangian. 
 We discuss the settings for $\kappa$ and $\epsilon_t$ in Theorems~\ref{thm:PDexp_fixed_epsilon}-\ref{thm:PDexp_varying_epsilon}.

\begin{algorithm}[t]
\caption{$\PDexp$ for one-dimensional packing}
\label{alg:PDexp_level}
\algsetblock[Name]{Parameters}{}{0}{}
\algsetblock[Name]{Initialize}{}{0}{}
\algsetblock[Name]{Define}{}{0}{}
\begin{algorithmic}[1]
\Parameters { A sequence $\{\epsilon_t\}$, scalar $\kappa$}
\Initialize { $\bN_0 = \{N_0(1), \ldots, N_0(B)\} = \mathbf{0}$}
\Define { Penalized Lagrangian}
\[ \mL^{exp}_t(\bN) := \sum_{h=1}^B N(h) + \frac{\kappa}{\epsilon_t} \sum_{h=1}^{B-1} e^{-\epsilon_t N(h)}
\]
\For {$t=1,2,\ldots$}
\State Observe item type $Y_t$
\State Choose action to minimize Lagrangian:
\begin{align*}
 \bU_t & \in \argmin_{\bU \in \mU(\bN_{t-1},Y_t)} \mL^{exp}_t(\bN_{t-1}+ \bC_{Y_t} \cdot \bU)
\end{align*}
\State Update state:
\[ \bN_t = \bN_{t-1} + \bC_{Y_t} \cdot \bU_t \]
\EndFor
\end{algorithmic}
\end{algorithm}


In informal terms, in Algorithm~\ref{alg:PDexp_level}, we find the level $h_t$ at which to place the item $Y_t$ where $U_t(h_t) = 1$; this increases $N(h+Y_t)$ by 1 and decreases $N(h)$ by 1 (unless $h=0$ in which case we open a new bin.

\paragraph{Running time:} The total time complexity of each round of $\PDexp$ is is $\mO(B)$ -- there are at most $B$ actions to evaluate, and each action in $\mU(\bN_{t-1},Y_t)$ can be evaluated in $\mO(1)$ time since it only changes two components of the potential next state $\bN_t$. Therefore, overall for $T$ items, the total time taken is $\mO(BT)$. In the bin packing literature on models similar to ours, it is typical to call an algorithm polynomial time if it is polynomial in the bin size $B$ and number of items $T$. In this sense, the $\PDexp$ algorithm is polynomial time.

\begin{figure}[ht]
\hspace{-0.3in}
\begin{minipage}{7.5in}
\subfigure[BW distribution]{\includegraphics[width=2.4in]{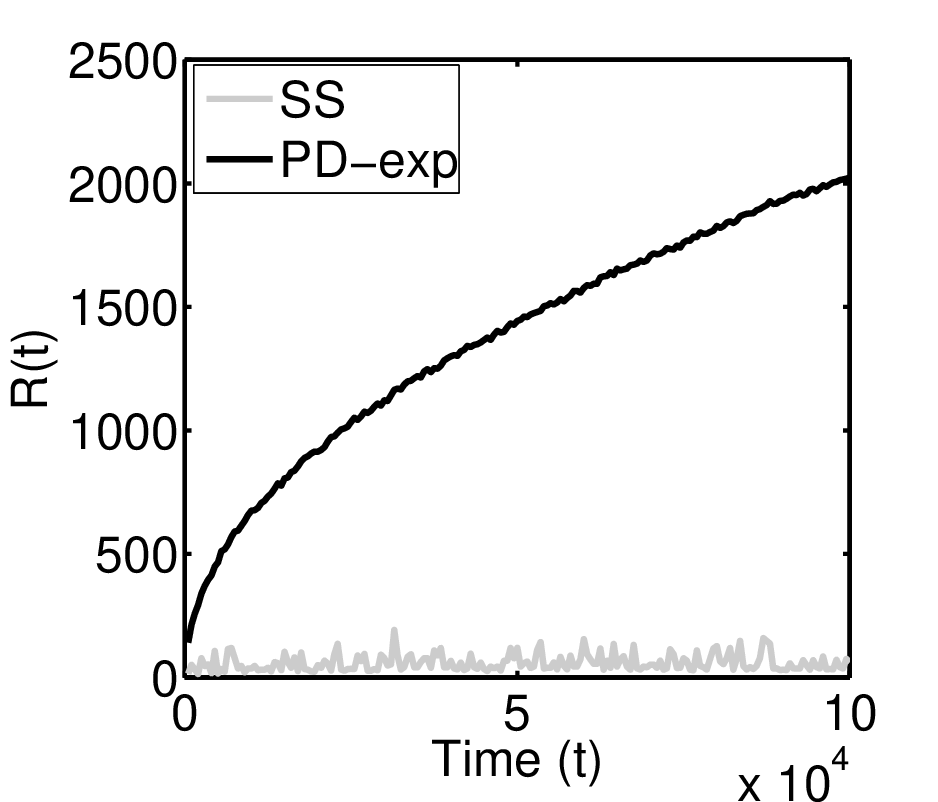}}
\hspace{-0.25in}
\subfigure[PP distribution]{\includegraphics[width=2.4in]{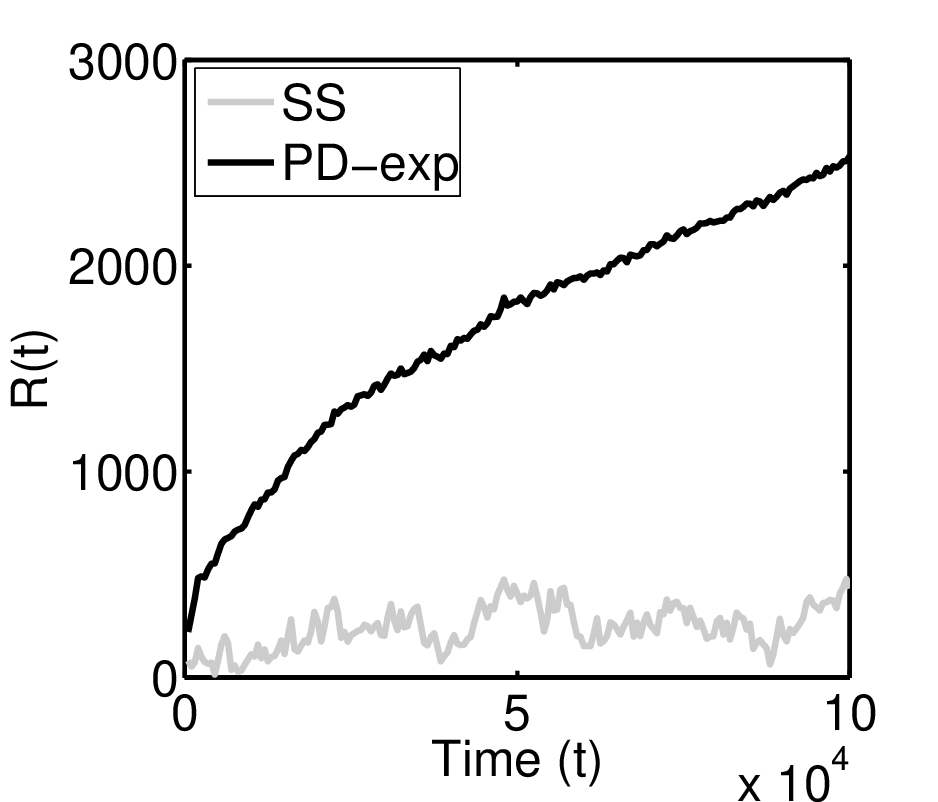}}
\hspace{-0.25in}
\subfigure[LW distribution]{\includegraphics[width=2.4in]{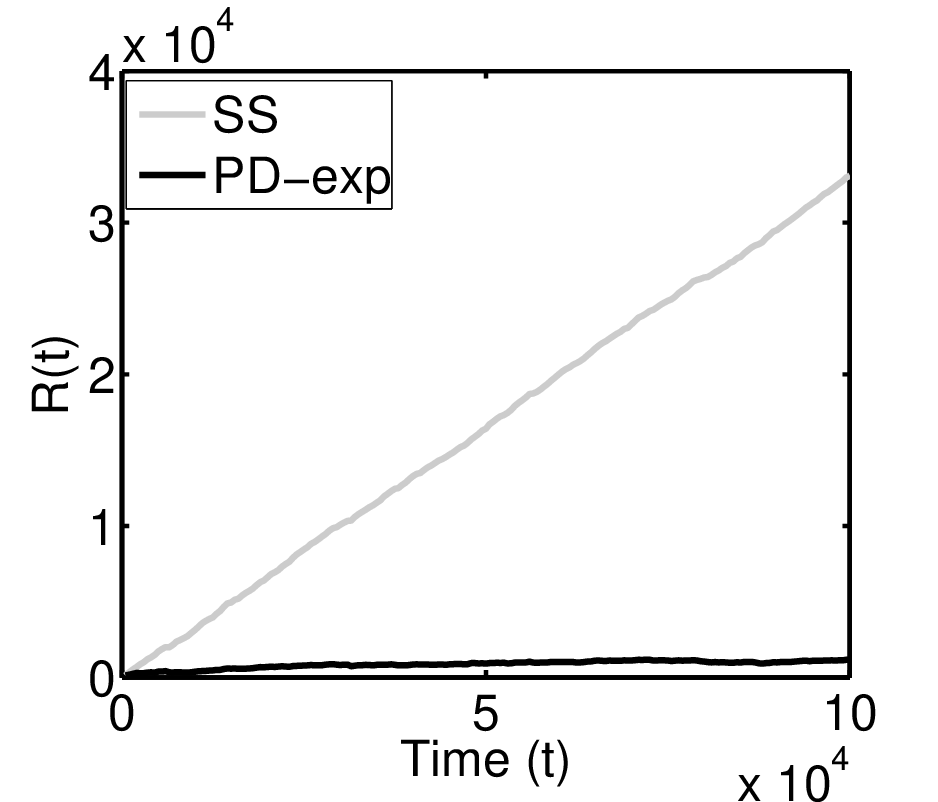}}
\end{minipage}
\caption{Simulation results comparing performance of $\SS$ and $\PDexp$ algorithms for Bounded Waste (BW), Perfectly Packable (PP), and Linear Waste (LW) distributions for one sample path. The $Y$-axis shows the difference between algorithms' waste $W^A_{F,t}$ and the waste of the optimal packing.}
\label{fig:SSvsPD}
\end{figure}

\paragraph{Simulation experiments:} 
Figure~\ref{fig:SSvsPD} shows a comparison of $\SS$ and $\PDexp$ Algorithm~\ref{alg:PDexp_level} for three distributions with the following parameters:
\begin{quote}
\emph{BW distribution}: $B=9$, $F = \left( p_2 = \frac{35}{48}, p_3 = \frac{13}{48}\right)$, \\
\emph{PP distribution}: $B=10$, $F = \left( p_1 = \frac{1}{4},
 p_3 = \frac{1}{4},  p_4 = \frac{1}{8},  p_5 = \frac{1}{4},  p_8 = \frac{1}{8} \right)$ (example borrowed from \cite{SumOfSquares_JACM_CsirikJKOSW06}), \\
\emph{LW distribution}: $B=10$, $F = \left( p_3 = \frac{1}{4}, p_4 = \frac{1}{4},  p_5 = \frac{1}{4},  p_8 = \frac{1}{4} \right)$.
\end{quote}
To highlight $\mO(\sqrt{T})$ regret of $\PDexp$, we have plotted the ``regret'' for the waste metric:
\[ R^A_{F,t}  \doteq W^A_{F,t} - t  \left( b(F) - \sum_j \frac{j}{B} \cdot p_j \right). \]
For BW distribution, recall that $\expct{R^{\SS}_{F,t}} = \expct{W^{\SS}_{F,t}} = \mO(\log t)$, whereas our Primal-Dual algorithm only gets an expected regret $\expct{R^{\PDexp}_{F,t}} = \mO(\sqrt{t})$. However, Primal-Dual is still asymptotically optimal for the metric of number of bins used. For PP distribution (middle figure), both $\SS$ and $\PDexp$ get $\mO(\sqrt{t})$ regret, but $\SS$ outperforms Primal-Dual. For the simulated LW distribution, the regret of $\SS$ is $\Theta(t)$, while it is still $\mO(\sqrt{t})$ for Primal-Dual. Therefore, while $\SS$ is not asymptotically optimal, Primal-Dual is.

\paragraph{Performance analysis:} 
We first analyze the case where we know the total number of arrivals $T$ and $\epsilon_t$ is a constant independent of $t$ but dependent on the horizon $T$. Next we present the more general case when $T$ is not known (\emph{open-ended} bin packing) and $\epsilon_t$ varies with $t$. In both cases, the expected number of bins used by the $\PDexp$ algorithm are $\mO(\sqrt{T})$ larger than the optimal-in-hindsight packing. In Appendix~\ref{sec:generalizedPD}, we prove that the $\mO(\sqrt{T})$ guarantee also holds when we use the quadratic and translated-quadratic penalties in the penalized Lagrangian. We have chosen to focus the presentation on $\PDexp$ algorithm because our experiments suggest that the $\PDexp$ algorithm also gives good performance in the dynamic packing setting where items arrive and leave and the goal is to minimize the steady state number of bins in the packing, while using the quadratic or the translated quadratic penalty does not.

\begin{theorem}[Known horizon $T$]  \label{thm:PDexp_fixed_epsilon}
For $i.i.d.$ item size sequence from distribution $F$, the $\PDexp$ algorithm with $\epsilon_t = \sqrt{\frac{B}{T}}$, $\kappa=1$, guarantees
\[ \expct{N^{\PDexp}_{F,T}} \leq T \cdot b(F) + \sqrt{4 BT } .\]
\end{theorem}

\begin{theorem}[Unknown $T$ (Open Ended)] \label{thm:PDexp_varying_epsilon}
For $i.i.d.$ item size sequence from distribution $F$, the $\PDexp$ algorithm with $\epsilon_t = \sqrt{\frac{B}{2(t+1)}}$, $\kappa=1$, guarantees
\[ \expct{N^{\PDexp}_{F,T}} \leq T \cdot b(F) + \sqrt{8 B T} .\]
\end{theorem}

Since $\expct{N^{\OPT}_{F,T}} \geq T \cdot b(F)$ by Proposition~\ref{prop:flowLP_props}, the above theorems also imply an expected regret bound compared to optimal offline packing. An identical result also holds for waste $\expct{W^{\PDexp}_{F,T}}$. The proofs are follow a potential function analysis approach -- we outline the approach in Section~\ref{sec:proofsketch} where we prove Theorem~\ref{thm:PDexp_fixed_epsilon}. Remainder of the proofs appear in Appendix~\ref{sec:proofs}. A simple application of Azuma-Hoeffding inequality extends our bounds on expected suboptimality in Theorem~\ref{thm:PDexp_varying_epsilon} to high probability bounds.

\begin{theorem}[Martingale concentration] \label{thm:martingale} 
For $i.i.d.$ item size sequence from distribution $F$, the $\PDexp$ algorithm with $\epsilon_t = \sqrt{\frac{B}{2(t+1)}}$, $\kappa=1$, guarantees
for all $t$, and for any constant $\lambda > 0$:
\[ \prob{ N^{\PDexp}_{F,t} \geq t \cdot b(F) +  \sqrt{8Bt} +  \sqrt{ 2 \lambda \delta t \log t} } \leq \frac{1}{t^\lambda} .\]
where $\delta = 4 + \frac{5}{2}\sqrt{B}$.
\end{theorem}

\subsection{Proof of Theorem~\ref{thm:PDexp_fixed_epsilon}}
\label{sec:proofsketch}
Our proof largely follows the analysis technique of {\it drift-plus-penalty} algorithms for control of stochastic networks typified in the monograph by \cite{neely2010stochastic}. 
To illustrate the basic proof technique, we prove Theorem~\ref{thm:PDexp_fixed_epsilon} here. All other proofs are deferred to the appendix. Recall, that for $\PDexp$ algorithm with $\epsilon_t = \epsilon$, the penalized-Lagrangian function is:
\begin{align}
\label{eqn:PDexpL_again}
\mL^{exp}(\bN_t) &= \underbrace{\sum_{h=1}^{B} N_t(h)}_{N_t} + \underbrace{\frac{\kappa}{\epsilon} \sum_{h=1}^{B-1} e^{-\epsilon N_t(h)}}_{V_t}
\end{align}
We will suppress the subscript $t$ in $\mL^{exp}_t$ because in this case $\epsilon_t$ is independent of $t$. We will call $N_t$ the \emph{objective function} term, and $V_t$ the \emph{potential function} term. 
Let $\bN_{t-1}$ be any arbitrary packing of items $\{Y_1,\ldots, Y_{t-1}\}$. Given the arriving item $Y_t$ at time $t$, the action by $\PDexp$ (Algorithm~\ref{alg:PDexp_level}) places the incoming item to minimize the Lagrangian \eqref{eqn:PDexpL_again}:
\begin{align}
\bU_t^{\PDexp} \in \argmin_{\bU \in \mU(\bN_{t-1}, Y_t)} \mL^{exp} \left( \bN_{t-1} + \bC_{Y_t}\cdot \bU \right)
\end{align}
and therefore,
\begin{align}
\mL^{exp}(\bN_{t-1} + \bC_{Y_t} \cdot \bU^{\PDexp}_t) \leq  \mL^{exp}(\bN_{t-1} + \bC_{Y_t} \cdot \bU^A_t)
\end{align}
where $\bU^A_t$ is the action taken by an arbitrary policy $\pi^A$. Subtracting $\mL^{exp}(\bN_{t-1})$ from both sides, we get:
\begin{align}
\label{eqn:Lyapgreedy}
\mL^{exp}(\bN_{t-1} + \bC_{Y_t} \cdot \bU^{\PDexp}_t) - \mL^{exp}(\bN_{t-1}) \leq  \mL^{exp}(\bN_{t-1} + \bC_{Y_t} \cdot \bU^A_t) - \mL^{exp}(\bN_{t-1})
\end{align}

Therefore, it remains to show that for any initial state, there exists a good single step action -- that is, it gives a small value for the right hand side above. The next crucial Lemma proves the existence of a randomized policy $A_F$ defining such a good action. We defer the proof of Lemma~\ref{lem:AFbound} to Appendix. 

\begin{lemma}
\label{lem:AFbound}
For $\{\epsilon_t\} < 1$ non-increasing in $t$, $\kappa \geq 1$, and an arbitrary item size distribution $F$, there exists a distribution-dependent algorithm $A_F$ which for any arbitrary initial packing $\bN$ and item $Y \in \supp(F)$ defines a random single-step action $\bU^{A_F}$ such that:
\begin{align} \label{eqn:AFbound}
\expctsub{ Y  \sim F, \bU^{A_F} \sim \pi^{A_F}(\bN,Y )}{ \mL^{exp}_t( \bN + \bC_{Y} \cdot \bU^{A_F}) } - \mL^{exp}_{t-1}(\bN) &\leq b(F) + \kappa \epsilon_{t-1} + (B-1) \kappa \left(\frac{1}{\epsilon_t} - \frac{1}{\epsilon_{t-1}} \right),
\end{align}
where $b(F)$ is the optimal value of the LP $\mathbf{(P_{1d-level})}$ giving the number of bins used per item under the optimal fractional packing of $F$.
\end{lemma}

With Lemma~\ref{lem:AFbound} in hand, the theorem follows easily. Denoting by $\bN^{\PDexp}_t$ the random packing obtained after packing $\{Y_1,\ldots, Y_t\}$ (suppressing the subscript $F$), we get
\begin{align}
\nonumber & \expctsub{Y_t \sim F}{ \mL^{exp}(\bN^{\PDexp}_t ) - \mL^{exp}(\bN^{\PDexp}_{t-1}) | \bN^{\PDexp}_{t-1} } \\
= & \   \expctsub{Y_t \sim F}{
\mL^{exp}(\bN^{\PDexp}_{t-1} + \bC_{Y_t} \cdot \bU^{\PDexp}_t) - \mL^{exp}(\bN^{\PDexp}_{t-1}) | \bN^{\PDexp}_{t-1}} \\
 \stackrel{\eqref{eqn:Lyapgreedy}}{\leq}  & \
\expctsub{Y_t \sim F, \bU^{A_F}_t}{\mL^{exp}(\bN^{\PDexp}_{t-1} + \bC_{Y_t} \cdot \bU^{A_F}_t) - \mL^{exp}(\bN^{\PDexp}_{t-1} ) | \bN^{\PDexp}_{t-1} } \\
\stackrel{\eqref{eqn:AFbound}}{\leq} & \ 
b(F) + \kappa \epsilon.
\end{align}
A telescoping sum of the above from $t=1$ to $T$ gives,
\begin{align}
\expctsub{\bY}{ \mL^{exp}(\bN^{\PDexp}_T)} - \mL^{exp}(\bN^{\PDexp}_0)
&= \sum_{t=1}^T 
\expctsub{Y_1, \ldots, Y_{t-1}}{ 
\expctsub{Y_{t} \sim F}{ 
\mL^{exp}(\bN^{\PDexp}_t)  
- \mL^{exp}(\bN^{\PDexp}_{t-1})  | \bN^{\PDexp}_{t-1} 
}}   \\
& \leq \sum_{t=1}^{T} \left( b(F) +  \kappa \epsilon \right ) = T \cdot b(F) + T \kappa \epsilon
\end{align}
Therefore,
\begin{align}
\nonumber \expct{N^{\PDexp}_T} = \expct{ \sum_{h=1}^B N^{\PDexp}_T(h) } & =  \expct{\mL^{exp}(\bN^{\PDexp}_T) - \frac{\kappa}{\epsilon} \sum_{h=1}^{B-1} e^{-\epsilon N^{\PDexp}_T(h)}} \\
& \leq T b(F) + T \kappa \epsilon + \mL^{exp}(\bN^{\PDexp}_0) \\
& = T b(F) + T \kappa \epsilon + \sum_{h=1}^B N^{\PDexp}_0(h) + \frac{\kappa}{\epsilon} \sum_{h=1}^{B-1} e^{-\epsilon N^{\PDexp}_0(h)} \\
\intertext{which since $\bN^{\PDexp}_0 = \mathbf{0}$,}
\label{eqn:final_general_bound}  & \leq T b(F) + T \kappa \epsilon + \frac{B \kappa}{\epsilon}
\intertext{Setting $\kappa = 1$ and $\epsilon = \sqrt{\frac{B}{T}}$ we get:}
\nonumber \expct{N^{\PDexp}_T} & \leq T \cdot b(F) + \sqrt{ 4 B T}
\end{align}
as in the theorem statement. $~\Box$

\subsection{Bounded inventory guarantees}
\label{sec:bounded_inventory}

The Primal-Dual algorithm described in Section~\ref{sec:PDlevel} keeps all bins open (potentially $\Theta(T)$ many) for future use, which might be undesirable. We now prove that if instead of $\mO(\sqrt{T})$ regret, we only desire $(1+\delta)$ competitive ratio, then it is sufficient to keep $\mO\left(B/\delta \right)$ bins open in inventory-at-hand. We also prove a lower bound of $\Omega(1/\delta)$ on the number of open bins necessary. 

\begin{definition}[$\eta$-bounded and $\eta$-per-level-bounded inventory algorithms]
Recall that $\widetilde{N}(h) \leq N(h)$ denotes the number of \emph{open} level $h$ bins (eligible for receiving new items). An algorithm is an $\eta$-per-level-bounded inventory algorithm if the number of open bins of each level  is bounded  by $\eta$: $\widetilde{N}_t(h) \leq \eta$ for all $h=\{1,\ldots,B-1\}, t\geq 1$. If a level $h$ bin is created at time $t$ when $\widetilde{N}_{t-1}(h) = \eta$ then the new bin is considered closed. 
An algorithm is $\eta$-bounded inventory algorithm if the total number of open bins is bounded by $\eta$: $\sum_{h=1}^{B-1} \widetilde{N}_t(h) \leq \eta$ for all $t \geq 1$. Therefore, an $\eta$-per-level-bounded algorithm is $\eta (B-1)$-bounded inventory algorithm.
\end{definition}

\begin{algorithm}[t]
\caption{$\eta$-per-level-bounded $\PDtquad$ for one-dimensional packing}
\label{alg:etabounded}
\algsetblock[Name]{Parameters}{}{0}{}
\algsetblock[Name]{Initialize}{}{0}{}
\algsetblock[Name]{Define}{}{0}{}
\begin{algorithmic}[1]
\Parameters { Scalar $\eta$}
\Initialize { Packing state $\bN_0 = \mathbf{0}$; Open bins $\widetilde{\bN}_0 = \mathbf{0}$}
\Define { Penalized Lagrangian}
\begin{align*}
\mathcal{L}_\eta^{tquad}(\bN, \widetilde{\bN}) &= \sum_{h=1}^{B} N(h) + \frac{1}{2\eta} \sum_{h=1}^{B-1} \left( (\eta - \widetilde{N}(h))^+ \right)^2
\end{align*}
\For {$t=1,2,\ldots$}
\State Observe item type $Y_t$
\State Choose action to minimize Lagrangian:
\begin{align*}
 \bU_t & \in \argmin_{\bU \in \mU(\widetilde{\bN}_{t-1},Y_t)} \mL^{tquad}_\eta( \bN_{t-1} + \bC_{Y_t}\cdot \bU , \widetilde{\bN}_{t-1}+ \bC_{Y_t} \cdot \bU)
\end{align*}
\State Update packing state
\[ \bN_t = \bN_{t-1} + \bC_{Y_t} \cdot \bU_t \]
\State Update open bin state
\[ \widetilde{\bN}_t = \left( \widetilde{\bN}_{t-1} + \bC_{Y_t} \cdot \bU_t   \right) \wedge \left( \eta \cdot \mathbf{1} \right) \]
\EndFor
\end{algorithmic}
\end{algorithm}

Algorithm~\ref{alg:etabounded} describes the  $\eta$-per-level-bounded inventory algorithm using translated quadratic penalty function. Informally, Algorithm~\ref{alg:etabounded} finds the level $h_t$ at which the arriving job is placed in one of the open bins by using the translated-quadratic Lagrangian $\mL^{tquad}_\eta$, and then increments the number of bins of level $h_t+Y_t$ by 1. If this causes $\eta+1$ bins to be open at level $h_t+Y_t$, then one of the open bins is irrevocably closed. Algorithm~\ref{alg:etabounded} is written purely in terms of the dynamics of the level summary; indeed which level $h_t+Y_t$ bin is closed, if needed, is immaterial. 

\begin{theorem}[Bounded inventory]\label{thm:bounded_upperbound} For i.i.d. item size sequence from distribution $F$, the $\eta$-per-level-bounded $\PDtquad$ algorithm guarantees
\[ \expct{N^{\PDtquad}_{F,T}}  \leq T \cdot \left( b(F) + \frac{1}{\eta} \right) + \frac{B \eta }{2}.\]
Therefore, $\PDtquad$ is asymptotically $\left( 1+\mO( \frac{1}{\eta}) \right)$ competitive.
\end{theorem}

The next theorem proves that for a general distribution, any $(1+\delta)$ competitive online algorithm must keep $\Omega\left( 1/\delta \right)$ bins open.

\begin{theorem}\label{thm:bounded_lowerbound} For the 1-dimensional bin packing instance with $B=3$, and items of size $1$ and $2$ with probability $p_{1}=p_2 = \frac{1}{2}$, any online $\eta$-bounded algorithm bins must have asymptotic competitive ratio of at least $\left(1+ \frac{1}{12 \eta}\right)$.
\end{theorem}

It would be interesting to close the gap between our upper bound of $\mO(B/\delta)$ open bins and the lower bound of $\mO(1/\delta)$ bins. 

\subsection{Packing continuous item size distributions}
\label{sec:PDcontinuous}

The algorithms demonstrated so far are only valid if the item size distribution has support on integers. In this section we present a simple extension of our Primal-Dual algorithm which gives $\mO(T^{2/3})$ additive suboptimality for distributions which are mixtures of a discrete distribution with integer support and a  continuous distribution with bounded density.

\begin{assumption} The item size distribution $F = F_{disc}+F_{cont}$ where $F_{disc}$ is a discrete distribution function (not necessarily a probability distribution) with support on $\{1,2,\ldots, B-1\}$, and $F_{cont}$ is a continuous distribution function with density bounded by $D$.
\end{assumption}

\begin{definition}
For a discretization parameter $\delta$ and a size $s$, define:
\[ s^\delta = \delta \ceil{s/\delta}  \ , \quad s_{\delta} = \delta \floor{s/\delta},\]
as the rounded-up and rounded-down versions, respectively, of $s$ to the nearest integer multiple of $\delta$. Note that if $\frac{1}{\delta}$ is an integer then for any integer size $s$, $s_\delta=s^\delta = s$.
With $S$ denoting a random variable with distribution $F$, let $F^\delta$ (rounded-up distribution) denote the distribution of $S^\delta$ and $F_\delta$ (rounded-down distribution) denote the distribution of $S_\delta$.
\end{definition}

The extension of $\PDexp$ for packing continuous distributions $F$ satisfying the above assumption is presented in Algorithm~\ref{alg:PDexp_continuous}. The algorithm proceeds in phases of geometrically increasing duration, phase $i$ lasting for $\tau_i = 8^i$ arrivals. During phase $i$, we choose a discretization level of $\delta_i = \Theta\left( \frac{1}{\tau_i^{1/3}} \right)$ so that the number of discretizations in a bin is $B_i = \Theta\left( \tau_i^{1/3} \right)$. Whenever we see an item, we round up its size according to $\delta_i$ of the currently active phase (integer sized items preserve their original size due to our choice of $\delta_i$) and then use the PD-exp algorithm. At the end of a phase, we close all open bins, and start next phase with fresh empty bins.

\begin{algorithm}[t]
\caption{$\PDexpcont$ algorithm for continuous item size distributions}
\label{alg:PDexp_continuous}
\algsetblock[Name]{Parameters}{}{0}{}
\algsetblock[Name]{Initialize}{}{0}{}
\algsetblock[Name]{Define}{}{0}{}
\begin{algorithmic}[1]
\Initialize { Phase index:  $i=0$ }
\Initialize { Phase duration:  $\tau_i=0$ }
\Initialize { Phase start time: $I_i = 0$}
\Initialize { Phase discretization: $\delta_i = 1$ }
\Initialize { Phase bin size: $B_i = B $ }
\Initialize { $\PDexp$ parameter : $\epsilon_i = 1$}
\For {$t=1,2,\ldots$}
\If {$t \geq  I_i + \tau_i$}
\State $ i \leftarrow i+1$; $\tau_i=8^i$; $I_i = t$; $\delta_i = \frac{1}{2^i}$; $B_i = \frac{B}{\delta_i}$; $\epsilon_i = \sqrt{\frac{B}{4^i}}$
\State Close all currently open bins and start a new $\PDexp$ packing instance
\EndIf
\State Observe item size $Y_t$ and calculate $\frac{1}{\delta_i} \cdot Y_t^{\delta_i}$
\State Pack $ \frac{1}{\delta_i} \cdot Y_t^{\delta_i}$ according to $\PDexp$ (Algorithm~\ref{alg:PDexp_level}) with $\epsilon_t = \epsilon_i$, $\kappa = 1$,  and bin size $B_i$
\EndFor
\end{algorithmic}
\end{algorithm}

\begin{theorem}\label{thm:PDexp_continuous} Under the assumption stated on item size distribution $F$, $\PDexpcont$  (Algorithm~\ref{alg:PDexp_continuous}) achieves $\mO((D+\sqrt{B})T^{2/3})$ additive suboptimality compared to optimal offline packing:
\[ \expct{N^{\PDexpcont}_{F,T}}  \leq \expct{N^{\OPT}_{F,T}} + 6(D + 2 \sqrt{B})T^{2/3}. \]
\end{theorem}

The guarantee proved above is tight for our algorithm -- that is, our algorithm can not obtain regret smaller than $\mO(T^{2/3})$. The algorithm of \cite{RheeTalagrand93} for packing $T$ items sampled {\it i.i.d.} from an arbitrary distribution has a regret $\mO(T^{1/2} \log^{3/4}T)$ and therefore we do not offer the best known regret for packing {\it i.i.d.} items. 
We view the two results as somewhat incomparable and complementary. An advantage of \cite{RheeTalagrand93} is that they do not impose any assumption on the distribution of item sizes. On the negative side, they require the knowledge of the horizon $T$, and their guarantees degrade drastically for non-{\it i.i.d.} item sizes while our results for non-stationary inputs proved in Section~\ref{sec:nonstationary} extend to continuous item sizes. 
Obtaining $\widetilde{\mO}(T^{1/2})$ regret for open-ended bin packing with continuous item sizes and favorable performance under non-{\it i.i.d.} item sizes would be significant progress.

%% file: tex/nonstationary.tex
\section{Analysis for non-$i.i.d.$ input sequences}
\label{sec:nonstationary}

As we mentioned in the introduction, one of our motivations behind designing distribution-oblivious online packing algorithms is that we hope they will be robust to non-stationary input sequences.
Our goal in this Section is not to devise new algorithms for non-stationary input models, but show that Algorithm~\ref{alg:PDexp_level} by virtue of being distribution oblivious has good performance under non-stationary input.
 However, unlike performance guarantees proved for \emph{i.i.d.} input against the offline optimal $\expct{N^{\OPT}_{F,T}}$ in Theorem~\ref{thm:PDexp_fixed_epsilon}, the offline optimal turns out to be an unrealistic benchmark in the non-stationary case. Instead, our theorems take the form of a family of upper bounds on $\expct{N^{\PDexp}_T}$ all of which hold simultaneously. The tightest upper bound within this family will depend on the specific instance, and we will not pursue this step in full generality here. 

To keep the exposition simple, in this section we will prove guarantees of the $\PDexp$ heuristic for one specific model of non-{\it i.i.d.} sequences: the {\it independent but not identically distributed} model.


\textbf{Model: } First, an arbitrary sequence of distributions $\mathbf{F}= \{ F_1, F_2, \ldots, F_T \}$ is generated, possibly adversarially. Given $\mathbf{F}$, Nature generates the sequence of item sizes $\mathbf{Y} = \{Y_1, Y_2, \ldots, Y_T\}$ by sampling $Y_t$ from $F_t$ independent of $Y_{s}$ ($s \neq t$).

\noindent {\bf A naive analysis : } If we naively extend the proof of Theorem~\ref{thm:PDexp_fixed_epsilon}, we get that the number of bins used by $\PDexp$ algorithm is bounded by:
\begin{align}
\label{eqn:naivenonstationary}
\expct{N^{\PDexp}_{\{F_t\},T}} &\leq \sum_{t=1}^T b(F_t) + \sqrt{4BT}. 
\end{align}

The following example illustrates that the above bound on the performance of $\PDexp$ can be too pessimistic.

\noindent \textbf{Example:} Consider a packing instance with $B=9, J=3$. The sequence of distributions is $\mathbf{F} = \{ 
 \underbrace{G_2, \ldots , G_2}_{3 T^{1/4}},
 \underbrace{G_3, \ldots, G_3}_{T^{1/4}} ,
 \ldots,
 \underbrace{G_{2}, \ldots , G_{2}}_{3 T^{1/4}}, 
 \underbrace{G_{3}, \ldots , G_{3}}_{T^{1/4}}
 \}$ where $G_2 := \{ p_2 = 1\}$ and $G_3 := \{p_3=1\}$ are deterministic distributions. That is, the sequence $\mathbf{Y}$ has $T^{3/4}/4$ phases of length $4T^{1/4}$ each. Within each phase, the first $3T^{1/4}$ items are of size $2$ and the last $T^{1/4}$ items are of size $3$.

We have $b(G_2)= 1/4$ and $b(G_3) = 1/3$. Therefore, $ \bar{b}(\mathbf{F}) := \frac{1}{T}\sum_{t=1}^T b(F_t) = \frac{13}{48} \approx 0.271$. However, for the \emph{average} distribution $\bar{\bF} := \frac{1}{T} \sum_{t=1}^T F_t = ( p_2 = 3/4, p_3 = 1/4 )$, $b(\bar{\mathbf{F}}) = 1/4 = 0.25$. Therefore we should expect the entire instance to be packed with sublinear waste, even though the distributions for a constant fraction of time steps are Linear Waste distributions (i.e., $G_2$). We will prove that for this instance $\PDexp$ indeed gives this desired result. We should also point out that a naive learning based algorithm which uses change point detection, re-solves the LP for optimal packing, and then packs using the optimal solution of the LP, will incur linear waste.

\noindent {\bf Towards a more sophisticated analysis : } The reason the naive analysis yields a somewhat bad result is apparent from the proof of Theorem~\ref{thm:PDexp_fixed_epsilon}, where the change in Lagrangian at a given time $t$ is upper bounded by the change in Lagrangian due to a probabilistic policy $A_{F_t}$ which depends on the distribution from which the item at time $t$ is generated. Hence the term $b(F_t)$ in the upper bound given in \eqref{eqn:naivenonstationary}. To get a tighter bound it is not sufficient to bound the change in Lagrangian at a {\it deterministic time} $t$; we need to define random times. For example, consider the random time $S_1$ which has a uniform distribution on $[4 T^{1/4}]$. The distribution of the arriving item at the random time $S_1$ is $F_{S_1} = ( p_2 = 3/4, p_3=1/4 )$ with $b(F_{S_1})=0.25 = b(\bar{\mathbf{F}}) < \bar{b}(\mathbf{F})$. Our goal is to bound the change in Lagrangian from time $t=0$ to $t=T$ via such random times whose distributions are smeared out enough so that the average item size distribution at such random times $S_t$ are ``nicer'' -- that is, have small $b(F_{S_t})$. Since $b(F)$ is a convex function of $F$ (Proposition~\ref{prop:flowLP_props}), the larger the support of these random times, lower the $b(F_{S_t})$. But as our main theorem (Theorem~\ref{thm:adversarial_smoothed}) shows, there is a trade-off that must be optimized.

\noindent {\bf Analysis:} As we mentioned, the crux of analysis is to calculate the change in Lagrangian at a random time so that the ``smoothed'' item size distribution at this random time has a smaller objective function value than at a deterministic time. Therefore, we need to define a collection of $T$ distributions for $T$ random times $\{S_1, \ldots, S_T\}$ which must satisfy the condition that the total change in Lagrangian over $\{1,\ldots, T\}$ equals the total change in Lagrangian over $\{S_1, \ldots, S_T\}$. It turns out we do not need the joint distributions of these $T$ random times but only their marginals. Towards this, we begin by defining what we call \emph{$L$-bounded fractional partition} of the time horizon $[T]$.

\begin{definition}[$L$-bounded fractional partition] A collection $\pi = \{ q_1, q_2, \ldots, q_T \}$ of $T$ probability measures, each with support on $[T]$ is called an $L$-bounded fractional partition of the time horizon $\{1,2,\ldots, T\}$ if it satisfies:
\begin{enumerate}
\item {\bf Bounded $L$-Support:} For all $u \in [T]$, $\max \{u : q_t(u)>0 \}-\min\{u : q_t(u)>0\} \leq L-1 $
\item {\bf Unit coverage:} For all $u \in [T]: \ \sum_t q_t(u) = 1$. Equivalently, the matrix $Q = [Q_{ij}] = [q_i(j)]$ is doubly stochastic.
\end{enumerate}
Let $\ell_t = \min\{u: q_t(u) > 0\}$ denote the earliest time $u$ in the support of $t$th window, and without loss of generality assume the windows are indexed so that $\ell_t$ is non-decreasing in $t$. Let $\Pi(L,T)$ denote the set of all $L$-bounded fractional partitions of $[T]$.
\end{definition}

\begin{figure}[ht!]
\centering
\includegraphics[width=5.5in]{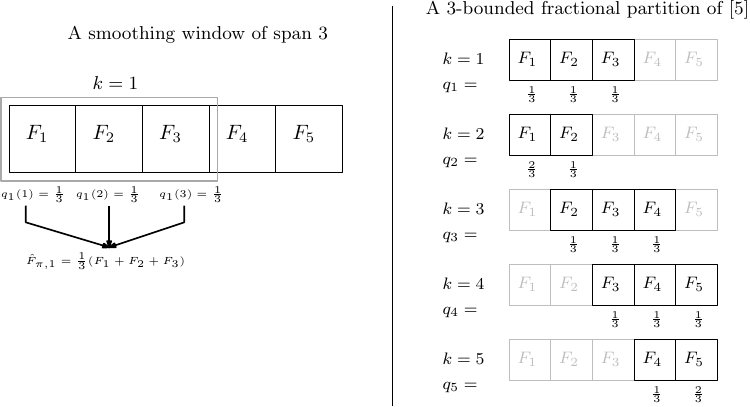}
\caption{An illustration of $L$-bounded fractional partitions of $[T]=\{1,2,\ldots,T\}$.}
\label{fig:partition_example}
\end{figure}

See Figure~\ref{fig:partition_example} for an illustration. In words, window $t$ is defined by a probability measure $q_t$ on $\{1,\ldots, T\}$ which gives the distribution of random time $S_t$ (which we will interchangeably also call the $t$th smoothing window). The $L$-boundedness condition says that no smoothing window can have support on time instants more than $L$ apart, and the distribution for each time instant $t$ is completely partitioned across the $T$ windows in $\pi$. We will assume that the partition is deterministic, that is independent of $\{Y_1,\ldots, Y_T\}$. Indeed, the simplest example is \emph{blocked partitioning} where $q_{iL+1} = q_{iL+2} = \cdots = q_{(i+1)L} = \mbox{Unif}\{iL+1,iL+2, \ldots, (i+1)L\}$, $i=1,\ldots, \frac{T}{L}$.

\begin{definition} Given an $L$-bounded fractional partition $\pi = \{q_t\}_{t\in [T]}$, the $t$th \emph{smoothed item size distribution} is defined to be:
\begin{align}
\hat{F}_{\pi, t} := \sum_{u=1}^T q_t(u) F_t 
\end{align}
Denote the optimal bin rate of $\mathbf{P_{1d-level}}$ with distribution $\hat{F}_{\pi, t }$ by $\hat{b}_{\pi, t} := b(\hat{F}_{\pi,t})$.
\end{definition}

We are now ready to state our main result of this section.

\begin{theorem} \label{thm:adversarial_smoothed}
Consider the independent but non-identical arrival model described above with a packing horizon of $T$ items. 
Let $\epsilon_t = \epsilon$ and $\kappa$ denote the parameters of the penalized-Lagrangian function in Algorithm~\ref{alg:PDexp_level}. 

If $L, \epsilon, \kappa, B$ satisfy: $\epsilon L < 1$, and $L \leq \frac{1}{\epsilon} \log \frac{\kappa B}{B-1}$, then for all $L$-bounded fractional partitions $\pi \in \Pi(L,T)$, the expected number of bins used in packing the sequence $\mathbf{Y}=\{Y_1,\ldots, Y_T\}$ by Algorithm~\ref{alg:PDexp_level} is upper bounded by
\[ \expct{ N^{\PDexp}_T } \leq \sum_{t=1}^T \hat{b}_{\pi,t} +  T \kappa \epsilon \left( L + \frac{3}{2} \right) + \frac{B \kappa }{\epsilon} . \]
As a corollary, setting $\kappa=3$, for $\epsilon L < 1$.
\[ \expct{ N^{\PDexp}_T } \leq \min_{\pi \in \Pi(L,T)} \sum_{t=1}^T \hat{b}_{\pi,t} +  9 T \epsilon L + \frac{3B}{\epsilon} . \]
\end{theorem}

The Theorem gives a family of bounds, one for each choice of $\epsilon, L$ and $L$-bounded fractional partition $\pi$. We do not discuss optimizing the bound over partitions $\pi$ as this will depend on the particular instance, but make two observations: First, since $\Pi(L_1,T) \subset \Pi(L_2, T)$ for $L_1 < L_2$, for any setting of $\epsilon$ as the smoothing window bound $L$ is increased from $\mO(1)$ to $o(1/\epsilon)$ the first term $\sum_t \hat{b}_{\pi,t }$ for the optimal partition $\pi \in \Pi(L,T)$ is non-increasing in $L$. However this improvement comes at the cost of an increased value for the second term $9T \epsilon L$ which is increasing in $L$. Second, while $L$ and $\pi \in \Pi(L,T)$ are purely part of analysis, the choice of $\epsilon$ is part of Algorithm~\ref{alg:PDexp_level} and controls the achievable family of bounds through the constraints $L < \frac{1}{\epsilon}$. A smaller $\epsilon$ thus allows smoothing over larger windows, but comes at a cost of $\frac{3 B}{\epsilon}$ in the performance bound in Theorem~\ref{thm:adversarial_smoothed}.

The proof of Theorem~\ref{thm:adversarial_smoothed} appears in Appendix~\ref{sec:proofs}.
The main idea in a nutshell, again, is to calculate the change in the Lagrangian over a random time distributed according to the measure $q_t$. We can view this as calculating the change in Lagrangian in a single time step with item size distribution being the smoothed distribution $\hat{F}_{\pi,t}$ but where the packing seen by the arriving job is random and potentially correlated with the type of the arriving item. Since the span of the support of the random time is bounded by $L$, the support of the random packing is ``narrow''. This allows us a graceful increase in suboptimality. This argument is similar in spirit to the one carried out by \cite{Neely2009} in a queueing control setting where the environment process is driven by a Markov chain. However, there the author considers a deterministically delayed Lyapunov function to allow the environment at time $t$ to ``de-correlate'' with the state of queues at time $t-L$, and then uses the fact that state at time $t$ is close to the state at time $t-L$.

Returning to our \textbf{example} from the beginning of this Section for illustration, Theorem~\ref{thm:adversarial_smoothed} gives $\expct{N^{\PDexp}_T} \leq 0.25 T + \mO(T^{3/4})$ with $\epsilon = \Theta(T^{-1/2})$. 
Here $0.25 = b(\bar{\bF})$ and hence $\expct{N^{\OPT}_T} \geq 0.25 T$.  
To apply the Theorem, we need to decide the value of $L = o(1/\epsilon)$ as well as a specific $L$-bounded fractional partition. A choice of $L=4T^{1/4}$ allows us to average over the distributions in each phase to get $\hat{F}_{\pi,t} = \bar{\bF}$ for all $t \in [T]$. For the latter, we will use the blocked partitioning scheme mentioned earlier. Substituting in Theorem~\ref{thm:adversarial_smoothed} we get the bound mentioned above. A still smaller regret can be optimized if we can optimize over the choice of $\epsilon$.

The following corollary gives another example where our algorithm gives sublinear regret compared to the offline optimal with only mild sample path assumptions.

\begin{corollary}
\label{cor:CLT_sample_path}
Let the item sequence $\{ Y_1, \ldots, Y_T \}$ be an adversarially generated deterministic sequence, that is, $F_t = \{ p_{Y_t}=1 \}$. Define 
\[ \bar{F}_{s:t} = \frac{1}{t-s+1} \sum_{\tau = s}^{t} F_\tau \] 
as the empirical distribution in the window [t,s]. If there exists a distribution $F$ such that
\[ 
|| F - \bar{F}_{s:s+w-1} ||_{TV} \leq \frac{K}{w^{\alpha}}, \quad  
\forall 1 \leq w \leq T, \  \forall s \leq T-w+1,
\]
for some $\alpha \geq 0$ and constant $K$, then
\begin{enumerate}
\item $\alpha$ unknown: Setting $\epsilon = \Theta \left( T^{-1/2} \right)$ gives $N^{\PDexp}_T \leq N^{\OPT}_T + \mO\left( T^{\frac{2+\alpha}{2(1+\alpha)}} \right)$,
\item $\alpha$ known: Setting $\epsilon = \Theta \left( T^{-\frac{1+\alpha}{1+2 \alpha}} \right)$ gives $N^{\PDexp}_T \leq N^{\OPT}_T + \mO\left(T^{\frac{1+\alpha}{1+2\alpha}}\right)$.
\end{enumerate}
\end{corollary}

\begin{figure}
  \begin{center}
	\includegraphics[width=2.5in,height=1.7in]{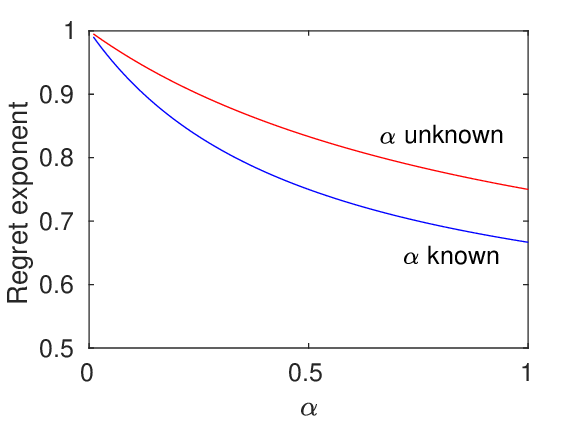}
  \end{center}
 \caption{Regret exponent vs. $\alpha$ for Corollary~\ref{cor:CLT_sample_path}.}
\label{fig:CLT_sample_path}
\end{figure}
\proof{Proof:} For any $L$-bounded blocked partitioning, we get $\hat{b}_{\pi,t} \leq b(F) + \frac{K}{L^{\alpha}}$. In the former case, setting $\epsilon = \Theta(T^{-1/2})$ gives a total regret of $\mO\left( \frac{T}{L^{\alpha}} + L \sqrt{T}  \right)$ which is optimized with $L = \mO\left( T^{\frac{1}{2(1+\alpha)}} \right)$. Since $\epsilon L < 1$ for this choice, we can apply Theorem~\ref{thm:adversarial_smoothed} giving the first result. In the latter case, we will choose $\epsilon = \mO(1/\sqrt{LT})$ and optimize over $L$. With this parametrized choice of $\epsilon$, the total regret is $\mO\left( \frac{T}{L^{\alpha}} + \sqrt{LT} \right)$ which is optimized with $L = \mO\left( T^{\frac{1+\alpha}{1+2\alpha}}\right)$ giving the second result. Figure~\ref{fig:CLT_sample_path} shows the exponent of regret as a function of $\alpha$. Even for very small $\alpha$, which corresponds to sample paths quite far from {\it i.i.d.} input, the Primal-Dual algorithm exhibits $o(T)$ regret.
\Halmos \endproof

%% file: tex/conclusions.tex
\section{Discussion and Open Questions} \label{sec:summary}

Our driving goal in this paper was to design simple, distribution-oblivious and asymptotically optimal algorithms for online stochastic packing. We proved that the bin packing Linear Program can be turned into online packing algorithms which achieve $\mO(\sqrt{T})$ regret for all distribution. The algorithms are simple greedy algorithms which use a penalized Lagrangian of the offline math program similar to interior-point methods. 
Our algorithms can also be viewed as Lyapunov function based control algorithms, and indeed we leverage this interpretation to prove guarantees for non-stationary inputs, which appear to be novel to the best of our knowledge. At the same time, our guarantees are quite naive, and tightening the upper bounds is an interesting line of future work.\\
{\bf Open Questions}
\begin{enumerate}
\item {\bf A poly-logarithmic regret conjecture for $\PDexp$ : } The characterization of distributions as Linear Waste, Perfectly Packable, and Bounded Waste seems discouraging: most distributions have Linear Waste which are bad instances for online packing. We conjecture that from an algorithm design viewpoint, the opposite situation is true -- most distributions are in fact easy to pack and Algorithm~\ref{alg:PDexp_level} can pack them with $\mO(\log^{1+\beta} T)$ regret (for any $\beta>0$) compared to the optimal in hindsight. We first define the distributions for which we conjecture this small regret is achievable:
\begin{definition}[Vertex-Dual distribution] A distribution $F$ is called a vertex-dual distribution if the optimal solution $\bz^*(F)$ to the dual program $\mathbf{D'_{1d-level}}$ (equation~\eqref{eqn:Dprime_1dlevel}) is unique (where we impose $z_j^*(F) = 0$ for $j \notin \supp(F)$). 
\end{definition}
Vertex-Dual distributions are universal in the sense that any random perturbation of $F$ (with no atoms in the perturbing noise) gives a Vertex-Dual distribution with probability 1. Further, they are robust in the sense that if we start with a Vertex-Dual distribution, then a small enough perturbation keeps the distribution Vertex-Dual with the same vertex $\bz^*$. 

For an extreme point $\bz^\dagger$,  let $\mathcal{P}(\bz^\dagger)$ denote the set of all Vertex-Dual distributions with unique optimal solution to the dual as $\bz^\dagger$:
\[ \mathcal{P}(\bz^\dagger) = \{ F |  \bz^*(F) = \bz^\dagger \mbox{ is unique}\}. \]

Let $\mathcal{P}(\bz^{\dagger},\delta)$ denote the set of all Vertex-Dual distributions whose $\delta$ ball is contained in $\mathcal{P}(\bz^\dagger)$:
\[ \mathcal{P}(\bz^\dagger, \delta) = \{ F | F' \in \mathcal{P}(\bz^\dagger) \ \forall\  ||F'-F||_{\infty}\leq \delta \}. \]

We next conjecture that Vertex-Dual distributions are in fact easy to pack online. The conjecture is backed by simulation experiments which we omit here in the interest of space.

\begin{conjecture}[$\mO(\log^{1+\beta} T)$ regret for Vertex-Dual distributions]
\label{conj:cornerpoint}
Let the sequence $\bY = \{ Y_1, \ldots, Y_T\}$ be generated by sampling $Y_t$ from $F_t$ independently, where each $F_t \in \mathcal{P}(\bz^\dagger, \delta)$ for some $\delta> 0$, and some extreme point $\bz^\dagger$ of $\mZ_J$. Then, applying the $\PDexp$ Algorithm~\ref{alg:PDexp_level} with $\epsilon_t = \Theta\left( \frac{1}{\log^{1+\beta} t} \right)$ for any $\beta > 0$,
\[ \expct{N^{\PDexp}_T - N^{\OPT}_T} \leq \mO(\log^{1+\beta} T). \]
\end{conjecture}
\item {\bf Parameter-free online algorithms:} Assuming Conjecture~\ref{conj:cornerpoint}, our distribution oblivious algorithms still require the knowledge of whether the input distribution has a unique dual or not to get the improved $\mO(\polylog T)$ regret for vertex-dual distributions. Whether there are algorithms which can without any tunable parameters simultaneously achieve $\mO(\polylog T)$ regret for vertex-dual distributions, and $\mO(\sqrt{T})$ regret otherwise is also an interesting question.
\item {\bf Mixing learning with $\PDexp$: } Given the promising performance of our distribution-oblivious heuristics, one natural question is \emph{what is the cost of obliviousness, or the value of learning?}
As mentioned earlier, our algorithms do not always achieve the best known regret for {\it i.i.d.} items obtained in the literature by algorithms which use explicit learning, even though the regret we get is always $o(T)$.
We adopted a philosophical stance of distribution-obliviousness because the resulting algorithms are simple, and turn out to be robust to non-stationary input sequences. Quite clearly, the best learning-based approach can only be better than any distribution-oblivious approach since the former can always discard what it has learned. 
Exploring the best marriage of learning-based algorithms with Lyapunov-based control to achieve favorable performance for non-stationary as well as {\it i.i.d.} input sequences is an interesting avenue for future research.
\item {\bf Alternate notions of non-stationary instances:} There is substantial literature on bin packing under adversarial sequences, as well as for \emph{i.i.d.} input sequences. In literature on control of queueing systems and in machine learning, researchers have started exploring input sequences which come from a Markov chain, or which exhibit mixing, so the auto-correlation function decays exponentially fast. In the algorithm design community, in addition to results for random permutation model, other notions of semi-random inputs have been proposed: $t$-bounded adversary of \cite{guha2006approximate},
$\epsilon$-generated random order of \cite{guha2007lower}, and \emph{i.i.d.} input with a fraction corrupted adversarially (e.g., \cite{esfandiari2015online}), to name a few.
One flavor of result we have presented in this paper is for locally perturbed \emph{i.i.d.} sequences. Two questions to explore here are: \emph{(a)} tightening our regret analysis, and \emph{(b)} whether locally perturbed \emph{i.i.d.} sequences are strictly harder than Markov/mixing input sequences (in terms of regret).
However our results also suggest another alternative to analyzing non-stationary instances -- rather than varying the strength of the adversarial instance, we vary the strength of the benchmark we compare the performance against.
In this vein we presented a regret result with respect to sum of LP solutions of local smoothing of the input sequence. A challenging open question to explore here is whether similar regret results can be obtained against online packing oracles which can look ahead some finite time steps into the future.
\item {\bf Boundary between learning-based vs. distribution-oblivious algorithms:} In the introduction we presented what we believe to be fairly weak conditions to call an online algorithm learning-based vs. distribution-oblivious. However, we admit that this distinction continues to be quite subjective and formalizing the notion of learning remains an interesting question. For example, while we classify Sum-of-Squares as learning-based because it uses the empirical distribution to solve a waste LP and inject phantom size 1 items at an optimal rate, presumably this rate can be learned by a procedure that does not track the empirical frequency. Similarly, while the Primal-Dual algorithm does not track the item size distribution, the system state memorizes the duals of the bin packing LP when item sizes are {\it i.i.d.}. At the end of the day, we desire algorithms which are robust to non-stationary inputs as we mention in point 3 above, and the distinction may only be a matter of semantics.
\end{enumerate}

%% file: tex/appendix.tex
\section{Proofs} \label{sec:proofs}


\subsection{Proof of Proposition~\ref{prop:flowLP_props}}

We begin by writing the dual for problem $\mathbf{P_{1d-level}}$ by introducing dual variables $\gamma_h$ for the \emph{no floating items} constraint for level $h$ ($h=\{1,\ldots,B-1\}$), and $z_j$ for \emph{mass balance} for item $j$ ($j \in [J]$).
\begin{subequations}
\begin{align}
\nonumber b(F) &= \max_{ \mathbf{\gamma, z}} \sum_{j=1}^J p_j \cdot z_j & \mathbf{(D_{1d-level})} \\
\label{eqn:Z1} \mbox{subject to} \quad & \gamma_0 := 1 \ , \ \gamma_B := 0  \ , \\
\label{eqn:Z2} & \forall j \in [J], h \in \{0,\ldots, B-j\} \ : \ z_j \leq \gamma_h - \gamma_{h+j}  \ , \\
\label{eqn:Z3} & \forall h \in [1,B-1] \ : \ \gamma_h \geq 0. 
\end{align}
\end{subequations}
We can express the above math program more abstractly as 
\begin{align}
\label{eqn:Dprime_1dlevel}
b(F) &= \max_{ \mathbf{z} \in \mZ_J} \sum_{j \in [J]} p_j z_j  & (\mathbf{D'_{1d-level}})
\end{align}
where the feasible polytope $\mZ_J$ is given by \eqref{eqn:Z1}-\eqref{eqn:Z3} and is independent of the distribution $F=(p_1,\ldots,p_J )$. Since for each $\mathbf{z} \in \mZ_J$, $\sum_{j \in [J]} p_j z_j$ is a linear function of $F$, and the supremum of linear functions is convex, $b(F)$ is a convex function of $F$ proving the first part.

For the second part, consider the optimal packing of $T$ items with $P_j$ items of size $j$, $\sum_{j \in [J]} P_j = T$. The number of bins in the optimal packing is lower bounded by the LP relaxation $b( P_1, \ldots, P_J)$. Now assume, $(P_1, \ldots, P_J)$ is obtained via $T$ i.i.d. samples from $F = (p_1, \ldots, p_J)$. Then,
\[ \expct{N^{\OPT}_{F,T}} \geq \expct{ b( P_1, \ldots, P_J) } \geq b(\expct{(P_1, \ldots, P_J)}) = T\cdot b(F),  \] 
where the first inequality follows from the fact that $b(F)$ is the optimal value of an LP relaxation (in fact this is an equality since there exists and optimal solution to $\mathbf{P_{1d-level}}$ with integral $\mathbf{v}$ in this case), and the second inequality follows from convexity of $b(\cdot)$.


\subsection{Proof of Lemma~\ref{lem:AFbound}}

\textbf{Construction of policy $A_F$ : } To define $A_F$, we will modify the construction used in \cite{SumOfSquares_JACM_CsirikJKOSW06} which was given only for the case when the distribution $F$ is perfectly packable or bounded waste. The $A_F$ policy is constructed in five steps, only step 4 differs from \cite{SumOfSquares_JACM_CsirikJKOSW06}:

\begin{enumerate}
\item {\bf Solving LP  $(\mathbf{P_{conf}})$ : } Before the item to be packed is presented, solve for an optimal fractional packing of $F$ via the following Linear Program:
\begin{align*}
\balpha^* &\in \argmin_{ \balpha \in [0,1]^\mC } \sum_{c \in \mC} \alpha_c & \mathbf{(P_{conf})}\\
\mbox{subject to } \quad 
 & \forall j \in [J] \ : \ \sum_{c} \alpha_c x_{c}(j) = p_j.
\end{align*}
Recall that $\mC$ denotes the set of feasible configurations of a bin of capacity $B$, and $c\in \mC$ is represented by $\bx_c = ( x_c(1),\ldots, x_c(J) )$ with $x_c(j)$ representing the number of items of size $j$ in $c$. 
The variable $\alpha^*_c$ in the optimal solution represents the fractional number of bins of configuration $c$ in an optimal fractional packing of $F=(p_1, \ldots, p_J)$. While the optimal solution may be non-unique, we have that for any optimal solution $\balpha^*$
\[  b(F) = \sum_{c \in \mC} \alpha^*_c. \]
Our Primal-Dual algorithm never actually solves for this packing $\balpha^*$. We only use it as an analysis tool to bound the change in the Lagrangian via policy $A_F$. \\
\textbf{Example:} Consider $B=9, J=3$ and $F = \{p_2=p_3=1/2\}$. In this case the (unique) optimal fractional packing has support on two configurations $x_1 = (0,0,3)$ and $x_2=(0,3,1)$ corresponding to bin with three size 3 items and bin with three size 2 and one size 3 item, respectively. The optimal solution to $\mathbf{(P_{conf})}$ will be $\balpha^* = ( \alpha^*_1 = 1/9, \alpha^*_2 = 1/6 )$.

\item {\bf Sampling item $Y$:} Once an optimal solution to $\mathbf{(P_{conf})}$ is found, the item $Y$ sampled from $F$ is presented to the online algorithm to pack. It will be crucial that $\balpha^*$ and $Y$ are mutually independent if $\mathbf{(P_{conf})}$ does not have a unique optimal solution. \\
{\bf Example (contd.):} For illustrative purposes, let $Y=3$.
\item {\bf Sampling configuration $X$:} On seeing item $Y$, the policy maps it to a random configuration $X \in \mC$ which is part of the optimal fractional packing such that
\begin{align}
\label{eqn:Xt_given_Yt}
\prob{ X = c | Y = j} = \frac{ \alpha^*_c \cdot x_c(j)  }{p_j}.  
\end{align}
Intuitively, assuming all $\alpha^*_c$ are rational numbers, if one imagines a packing with $n\cdot b(F)$ bins (for $n$ large enough) $\alpha^*_c/b(F)$ fraction of which are in configuration $c$ (for all $c\in \mC$) and then samples a random item out of the $n$ total items in the packing, then conditioned on the item being of type $Y$, the probability that the item was packed in a bin in configuration $X=c$ is given by \eqref{eqn:Xt_given_Yt}.

To see that for any $Y$, this gives a distribution on $\mC$, 
\begin{align}
\sum_{c \in \mC} \prob{ X = c | Y = j} = \frac{ \sum_{c \in \mC}  \alpha^*_c \cdot x_c(j)  }{p_j} = 1,
\end{align}
where the last equality follows from the constraints of $\mathbf{P_{conf}}$.

{\bf Example (contd.):} Given $Y=3$, the $A_F$ policy samples a configuration $X$ with the probability of picking configuration 1 equal to $\frac{x_1(3) \cdot \alpha^*_1}{p_3}= \frac{3 \cdot 1/9}{1/2} = 2/3$, probability of picking configuration 2 equal to $\frac{x_2(3) \cdot \alpha^*_2}{p_3} = \frac{1 \cdot 1/6}{1/2}=1/3$.
\item {\bf Ordering items in $X$:} Let $|X| = \sum_{j \in [J]} {x_{X}(j)}$ denote the total number of items in configuration $X$. Given the current arbitrary packing $\bN$ and the configuration $X$, find an ordering $\mJ_X=(j_1,j_2,\ldots, j_{|X|})$ of the items in $X$, and a threshold index $last(X), \ 0 \leq last(X) \leq |X|$, such that :
\begin{enumerate}
\item $\bN$ has bins with level $j_1$, $j_1+j_2$, $\ldots$, $j_1+j_2+\cdots+j_{last(X)}$.\\
(Except: if $N(B)=0$ and $\sum_{k=1}^{|X|} j_k=B$, then it is acceptable to have $last(X) = |X|$.)
\item $\bN$ has no bins of levels $j_1+\cdots+j_{last(X)}+j_k$ for any $k > last(X)$.
\end{enumerate}
We point out the differences compared to \cite{SumOfSquares_JACM_CsirikJKOSW06}: since they only consider the case of perfectly packable distributions, necessarily $\sum_{k=1}^{|X|} j_k = B$. The definition of  $last(X)$ there is almost exactly the same as (a) and (b) above except that ``$\bN$ has (has no) bins'' is replaced by ``$\bN$ has (has no) {\it partially filled} bins.'' Therefore $last(X) < |X|$ in \cite{SumOfSquares_JACM_CsirikJKOSW06}. To study linear waste distributions where bins in optimal packing can be partially full, we must consider configurations $X$ where $\sum_{k=1}^{|X|} j_k < B$ as we do above.

To aid subsequent analysis, define $h_k$ for $0 \leq k \leq last(X)$ as:
\[h_0=0 ; \quad h_k = h_{k-1} + j_k. \]
The ordering can be achieved as follows: Start with $h_0=0$. Pick any item $j'$ in $X$ such that $N(j')>0$. If there is such an item, then $j_1=j$ and $h_1=h_0+j_1$. Otherwise $last(X)=0$, we pick any random ordering of remaining items $j_1,\ldots, j_{|X|}$ and terminate. If a $j_1$ was found, we continue by picking any item $j'$ in $X$ excluding $j_1$ such that $N(j_1+j') > 0$. If there is such an item, then $j_2=j'$, $h_2=h_1+j_2$. Otherwise $last(X)=1$ and we pick an arbitrary  ordering $j_2, \ldots, j_{|X|}$ of the remaining items in $X$ and terminate. We continue in this manner until either all item in $X$ have been used (in which case $last(X)=|X|$), or the second condition above is met. \\
{\bf Example (contd.):} Suppose $Y=3$, and the configuration $X$ is configuration 2 ($\bx_2=(0,3,1)$). Let $\bN = (0, 1, 1, 1, 1, 1, 0, 1, 1)$. That is, the initial packing has non-zero bins (in fact 1 bin) for levels $2,3,4,5,6,8,9$. One valid ordering of the items in configuration $X$ is $(2,2,2,3)$ with $last(X)=4$ -- $\bN$ has bins of levels $2,2+2,2+2+2,2+2+2+3$. Another valid ordering is $(2,3,2,2)$ with $last(X)=2$ -- $\bN$ has bins of levels $2, 2+3$ but no bins of level $2+3+2$. The policy $A_F$ can pick any valid ordering as long as it obeys the two conditions mentioned above. Let us assume that in this case we pick the second ordering so that $h_0=0, h_1 = 2, h_2 = 5$ and $last(X)=2$.

\item {\bf Placing item $Y$:} Map the arriving item $Y$ to one of the $x_X(Y)$ items in the ordering $\mJ_X = ( j_1,\ldots, j_{|X|} )$ of size $Y$ uniformly at random. Let $K$ denote the index of this item in the ordering $\mJ_X$. If $K \leq last(X)$, the arriving item is placed in a bin of level $h_{K-1}$ to create a bin of level $h_{K} = h_{K-1}+j$. If $K > last(X)$, then the arriving item is placed in a bin of level $h_{last(X)}$. That is,
\begin{align*}
\bU^{A_F} &=  \begin{cases}
\be_{h_{K-1}} , &   K \leq last(X), \\
\be_{h_{last(X)}} ,  & \mbox{otherwise}.
\end{cases}
\end{align*}
or more succinctly: $\bU^{A_F} = \be_{ \min\{ h_{K-1} , h_{last(X)} \} }$.\\
{\bf Example (contd.):} There is only one item of size $3$ in configuration 2. The arriving item is therefore mapped to the second item in ordering $(2,3,2,2)$. Since $K=2 \leq last(X)$, there is a bin of level $h_{K-1}=h_1 = 2$ in $\bN$. The arriving item of size $3$ is placed in a bin of level $2$ to increase its level to $5$. The new packing would be $ \bN' = \bN + \bC_{3} \cdot \be_{2} = (0, 0, 1, 1, 2, 1, 0, 1, 1)$.
\end{enumerate}
\ \\
{\bf An equivalent view for generating $(Y,X)$} \\
In the construction of $A_F$ policy above, the configuration $X$ was sampled after the item size $Y$ was observed. We now present an alternate method to generate the pair $(Y,X)$ with the same joint distribution as above. We will find this view helpful in analyzing the policy $A_F.$
\begin{enumerate}
\item {\bf Solving LP $\mathbf{(P_{conf})}$:}  As before, this produces an optimal fractional packing $\balpha^*$ \\
{\bf Example (contd.):} For our running example, $\balpha^* = (\alpha_1^*=1/9, \alpha^*_2 = 1/6)$ where configuration 1 is $x_1= (0,0,3)$ and configuration 2 is $x_2 = (0,3,1)$.
\item {\bf Sampling configuration $X$:} Configuration $X$ is sampled (independent of the initial packing $\bN$) so that:
\begin{align}
\label{eqn:Xt}
\prob{ X = c} = \alpha^*_c \left( \sum_{j \in [J]} x_{c}(j) \right).
\end{align}
{\bf Example (contd.):} The probability of sampling configuration 1 is $\alpha^*_1 \times 3 = 1/3$, and the probability of sampling configuration 2 is $\alpha^*_2 \times 4 = 2/3$. Let us assume that this step generates configuration 2.
\item {\bf Ordering items in $X$:} As before, given the initial packing $\bN$ and configuration $X$, we order the items in $X$ to generate $\mJ_X = ( j_1, \ldots , j_{|X|})$ and index $0 \leq last(X)\leq |X|$ satisfying the two properties mentioned in Step 4 previously. \\
{\bf Example (contd.):} We assume as before that $\bN=(0,1,1,1,1,1,0,1,1)$ so that one valid ordering of items in configuration 2 is $(2,3,2,2)$ with $last(X) = 2$.
\item {\bf Sampling item $Y$:} Next, we sample $K$ uniformly in the set $\{1,\ldots,|X|\}$ and item $Y$ is chosen as $j_K$; the $K$th item in the ordering $(j_1,\ldots, j_{|X|})$ of configuration $X$, so that:
\begin{align}
\label{eqn:Yt_given_Xt}
\prob{Y = j | X = c} = \frac{x_c(j)}{|X|}. 
\end{align}
The above procedure ensures that the marginal distribution of $Y$ is indeed $F$:
\[ \prob{ Y = j} = \sum_{c \in \mC} \prob{Y =j | X = c} \cdot \prob{X = c} = \sum_{c \in \mC} \frac{x_{c}(j)}{ \sum_{k \in [J]} x_{c}(k) } \cdot \alpha^*_c \left( \sum_{k \in [J]} x_{c}(k) \right) = \sum_{c \in \mC} \alpha^*_c x_{c}(j) = p_j.  \]
{\bf Example (contd.):} Given $X$ was configuration 2, and the ordering is $(2,3,2,2)$, $K$ is sampled uniformly in $[4]$. If $K=2$, this gives $Y = j_2 = 3$.
\item {\bf Placing item $Y$:} Given configuration $X$ and ordering $\mJ_X$, index $K$ and item $Y = j_K$ in ordering $\{j_1,\ldots, j_{|X|}\}$, the item is placed as described in Step 5 previously: $\bU = \be_{ \min\{ h_{K-1}, h_{last(X)}\}}$.
\end{enumerate}

Both views above, either sampling $Y \sim F$ first and then $X$ according to \eqref{eqn:Xt_given_Yt} or sampling configuration $X$ first according to \eqref{eqn:Xt} and then $Y$ according to \eqref{eqn:Yt_given_Xt}, produce exactly the same joint distribution on $(Y,X)$. For the former, we get
\begin{align*}
\prob{ X = c , Y = j } = \prob{X = c | Y = j} \cdot \prob{Y=j} \stackrel{\tiny \eqref{eqn:Xt_given_Yt}}{=} \frac{ \alpha^*_c \cdot x_c(j) }{p_j} \cdot p_j = \alpha^*_c \cdot x_c(j),
\end{align*}
and for the latter,
\begin{align*}
\prob{X = c , Y=j} = \prob{ Y=j | X = c } \cdot \prob{X=c} \stackrel{\tiny \eqref{eqn:Xt},\eqref{eqn:Yt_given_Xt}}{=}  \frac{x_c(j)}{|X|} \cdot \alpha^*_c \left( \sum_{j \in [J]} x_c(j) \right) = x_c(j) \cdot \alpha^*_c.
\end{align*}
Crucially, the marginal distribution of $Y$ is $F$, and $Y$ is independent of the initial packing $\bN$. We will use the latter view for analysis of $A_F$, and we reiterate, adopting the latter view has no bearing on the algorithm at all and is purely for ease of analysis.

\ \\
{\bf Analysis of policy $A_F$ : }
Denote
\[ \bN' = \bN + \bC_Y \cdot \bU \]
as the state after item $Y$ is packed in $\bN$ using action $\bU$.  
Our goal is to bound for $Y\sim F$, $\bU \sim \pi^{A_F}(\bN,Y)$ the difference:
\begin{align}
\mL^{exp}_{t}(\bN') - \mL^{exp}_{t-1}(\bN) 
&= \sum_{h=1}^B (N'(h) - N(h)) + \sum_{h=1}^{B-1} \left(\frac{\kappa}{\epsilon_t} e^{-\epsilon_t N'(h)} - \frac{\kappa}{\epsilon_{t-1}} e^{-\epsilon_{t-1} N(h)} \right) \\ 
&= \sum_{h=1}^B (N'(h) - N(h)) 
+ \sum_{h=1}^{B-1} \left(\frac{\kappa}{\epsilon_{t-1}} e^{-\epsilon_{t-1} N'(h)} - \frac{\kappa}{\epsilon_{t-1}} e^{-\epsilon_{t-1} N(h)} \right) \\ 
& \qquad + \sum_{h=1}^{B-1} \left(\frac{\kappa}{\epsilon_t} e^{-\epsilon_t N'(h)} - \frac{\kappa}{\epsilon_{t-1}} e^{-\epsilon_{t-1} N'(h)} \right) 
\end{align}
Since $\frac{\partial}{\partial n}\left( \frac{e^{- \epsilon_t n}}{\epsilon_t} - \frac{e^{-\epsilon_{t-1} n}}{\epsilon_{t-1}} \right) = e^{-\epsilon_{t-1}n} - e^{-\epsilon_t n } \leq 0$ when $\epsilon_t \leq \epsilon_{t-1}$ as we have assumed, it is a decreasing function in $n$ maximized for $n=0$ and thus at most $ \frac{1}{\epsilon_{t}}-\frac{1}{\epsilon_{t-1}}$. Therefore,
\begin{align}
\mL^{exp}_{t}(\bN') - \mL^{exp}_{t-1}(\bN) 
& \leq \sum_{h=1}^B (N'(h) - N(h)) 
+ \sum_{h=1}^{B-1} \left(\frac{\kappa}{\epsilon_{t-1}} e^{-\epsilon_{t-1} N'(h)} - \frac{\kappa}{\epsilon_{t-1}} e^{-\epsilon_{t-1} N(h)} \right) \\ 
\nonumber & \qquad + (B-1)\kappa \left( \frac{1}{\epsilon_{t} } - \frac{1}{\epsilon_{t-1}} \right) \\
&= \mL^{exp}_{t-1}(\bN')-\mL^{exp}_{t-1}(\bN) + (B-1) \kappa \left( \frac{1}{\epsilon_{t}} - \frac{1}{\epsilon_{t-1}} \right).
\end{align}

In the rest of the proof, we will bound the first two terms in the expression for the policy $A_F$. Since these terms only depend on $\epsilon_{t-1}$, let $\epsilon_{t-1} = \epsilon$, and define:
\begin{align}
\Delta \mL(Y,\bU) &:= \mL^{exp}_{t-1}( \bN')  - \mL^{exp}_{t-1}(\bN)  \\
&= \underbrace{N' - N}_{=: \Delta N} + \sum_{h=1}^{B-1} \underbrace{ \frac{\kappa}{\epsilon} \left( e^{ - \epsilon N'(h)}  - e^{-\epsilon N(h)}\right) }_{=: \Delta V_h}
\end{align}

Our goal is to prove 
\begin{align}
\label{eqn:AFbound2}
\expctsub{ Y \sim F, \bU \sim \pi^{A_F}(\bN,Y )}{ \Delta \mL(Y, \bU) } &\leq b(F) + \kappa \epsilon.
\end{align}
Recalling the construction of policy $A_F$ which induces a joint distribution on (item, configuration) $(Y,X)$, we will write the above by first conditioning on the configuration $X$ and the ordering $\mJ_X = (j_1,\ldots, j_{|X|})$ of items in $X$, and then the item type $Y = j_K$ for $K \sim \mathsf{Unif}[|X|]$:
\begin{align}
\label{eqn:alt_bound}
\expctsub{ Y \sim F , \bU \sim \pi^{A_F}}{ \Delta \mL(Y, \bU) } &= 
\expctsub{X, \mJ_X}{  \expctsub{ K \sim \mathsf{Unif}[|X|] } { \left. \Delta \mL(j_K, e_{\min\{ h_{K-1}, h_{last(X)}\}}) \right| X , \mJ_X } } 
\end{align}
We will prove that:
\begin{align}
\label{eqn:perbinbound}
\expctsub{ K \sim \mathsf{Unif}[|X|] }{ \left. \Delta \mL(j_K, e_{\min\{h_{K-1} , h_{last(X)}\}}) \right| X , \mJ_X  } & \leq \frac{1}{ |X| } + \kappa \epsilon = \frac{1}{\sum_{j \in [J]} x_X(j) } + \kappa \epsilon.
\end{align}
Given \eqref{eqn:alt_bound}-\eqref{eqn:perbinbound}, the bound \eqref{eqn:AFbound2} follows by unconditioning on the configuration $X$: Noting that $\prob{X=c} = \alpha^*_c \cdot \sum_{j \in [J]} x_{c}(j) $:
\begin{align}
\expctsub{ Y \sim F , \bU \sim \pi^{A_F}}{ \Delta \mL(Y, \bU) } 
& \leq \expctsub{X}{ 
	\frac{1}{ \sum_{j \in [j]} x_{X}(j) } + \kappa \epsilon 
} \\
 &= \sum_{c \in \mC } \prob{X = c} \cdot \left(  
	\frac{1}{ \sum_{j \in [j]} x_{c}(j) } \right) + \kappa \epsilon  \\
 & = \sum_{c \in \mC} 
\alpha^*_i \sum_{j \in [J]} x_{c}(j) \cdot \left( \frac{1}{\sum_{j \in [J]} x_{c}(j)} \right) + \kappa \epsilon
\\
& = \sum_{c \in \mC} \alpha^*_c + \kappa \epsilon \\
&= b(F) + \kappa \epsilon
\end{align}
which gives the required bound claimed in \eqref{eqn:AFbound}.


To prove inequality \eqref{eqn:perbinbound}, fix a configuration $X$, an ordering $\mJ_X = (j_1, \ldots, j_{|X|})$ along with $last(X)$. Let $K$ be uniformly distributed in $\{1, \ldots, |X|\}$, and $Y = j_K$. 


We split the analysis depending on whether $last(X)=0$ or $last(X) > 0$:
\begin{enumerate}
\item Case $last(X) = 0$ : Recall that this implies that there is no bin of height $j_1, j_2, \ldots, j_{|X|}$. Thus a new bin is opened with probability $1$, $Y = j_K$ is packed at level 0 in this bin, and the level of the newly opened bin becomes $j_K<B$. Crucially, $N(j_K)=0$ before the new arrival was packed. Therefore further conditioning on $K$, the change in Lagrangian is upper bounded by:
\begin{align}
\nonumber \expct{\Delta \mL | X, \mJ_X, K} = 1 + \frac{\kappa}{\epsilon}\left( e^{-\epsilon N'(j_K)} - e^{-\epsilon N(j_K)}\right) & =  1 + \frac{\kappa}{\epsilon} (e^{-\epsilon} - 1)   = 1 - \kappa \left(1 - \frac{\epsilon}{2!} + \frac{\epsilon^2}{3!} - \ldots \right) \\
 & \leq 1 - \kappa \left( 1 - \frac{\epsilon}{2}  \right) \ \leq \ \kappa \frac{\epsilon}{2}
\label{eqn:case1subcase1} 
\end{align}
where the first inequality follows under the assumption $\epsilon < 1$, and then second under the assumption $\kappa \geq 1$. Finally, by removing the conditioning on $K$, when $last(X)=0$:
\begin{align}
\label{eqn:DeltaLcase1}
\expctsub{K \sim \mathsf{Unif}[|X|]}{\Delta \mL  | X, \mJ_X } & \leq \kappa \frac{\epsilon}{2}.
\end{align}
\item Case $last(X) > 0$ : In this case, there are bins of levels $j_1, j_1 + j_2, \ldots, j_{1}+ \cdots + j_{last(X)}$ in the existing packing. The change in Lagrangian depends on whether or not the index $K$ of the arriving item is strictly larger than $last(X)$:
\begin{enumerate}
\item Subcase $K > last(X)$: In this case, the item is packed at level $ h_{last(X)}$, with $ n := N(h_{last(X)}) \geq 1$ so $\Delta N = 0$ since no new bin is created. Further, since $N(h_{last(X)} + j_K) = 0$, we next prove that the potential function term does not increase either: 
\begin{align}
\nonumber
& \expct{\Delta \mL | X, \mJ_X, K > last(X)}  \\
\nonumber
= & \ \frac{\kappa}{\epsilon}
\left[ 
\left( 
e^{-\epsilon (N(h_{last(X)}+j_K)+1)}  
- e^{-\epsilon N(h_{last(X)}+j_K)}  
 \right)
+ \left( 
e^{-\epsilon (N(h_{last(X)})-1)} 
- e^{-\epsilon N(h_{last(X)}) } 
 \right) \right] \\
\nonumber
 = & \ \frac{\kappa}{\epsilon} \left[ \left(e^{-\epsilon} -1 \right) + \left(e^{- (n-1) \epsilon} - e^{n \epsilon }\right)  \right] \\
\nonumber
 = & \ \frac{\kappa}{\epsilon}  \left(e^{ \epsilon} - 1\right)  \left(  e^{-n\epsilon} - e^{-\epsilon} \right) \\
\leq & \ 0
\label{eqn:case2subcase1}
\end{align}
\item Subcase $ 1 \leq K \leq last(X)$ : Recall that in this case, the item is packed in a level $h_{K-1}$ level bin to create a level $h_{K}$ level bin. If $K=1$, then $h_{K-1}=0$, so that a new bin must be opened and the item is packed at level 0. Therefore, the objective function term increases by $1$:
\begin{align}
\label{eqn:objfun1}
\expct{ \Delta N | X, \mJ_X , K = 1  } &= 1   
\end{align}
If $2\leq K \leq last(X)$, then the item is packed in an existing bin of height $h_{K-1}$ so that the objective function term does not change:
\begin{align}
\label{eqn:objfun2}
\expct{ \Delta N | X, \mJ_X , 2 \leq  K \leq last(X)  } &= 0   
\end{align}

To bound the change in potential function term, first observe that the contribution due to levels $h$ which are not $h_{k}$ for some $1\leq k \leq last(X)$ because of items with $1\leq K\leq last(X)$ is 0:
\begin{align}
\label{eqn:potfun1}
 \expct{ \Delta V_h | X, \mJ_X , 1\leq K \leq last(X)}  &= 0,  \qquad ( \nexists  1\leq k\leq last(X), h = h_{k}).
\end{align}
The contribution due to $h_{k}$ with $k \leq last(X)-1$ is (let $N_{h_k} = n$):
\begin{align}
\nonumber & \expct{ \Delta V_{h_k} | X, \mJ_X,  1\leq K \leq last(X)}  \\
\nonumber &= 
\frac{\kappa}{\epsilon} \left( e^{-\epsilon (n+1)} - e^{-\epsilon n} \right) \cdot \prob{K=k | 1 \leq K \leq last(X)} \\
\nonumber
& \qquad + 
\frac{\kappa}{\epsilon} \left( e^{-\epsilon (n-1)} - e^{-\epsilon n} \right) \cdot \prob{K=k+1 | 1 \leq K \leq last(X)} \\
\nonumber &  = \frac{\kappa}{last(X) \epsilon} \left[ \left( e^{-(n+1)\epsilon} - e^{-n\epsilon} \right)  + \left(e^{-(n-1)\epsilon} - e^{-n \epsilon} \right) \right] \\
\nonumber &= \kappa \frac{e^{-n \epsilon}}{ last(X) \epsilon} \left[ e^{-\epsilon} + e^{\epsilon} - 2  \right] \\
& \leq  \frac{\kappa \epsilon}{last(X)}.
\label{eqn:potfun2}
\end{align}
For the last inequality, we use the assumption $0 < \epsilon < 1$.

The contribution due to $h_{last(X)}$ is (let $N_{last(X)} = n$):
\begin{align}
\nonumber & \expct{ \Delta V_{h_{last(X)}} | X, \mJ_X, 1\leq K \leq last(X)}  \\
\nonumber &= 
\frac{\kappa}{\epsilon} \left( e^{-\epsilon (n+1)} - e^{-\epsilon n} \right) \cdot \prob{K= last(X) | 1 \leq K \leq last(X)} \\
& \leq  0.
\label{eqn:potfun3}
\end{align}
Combining \eqref{eqn:objfun1}-\eqref{eqn:potfun3}. 
\begin{align}
\nonumber
\expct{\Delta \mL | X, \mJ_X, 1 \leq K \leq last(X)}  
& \leq \frac{1}{last(X) } + \frac{\kappa \epsilon}{last(X)} \cdot last(X) \\
&= \frac{1}{last(X)} + \kappa \epsilon
\label{eqn:case2subcase2}
\end{align}
\end{enumerate}
Finally combining \eqref{eqn:case2subcase1} and \eqref{eqn:case2subcase2}, when $\bN,X,\mJ_X$ are such that $last(X) \geq 1$:
\begin{align}
\nonumber & 
\expct{\Delta \mL | X , \mJ_X} \\ 
\nonumber
&= 
\expct{\Delta \mL | X , \mJ_X, K > last(X) } \prob{K > last(X) | X, \mJ_X }
 \\ 
\nonumber & \qquad + \expct{\Delta \mL | X , \mJ_X, 1\leq K \leq  last(X) } \prob{1\leq K \leq last(X) | X , \mJ_X} \\
\nonumber & \leq 0 \cdot \frac{|X|-last(X)}{|X|} + \left( \frac{1}{last(X)} + \kappa \epsilon \right) \cdot \frac{last(X) }{|X|} \\
& \leq \frac{1}{|X|} + \kappa \epsilon  
\label{eqn:DeltaLcase2}
 \end{align}
\end{enumerate}
Combining  \eqref{eqn:DeltaLcase1} for $last(X)=0$ with \eqref{eqn:DeltaLcase2} for $last(X) \geq 1$ gives 
\begin{align}
\expct{\Delta \mL | X, \mJ_X} & \leq \frac{1}{|X|} + \kappa \epsilon,
\end{align}
thus proving \eqref{eqn:perbinbound}.

\subsection{Proof of Theorem~\ref{thm:PDexp_varying_epsilon}}
In the case of open-ended bin packing with time varying $\epsilon_t$, the penalized-Lagrangian is given by:
\begin{align}
\mL^{exp}_t(\bN_t) &= \sum_{h=1}^B N_t(h)+ \frac{\kappa}{\epsilon_t} \sum_{h=1}^{B-1} e^{-\epsilon_t N_t(h)}
\end{align}
The $\PDexp$ algorithm packs the item $Y_t$ arriving at time $t$ so that:
\[ \bU_t^{\PDexp} \in \argmin_{\bU \in \mU(\bN_{t-1}, Y_t)}  \mL^{exp}_t( \bN_{t-1} + \bC_{Y_t} \cdot \bU ).\]
Taking expectation with respect to $Y_t \sim F$, and the actions of the randomized $A_F$ policy, we get using Lemma~\ref{lem:AFbound},
\begin{align}
\nonumber
& \expctsub{Y_t\sim F}{\mL^{exp}_t(\bN^{\PDexp}_{t-1} + \bC_{Y_t} \cdot \bU^{\PDexp}_t) - \mL^{exp}_{t-1}(\bN^{\PDexp}_{t-1}) | \bN_{t-1}^{\PDexp}}  \\
\nonumber \leq &  \ \expctsub{Y_t \sim F, \bU_t^{A_F} \sim \pi^{A_F}}{\mL^{exp}_t(\bN^{\PDexp}_{t-1} + \bC_{Y_t} \cdot \bU^{A_F}_t) - \mL^{exp}_{t-1}(\bN^{\PDexp}_{t-1}) | \bN_{t-1}} \\
\label{eqn:PDtv_drift}
\leq & \  b(F) + \kappa \epsilon_{t-1} + (B-1) \kappa \left( \frac{1}{\epsilon_t} - \frac{1}{\epsilon_{t-1}} \right)
\end{align}

A telescoping sum of the above from $t=1$ to $T$ gives,
\begin{align}
\expctsub{\bY}{ \mL_T^{exp}(\bN^{\PDexp}_T)} - \mL^{exp}_0(\bN^{\PDexp}_0)
&= \sum_{t=1}^T 
\expctsub{Y_1, \ldots, Y_{t-1}}{ 
\expctsub{Y_{t} \sim F}{ 
\mL^{exp}_{t}(\bN^{\PDexp}_t)  
- \mL^{exp}_{t-1}(\bN^{\PDexp}_{t-1})  | \bN^{\PDexp}_{t-1} 
}}   \\
& \leq \sum_{t=1}^{T} \left( b(F) +  \kappa \epsilon_{t-1} + (B-1) \kappa\left( \frac{1}{\epsilon_t} - \frac{1}{\epsilon_{t-1}} \right ) \right) \\
& =  T \cdot b(F) + \kappa \sum_{t=1}^T \epsilon_t + (B-1)\kappa \left( \frac{1}{\epsilon_T} - \frac{1}{\epsilon_0} \right)
\end{align}
Therefore,
\begin{align}
\nonumber \expct{N^{\PDexp}_T} & \leq  \expctsub{\bY}{\mL^{exp}(\bN^{\PDexp}_T) - \frac{\kappa}{\epsilon} \sum_{h=1}^{B-1} e^{-\epsilon N^{\PDexp}_T(h)}} \\
& \leq T b(F) + T \kappa \sum_{t=1}^T \epsilon_{t-1} + (B-1)\kappa \left( \frac{1}{\epsilon_T} - \frac{1}{\epsilon_0} \right)
 + \mL^{exp}(\bN^{\PDexp}_0) \\
\intertext{which since $\bN^{\PDexp}_0 = \mathbf{0}$,}
\label{eqn:final_general_bound2}  & \leq T b(F) + T \kappa \sum_{t=1}^{T} \epsilon_{t-1} + (B-1)\kappa \frac{1}{\epsilon_T}.
\intertext{Choosing $\kappa=1$, $\epsilon_t = \frac{k}{\sqrt{t+a}}$ with $k,a$ to be decided}
& \leq T b(F) + \kappa k \sum_{t=1}^{T} \frac{1}{\sqrt{t+a-1}} + (B-1) \frac{\kappa}{k} \sqrt{T+a} \\
& \leq T b(F) + \kappa k \int_{t=0}^T \frac{dt}{\sqrt{t+a-1}} + (B-1) \frac{\kappa}{k} \sqrt{T+a} \\
& = T b(F) + 2 \kappa k \left( \sqrt{T+a-1} - \sqrt{a-1} \right) + (B-1) \frac{\kappa}{k} \sqrt{T+a} 
\intertext{Choosing $a=1$:}
& \leq T b(F) +  \left( 2  k  + \frac{B}{k} \right) \sqrt{T} 
\intertext{and further, $k = \sqrt{B/2}$}
 & \leq T b(F) + \sqrt{8BT} .
\end{align}
as in the theorem statement. $~\Box$


\subsection{Proof of Theorem~\ref{thm:martingale}}
We will use the following version of Azuma-Hoeffding inequality.
\begin{lemma}(Azuma-Hoeffding) Suppose $X_0, \ldots, X_T$ is a supermartingale with $|X_i - X_{i-1}| \leq C_i$ for each $i$. Then
\[ \prob{X_T - X_0  \geq a} \leq e^{-\frac{a^2}{2\sum C_i^2}} \]
\end{lemma}

We consider the setup of Theorem~\ref{thm:PDexp_varying_epsilon} with $\epsilon_t = \sqrt{\frac{B}{2(t+1)}}$, $\kappa = 1$. Suppressing the subscript $F$ to denote the distribution, define
\begin{align}
 X_t & = \mL^{exp}_t(\bN_t) - \sum_{s=1}^t \left[ b(F) + \epsilon_{s-1} + (B-1)\left( \frac{1}{\epsilon_s} - \frac{1}{\epsilon_{s-1}} \right) \ \right]  \\
\nonumber & = \sum_{h=1}^B N_t(h) + \frac{1}{\epsilon_t} \sum_{h=1}^{B-1} e^{-\epsilon_t N_t(h)} - 
\left[ t\cdot b(F) + \sum_{s=1}^{t} \epsilon_{s-1} + (B-1) \left( \frac{1}{\epsilon_t} - \frac{1}{\epsilon_0} \right) \right] 
\end{align}
Inequality \eqref{eqn:PDtv_drift} implies that $\{X_t\}$ is a supermartingale with differences bounded as:
\begin{align*}
 |X_t - X_{t-1}| & \leq |N^{\PDexp}_t - N^{\PDexp}_{t-1}| + |V_t - V_{t-1}| + b(F) + \sqrt{\frac{B}{2t}} + \sqrt{\frac{2(B-1)^2}{B}} \left( \sqrt{t+1} - \sqrt{t} \right)\\
& \leq  1 + 2 +  b(F) + \sqrt{B} +  \sqrt{2B} \\
& \leq 4 + \frac{5}{2} \sqrt{B} =: \delta
\end{align*}	
Which gives:
\begin{align*}
& \prob{ \mL^{exp}_t(\bN_t) \geq \left( t\cdot b(F) + \sqrt{8Bt} \right) + a } \\
& \leq \prob{  \mL^{exp}_t(\bN^{\PDexp}_t) \geq 
\left[ t\cdot b(F) + \sum_{s=1}^{t} \epsilon_{s-1} + (B-1) \left( \frac{1}{\epsilon_{t}} - \frac{1}{\epsilon_{0}} \right) \right] 
 + a } \\
&  \leq e^{-\frac{a^2}{2  t \delta }} 
\end{align*}
This in turn implies for any positive constant $\lambda$:
\begin{align}
\prob{N^{\PDexp}_t \geq t b(F) +  \sqrt{8Bt}  +  \sqrt{ 2 \lambda \delta t \log t} } \leq \frac{1}{t^\lambda} .
\end{align}


\subsection{Proof of Theorem~\ref{thm:bounded_upperbound}}
The penalized Lagrangian in this case is given by:
\begin{align}
\label{eqn:Ltquad_again}
\mathcal{L}^{tquad}_\eta(\mathbf{N}, \widetilde{\bN}) &= \sum_{h=1}^{B} N_h + \frac{1}{2\eta} \sum_{h=1}^{B-1} \left( (\eta - \widetilde{N}_h)^+ \right)^2
\end{align}
where $\mathbf{N}$ is the number of bins used in the current packing at each level, and $\mathbf{\widetilde{N}}$ is the number of bins open.

The proof is a very minor modification over Theorem~\ref{thm:PDexp_fixed_epsilon}. The following lemma (whose proof we defer to the end of this subsection) analogous to Lemma~\ref{lem:AFbound} establishes the existence of a randomized policy $A_F$ which will be used to bound the change in the penalized-Lagrangian for $\PDtquad$ algorithm.

\begin{lemma}
\label{lem:AFbound_tquad}
For a positive integer $ \eta $, and an arbitrary item size distribution $F$,
there exists a distribution-dependent algorithm $A_F$ which for each item size $Y \in \supp(F)$ and arbitrary initial packing state $(\bN, \widetilde{\bN})$ ($\bN$ denotes the level summary of the packing, and $\widetilde{\bN}$ ($\widetilde{\bN} \leq \bN$, $\widetilde{\bN} \leq \eta \cdot \mathbf{1}$) the open bins eligible for receiving new items) defines a random single-step action $\bU^{A_F}$ for packing item $Y$  such that:
\begin{align} \label{eqn:AFbound_tquad}
\expctsub{ Y  \sim F, \bU^{A_F} \sim \pi^{A_F}(\widetilde{\bN},Y )}{ \mL^{tquad}_\eta \left( \bN + \bC_{Y} \cdot \bU^{A_F}, \widetilde{\bN} + \bC_{Y} \cdot \bU^{A_F} \right)} - \mL^{tquad}_{\eta}(\bN, \widetilde{\bN}) &\leq b(F) + \frac{1}{\eta},
\end{align}
where $b(F)$ is the optimal value of the LP $\mathbf{(P_{1d-level})}$ giving the number of bins used per item under the optimal fractional packing of $F$.
\end{lemma}

The next lemma proves a useful relationship for the proof of the theorem:
\begin{lemma}
\label{lem:Ltquad_relationship}
Let $(\bN, \widetilde{\bN})$ be an arbitrary initial packing with $\widetilde{\bN} \leq \eta \cdot \mathbf{1}$ for a positive integer $\eta$, and for item size $Y$, let $\bU \in \mU(\widetilde{\bN},Y)$ be a valid action that packs item $Y$ in one of the open bins. Denoting $\bN' = \bN + \bC_Y \cdot \bU$ and $\widetilde{\bN}' = \widetilde{\bN} +\bC_Y \cdot \bU$:
Then,
\begin{align}
\mL^{tquad}_\eta( \bN' , \widetilde{\bN}' \wedge (\eta \cdot \mathbf{1}) ) = 
\mL^{tquad}_\eta( \bN' , \widetilde{\bN}'). 
\end{align}
\end{lemma}
\begin{proof}{Proof of Lemma~\ref{lem:Ltquad_relationship}:}
Noting the definition of $\mL^{tquad}_\eta$, it suffices to show that for $\widetilde{N}'(h) = n$, when $ 1 \leq n \leq \eta$,
\[  (\eta - ( n - 1))^+ = (\eta - \min\{n-1, \eta\})^+  \]
and when $0 \leq n \leq \eta$,
\[  (\eta - ( n + 1))^+ = (\eta - \min\{n+1 , \eta\})^+  \]
The first is true because $n-1 = \min\{ n - 1, \eta \}$ for $n \leq \eta$. The second is true because when $0 \leq n \leq \eta - 1$, $n+1 = \min\{n+1, \eta\}$. When $n = \eta$, 
\[ (\eta - (n+1))^+ = (-1)^+ = 0  
 = (\eta - \eta )^+ 
 = (\eta - \min\{ \eta+1, \eta\})^+ 
= (\eta - \min\{ n+1, \eta\})^+. \]
$~\Box$
\end{proof}

With Lemma~\ref{lem:AFbound_tquad} in hand, we carry out a similar analysis as in the proof of Theorem~\ref{thm:PDexp_fixed_epsilon}.  Denoting by $(\bN^{\PD}_t , \widetilde{\bN}^{\PD}_t ) $ the random packing obtained after packing $\{Y_1,\ldots, Y_t\}$ using the algorithm $\PDtquad$ (suppressing the subscript $F$, and using superscript $\bN^\PD$ in place of $\bN^\PDtquad$ to reduce clutter), and
\begin{align*}
 \bU^{\PD}_t & \in \argmin_{\bU \in \mU(\widetilde{\bN}^{\PD}_{t-1},Y_t)} \mL^{tquad}_\eta( \bN^{\PD}_{t-1} + \bC_{Y_t}\cdot \bU , \widetilde{\bN}^{\PD}_{t-1}+ \bC_{Y_t} \cdot \bU)
\end{align*}
we get
\begin{align*}
\nonumber & \expctsub{Y_t \sim F}{ \mL^{tquad}_\eta ( \bN^{\PD}_t , \widetilde{\bN}^{\PD}_{t} ) - \mL^{tquad}_\eta(\bN^{\PD}_{t-1} , \widetilde{\bN}^{\PD}_{t-1}) | \bN^{\PD}_{t-1} , \widetilde{\bN}^{\PD}_{t-1} } \\
\nonumber = \   & \expctsub{Y_t \sim F}{ \mL^{tquad}_\eta \left( \bN^{\PD}_{t-1} + \bC_{Y_t} \cdot \bU^{\PD}_t , \left(\widetilde{\bN}^{\PD}_{t-1} + \bC_{Y_t} \cdot \bU^{\PD}_t \right) \wedge (\eta \cdot \mathbf{1}) \right) - \mL^{tquad}_\eta(\bN^{\PD}_{t-1} , \widetilde{\bN}^{\PD}_{t-1}) | \bN^{\PD}_{t-1} , \widetilde{\bN}^{\PD}_{t-1} } \\
\nonumber = \ & \expctsub{Y_t \sim F}{ \mL^{tquad}_\eta(\bN^{\PD}_{t-1} + \bC_{Y_t} \cdot \bU^{\PD}_t , \widetilde{\bN}^{\PD}_{t-1} + \bC_{Y_t} \cdot \bU^{\PD}_t ) - \mL^{tquad}_\eta(\bN^{\PD}_{t-1} , \widetilde{\bN}^{\PD}_{t-1}) | \bN^{\PD}_{t-1} , \widetilde{\bN}^{\PD}_{t-1} } \\
\leq  \ & \expctsub{Y_t \sim F, \bU_t^{A_F}}{ \mL^{tquad}_\eta( \bN^{\PD}_{t-1} + \bC_{Y_t} \cdot \bU^{A_F}_t , \widetilde{\bN}^{\PD}_{t-1} + \bC_{Y_t} \cdot \bU^{A_F}_t ) - \mL^{tquad}_\eta(\bN^{\PD}_{t-1} , \widetilde{\bN}^{\PD}_{t-1}) | \bN^{\PD}_{t-1} , \widetilde{\bN}^{\PD}_{t-1} } \\
\leq & \ 
b(F) + \frac{1}{\eta}.
\end{align*}
The first equality is by definition of $(\bN^{\PD}_t, \widetilde{\bN}^{\PD}_t)$, second equality by Lemma~\ref{lem:Ltquad_relationship}, the subsequent inequality by the definition of $\PDexp$ as being greedy with respect to $\mL^{tquad}_\eta$, and the last by Lemma~\ref{lem:AFbound_tquad}. 

A telescoping sum of the above from $t=1$ to $T$ gives,
\begin{align*}
\expctsub{\bY}{ \mL^{tquad}_\eta(\bN^{\PD}_T , \widetilde{\bN}^{\PD}_T)} - \mL^{tquad}_\eta(\bN^{\PD}_0 , \widetilde{\bN}^{\PD}_0)
& \leq T \left( b(F) + \frac{1}{\eta} \right) 
\end{align*}
which implies,
\begin{align*}
\expct{ N^{\PD}_T } & \leq T \cdot b(F) \left( 1 + \frac{1}{b(F) \cdot \eta} \right) + \frac{B\eta}{2}. 
\end{align*}
Therefore, asymptotically as $T \to \infty$, the $\eta$-bounded Primal-Dual algorithm is $\left( 1 + \frac{1}{b(F) \eta} \right)$ competitive.
$~\Box$

\ \\
{\bf Proof of Lemma~\ref{lem:AFbound_tquad}} \\
We highlight the changes needed compared to Lemma~\ref{lem:AFbound}:
\begin{itemize}
\item \textbf{Modified $A_F$ policy}:  We will modify the policy $A_F$ to use the open inventory state $\widetilde{\bN}$ instead of the total number of bins ${\bN}$. In particular, given a configuration $X$ and the current open bins state $\widetilde{\bN}$, the ordering $\mJ_X = (j_1,j_2,\ldots, j_{|X|})$ of the items in $X$ and $last(X), \ 0 \leq last(X) \leq |X|$ are chosen such that :
\begin{enumerate}
\item $\widetilde{\bN}$ has bins with level $j_1$, $j_1+j_2$, $\ldots$, $j_1+j_2+\cdots+j_{last(X)}$.\\
(Except: if $\sum_{k=1}^{|X|} j_k=B$, then it is acceptable to have $last(X) = |X|$ irrespective of $\widetilde{N}(B)$.)
\item $\widetilde{\bN}$ has no bins of levels $j_1+\cdots+j_{last(X)}+j_k$ for any $k > last(X)$.
\end{enumerate}
Define $h_k$ for $0 \leq k \leq last(X)$ as:
\[h_0=0 ; \quad h_k = h_{k-1} + j_k. \]

To place the item, we follow exactly the same rule as in the policy constructed in Lemma~\ref{lem:AFbound}: Given an item $Y$, we sample a configuration $X$ according to 
\[ \prob{X=c | Y=j}  = \frac{\alpha^*_c \cdot x_c(j)}{p_j} \]
and then map item $Y$ to one of the $x_X(Y)$ items in the ordering $(j_1,\ldots, j_{|X|})$ of size $Y$ uniformly at random. If $K$ denotes the index of the item in the ordering, then if $K \leq last(X)$, the arriving item is placed in an open bin of level $h_{K-1}$ to create a bin of level $h_K$. Else the arriving item is placed in an open bin of level $h_{last(X)}$. That is,
\[ \bU_{A_F} = \be_{ \min\{  h_{K-1} , h_{last(X)}  \} } .\]
\item \textbf{Drift analysis of $A_F$ : } As in the proof of Lemma~\ref{lem:AFbound}, for analysis only, we view the policy $A_F$ as generating a pair $(Y,X)$ where the configuration $X$ is sampled first and an ordering $\mJ_X = (j_1,\ldots, j_{|X|})$ of the items in $X$ is determined. Then a positive integer $K$ is sampled uniformly between $1$ and $|X|$. Finally $Y$ is set to be the item $j_K$ in the ordering $(j_1, \ldots, j_{|X|})$. Denote
\begin{align}
\Delta \mL(Y,\bU) & := \mL^{tquad}_{\eta}( \underbrace{\bN + \bC_{Y} \cdot \bU}_{=: \bN'}, \underbrace{ \widetilde{\bN} + \cdot \bC_Y \cdot \bU }_{=: \widetilde{\bN}'})  - \mL^{tquad}_{\eta}(\bN, \widetilde{\bN}')  \\
& = \underbrace{N' - N}_{ =: \Delta N} + \sum_{h=1}^{B-1} \underbrace{ \frac{1}{2\eta} \left( (\eta - \widetilde{N}'(h))^+ \right)^2 - \left(  (\eta - \widetilde{N}(h))^+  \right)^2 }_{ =: \Delta V_h}  
\end{align}
as the change in the penalized Lagrangian after item $Y$ is packed in $\widetilde{\bN}$ using action $\bU$.

To analyze the drift of $A_F$, we will condition on the configuration $X$ sampled at the first step as well as the ordering of item $\mJ_X = (j_1,\ldots, j_{|X|})$ chosen based on $\widetilde{\bN}$, and prove (analogous to \eqref{eqn:perbinbound})
\begin{align}
\expctsub{K \sim \mathsf{Unif}[|X|]}{\Delta \mL(j_K, \be_{ \min\{h_{K-1} , h_{last(X)} \} } ) | X, \mJ_X } & \leq \frac{1}{|X|} + \frac{1}{\eta}
\label{eqn:perbinbound_tquad}
\end{align} 
and unconditioning on the bin $X$:
\begin{align}
\expctsub{Y\sim F, \bU \sim \pi^{A_F} }{\Delta \mathcal{L}(Y,\bU)} & \leq b(F) + \frac{1}{\eta}
\end{align}
as in the Theorem.

In the rest of the proof we prove \eqref{eqn:perbinbound_tquad} by going through the cases as in proof of Lemma~\ref{lem:AFbound}:
\begin{enumerate}
\item Case $last(X) = 0$: 
In this case a new bin is opened with probability 1. If a level $h = j_K$ bin is created, then we must have $\widetilde{N}(h) = 0$. Therefore,
\begin{align}
\label{eqn:DeltaLcase1_tquad}
\expct{\Delta \mathcal{L} | X, \mJ_X} = \Delta N + \Delta V_{j_K} =  1 + \frac{1}{2 \eta} \left( (\eta - 1)^2 - (\eta-0)^2 \right) = 1 - \frac{2\eta - 1}{2\eta} = \frac{1}{2\eta}.
\end{align}
\item Case $last(X) > 0$: 
\begin{enumerate}
\item Subcase $K > last(X)$: This step remains unchanged -- Since no new bin is created, the objective function term in the Lagrangian is unchanged. The item is packed at level $h_{last(X)}$ with $n := \widetilde{N}(h_{last(X)}) \geq 1$, and $\widetilde{N}(h_{last(X)+j_K})=0$. Convexity of the potential function terms then gives:
\begin{align}
\expct{\Delta \mathcal{L} | X, \mJ_X , K > last(X)}  &\leq 0
\end{align}
\item Subcase $1 \leq K \leq last(X)$:
The change in the objective function term is as before -- only $K=1$ causes a new bin to be opened:
\begin{align}
\label{eqn:objfun1_tquad}
\expct{ \Delta N | X, \mJ_X , K = 1  } &= 1 \\  
\label{eqn:objfun2_tquad}
\expct{ \Delta N | X, \mJ_X, 2 \leq  K \leq last(X)  } &= 0   
\end{align}

However, the analysis of the change in potential function term is more subtle. We still have
\begin{align}
\expct{ \Delta V_{h} | X, \mJ_X, 1\leq K \leq last(X)} &= 0, \qquad (\nexists  1 \leq k \leq last(X), h = h_k). 
\end{align}
For $h_{last(X)}$,
\begin{align}
\expct{\Delta V_{h_{last(X)}} | X , \mJ_X, 1 \leq K \leq last(X)  } \leq 0,
\end{align}
because conditioned on $1 \leq K \leq last(X)$, $\widetilde{N}'(h_{last(X)}) \geq \widetilde{N}(h_{last(X)})$, and the potential term is non-increasing in $\widetilde{N}$.\\

To analyze the contribution to the change in the potential term due to level $h_k$ ($1 \leq k \leq last(X)-1$), we have to split the analysis in two subcases: 
\begin{enumerate}
\item Subsubcase  $\widetilde{N}({h_k}) = n < \eta$ : In this case $\widetilde{N}'({h_k}) = \widetilde{N}(h_k) \pm 1$ with equal probability:
\begin{align*}
& \expct{ \Delta V_{h_k} | X, \mJ_X , 1 \leq K \leq last(X)}  \\
&= \sum_{a=1}^{last(X)} \expct{\Delta V_{h_k} | X, \mJ_X , K = a } \cdot \prob{ K = a | 1\leq K \leq last(X)} \\
& = \frac{1}{last(X)} \left( \expct{\Delta V_{h_k} | X, \mJ_X , K=k}  + \expct{\Delta V_{h_k} | X, \mJ_X , K=k+1} \right) \\
  &  \leq \frac{1}{2 \eta \cdot last(X) } \left[ \left( (\eta - n - 1)^2 - (\eta-n)^2 \right)  + \left( (\eta - n + 1)^2 -  (\eta - n)^2 \right) \right] \\
 & =  \frac{1}{\eta \cdot  last(X) } 
\end{align*}
\item Subsubcase $\widetilde{N}(h_k) =  \eta$ : In this case $\widetilde{N}'(h_k)$  is either $\eta \pm 1$ with equal probability. Therefore the potential term can increase, but not decrease. However, the change can be bounded as follows:
\begin{align*}
& \expct{\Delta V_{h_k} | X, \mJ_X , 1\leq K \leq last(X)}  \\
&= \sum_{a=1}^{last(X)} \expct{\Delta V_{h_k} | X, \mJ_X , K = a } \cdot \prob{ K = a | 1\leq K \leq last(X)} \\
& = \frac{1}{last(X)} \left( \expct{\Delta V_{h_k} | X, \mJ_X , K=k}  + \expct{\Delta V_{h_k} | X, \mJ_X , K=k+1} \right) \\
  &  \leq \frac{1}{2 \eta \cdot last(X) } \left[ \left( (\eta - \eta )^2 - (\eta-\eta)^2 \right)  + \left( (\eta - \eta + 1)^2 -  (\eta - \eta )^2 \right) \right] \\
 & =  \frac{1}{2 \eta \cdot last(X) } 
\end{align*}
\end{enumerate}
To summarize, when $last(X)>0$:
\begin{align*}
\expct{\Delta \mathcal{L} | X, \mJ_X , 1\leq K \leq last(X)} & = 
\expct{\Delta N | X, \mJ_X, 1\leq K \leq last(X)}  \\
& \qquad +\sum_{k=1}^{last(X)} \expct{\Delta V_{h_k} | X, \mJ_X, 1\leq K \leq last(X)} \\
& \leq  \frac{1}{last(X)} + \frac{1}{\eta}
\end{align*}
\end{enumerate}
Finally, for $last(X) > 0$
\begin{align}
\nonumber \expct{\Delta \mathcal{L} | X , \mJ_X} &= 
\expct{\Delta \mathcal{L} | X , \mJ_X , K > last(X) } \prob{K > last(X) | X, \mJ_X}
 \\ 
\nonumber & \qquad + \expct{\Delta \mathcal{L} | X , \mJ_X , K \leq  last(X) } \prob{K \leq last(X) | X, last(X)> 0} \\
\nonumber & \leq 0 \cdot \frac{|X|-last(X)}{|X|} + \left( \frac{1}{last(X)} +  \frac{1}{\eta} \right) \cdot \frac{last(X) }{|X|} \\
& \leq \frac{1}{|X|} + \frac{1}{\eta}
\label{eqn:DeltaLcase2_tquad}
 \end{align}
\end{enumerate}
Combining \eqref{eqn:DeltaLcase1_tquad} (for $last(X)=0$) with \eqref{eqn:DeltaLcase2_tquad} (for $last(X)>0$) gives
\begin{align*}
\expct{\Delta \mathcal{L} | X, \mJ_X} & \leq \frac{1}{|X|} + \frac{1}{\eta}.
\end{align*}
which is the desired inequality \eqref{eqn:perbinbound_tquad}.
\end{itemize}


\subsection{Proof of Theorem~\ref{thm:bounded_lowerbound}}

To prove a lower bound on the number of bins used by any bounded-inventory online packing algorithm, we consider a relaxed model where the online algorithm does not have to commit the items  to a physical bin as soon as they arrive. Instead, the algorithm can keep a set of items unpacked as long as these items can be packed in $\eta$ bins (the budget on open bins). Denoting by $\bM = (M(1), M(2))$ the number of size 1 and size 2 items, respectively, that are currently uncommitted to a physical bin, the set of allowed $(M(1), M(2))$ satisfies:
\begin{align*}
M(2) & \leq \eta, \quad 
M(1) + 2 \cdot M(2) \leq 3 \eta.
\end{align*}
The first constraint captures the observation that only one size 2 item can be packed in a bin, and hence their number must be bounded by the maximum number of bins that can be open. The second constraint is a volume constraint since $B=3$.

Whenever the arrival of an item causes the set of uncommitted items to violate the above constraints, some subset of these items must be packed in bins which are then closed and not available subsequently for packing. For example, suppose $\eta = 1$, and $M(1) = 2, M(2) = 0$. The arrival of a size 2 item causes the set of uncommitted items to be infeasible. In this case, the algorithm can pack a bin in configuration $(1,1)$ (one item of size 1 and 2 each), to make the state of uncommitted items $M(1)=1, M(2) = 0$. Another feasible action would be to pack in configuration $(2,0)$ (two items if size 1) to make the state of uncommitted items $M(1)=0, M(2)=1$. This relaxed model can simulate the original model where items must be packed online and at most $\eta$ bins are allowed to be open, and thus the optimal number of bins in this model lower bounds the optimal for the original model.

Since the item size distribution has a lot of structure, we can arrive at the optimal online algorithm under the relaxed online packing model. Lemma~\ref{lem:structure_opt} characterizes an optimal online policy that minimizes the number of bins closed on every sample path of arrivals, and Lemma~\ref{lem:analysis_opt} gives the asymptotic performance of this policy as the packing horizon $T$ grows. Proofs of Lemmas~\ref{lem:structure_opt} and \ref{lem:analysis_opt} are presented in Section~\ref{sec:structure_opt} and \ref{sec:analysis_opt}, respectively.

\begin{lemma}
\label{lem:structure_opt}
For $B=3$ and items of size $1$ or $2$, on every sample path of arrivals and initial state of uncommitted items, the policy $A^*$ with the properties described below minimizes the number of bins closed: If the arrival of an item causes the state of uncommitted items to become infeasible, then 
\begin{itemize}
\item[\bf P1:] If $M(1), M(2) \geq 1$, then pack and close a bin with one item each of size 1 and 2. 
\item[\bf P2:] Else, if $M(2)=0$, pack and close a bin with three items of size 1. 
\item[\bf P3:] Else, if $M(1) = 0$, then pack and close a bin with one item of size 2. 
\end{itemize}
In the above, $M(1),M(2)$ include the newly arriving item.
\end{lemma} 

Using the above lemma, we can simplify the policy $A^*$ even further. Since we always prefer to pack items 1 and 2 together, instead of deferring this action until the state of uncommitted items becomes infeasible we pack and close a bin as soon as we have any size 1 arrival while $M(2) \geq 1$, or we have a size 2 arrival while $M(1) \geq 1$. Therefore, at any point of time, only one of $M(1)$ or $M(2)$ are non-zero. We call this policy \emph{$A^*$ with early closing}.

\begin{lemma}
\label{lem:analysis_opt}
For the online policy $A^*$ with early closing described above, starting from any initial state $\bM_0 = (M_0(1) , M_0(2))$ of uncommitted items with $M_0(1) \cdot M_0(2) = 0$, the expected number of bins closed after packing $T$ items from the distribution $F=( p_1 = 0.5 , p_2 = 0.5\}$ is $\frac{T}{2}\left( 1 + \frac{1}{12 \eta} \right) + o(T)$.
\end{lemma}

Denoting by $\ONOPT$ the optimal online policy, we thus have $\expct{N^{\ONOPT}_{F,T}}$ is lower bounded by the expected number of closed bins under $A^*$ with early closing, which is $\frac{T}{2}\left( 1 + \frac{1}{12\eta}  \right) + o(T)$. The offline optimal packing uses $\expct{N^{\OPT}_{F,T}} = \frac{T}{2}+o(T)$ bins, thus giving a lower bound on the competitive ratio of $\left( 1 + \frac{1}{12\eta} \right)$ proving Theorem~\ref{thm:bounded_lowerbound}.

\subsubsection{Proof of Lemma~\ref{lem:structure_opt}}
\label{sec:structure_opt}
Property {\bf P3} is vacuously true since when $M(1)=0$, the only action that makes the state of uncommitted items feasible is to pack an item of size 2 in a bin and close it. Further, closing more than 1 bins is never a strictly better action -- packing 1 bin makes the state of uncommitted items feasible and closing more bins can be deferred.

Next observe that if an action $a$ packs and closes a bin with a set $S$ of items, and another action $a'$ packs and closes a bin with a set $S' \subset S $, then action $a$ is always weakly improving; that is the number of bins closed can never increase if $a'$ is replaced by $a$ whenever feasible. This is a straightforward sample path dominance argument which we omit.

When $M(2) = 0$, there are three actions which make the state of uncommitted items feasible: Pack and close a bin with either one, two, or three items of size 1. By the above observation, the action which packs three items is weakly optimal, hence proving property {\bf P2}.

For property {\bf P1} when both $M(1), M(2) \geq 1$, we consider three cases:
\begin{enumerate}
\item $M(1) = 1$: In this case packing in configuration $(1,1)$ (one item of size 1 and 2 each) dominates any other feasible action and hence is optimal by the observation above.
\item $M(1) = 2$: The two dominating configurations are $(2,0)$ (two items of size 1) and $(1,1)$. First, observe that action corresponding to configuration $(0,1)$ (packing one size 2 item) dominates packing in configuration $(2,0)$ -- we can emulate any policy $\hat{A}$ that uses configuration $(2,0)$ and does not pack a size 2 item by instead packing a size 2 item and simulating the size 2 item in $\hat{A}$ by the two size 1 items which are now still uncommitted. Finally, the configuration $(1,1)$ subset dominates configuration $(0,1)$, which we have just argued weakly dominates $(2,0)$ and hence is weakly optimal.
\item $M(1) \geq 3$: The two dominating configurations are $(3,0)$ and $(1,1)$. By the same argument as above, packing in configuration $(1,1)$ can simulate any policy that uses the configuration $(3,0)$ and hence $(1,1)$ is weakly dominating.
\end{enumerate}

\subsubsection{Proof of Lemma~\ref{lem:analysis_opt}}
\label{sec:analysis_opt}
We now analyze policy $A^*$ with early closing. The state space of this policy is $\mM = \{ \bM \geq 0 | M(1) \cdot M(2) = 0 , M(1) \leq 3\eta  , M(2) \leq \eta \})$, with the following transition diagram:

\begin{figure}[h!]
\begin{center}
\includegraphics[width=4in]{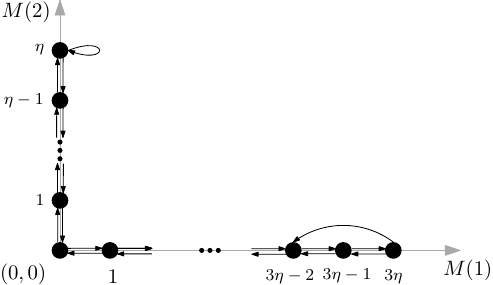}
\end{center}
\end{figure}

In the above diagram, all transitions have probabilities $0.5$. For example, the transition from state $(3\eta,0)$ to $(3\eta - 2, 0)$ represents an arrival of item of size 1, which increases $M(1)$ to $3\eta + 1$, from which a bin with three size 1 items is created and closed to reach the state $(3\eta - 2, 0)$ for uncommitted items. 

Let $\pi(a,b)$ denote the steady state probability that the above Markov Chain is in state $\bM = (a,b)$; and let $\delta := \pi(0,\eta)$.
 Flow balance equations imply the following:
\begin{align}
\pi(0,\eta) & = \pi(0,\eta-1) = \cdots = \pi(0,0) = \pi(1,0) = \cdots = \pi(3\eta - 2, 0) = \delta \\
\pi(3\eta,0) & = \frac{1}{2} \pi(3\eta - 1, 0) \\
\pi(3\eta-1, 0) &= \frac{2}{3} \pi(3\eta -2, 0) 
\end{align}
and hence $\pi(3\eta -1,0) = 2\delta/3$ and $\pi(3 \eta, 0) = \delta/3$. The normalization condition gives $\delta = \frac{1}{4\eta}$.

The policy $A^*$ with early closing creates bins of level 3 except when it is in state $\bM = (0,\eta)$ and an item of size 2 arrives giving the asymptotic rate of creation of bins of configuration $(0,1)$ as $\pi(0,\eta) \cdot p(2) = \delta/2 $. Then, for $T \gg \eta^2$ (the mixing time of the Markov chain above is $\Theta(\eta^2)$), the expected number of bins closed asymptotically will be at least 
\[ \underbrace{\frac{T}{2}(1- \delta)}_{\mbox{2's paired with 1}} + \underbrace{\frac{T}{2} \delta}_{\mbox{2's alone}} +  \underbrace{\frac{T}{2} \delta \cdot \frac{1}{3}}_{\mbox{remaining 1's}} + o(T) =  \frac{T}{2} \left( 1 + \frac{\delta}{3} \right) + o(T) = \frac{T}{2}\left( 1 + \frac{1}{12 \eta} \right) + o(T).\]


\subsection{Proof of Theorem~\ref{thm:PDexp_continuous}}

Recall our assumption that the item size distribution $F=F_{disc}+F_{cont}$ where $F_{cont}$ has density bounded by $D$.

\begin{lemma} For a list $L_G = \{Y_1, \ldots, Y_T\}$ of items sampled {\it i.i.d.} from some distribution $G$, let $OPT(L_G)$ denote the number of bins in the optimal offline packing and 
\[ b(G) = \lim_{T \to \infty} \frac{\expct{OPT(L_G)}}{T} \]
denote the asymptotically optimal number of bins needed per item. Further let $\frac{1}{\delta}$ be a positive integer. Then:
\begin{enumerate}
\item $b(F_\delta) \leq b(F) \leq b(F^\delta)$,
\item $ b(F^\delta) \leq b(F_\delta) + \delta D  \leq b(F) + \delta D$.
\end{enumerate}
\end{lemma}
\proof{Proof:}
The first statement is easy to see via a coupling argument: given a sequence of items $L = \{Y_1, \ldots, Y_T\}$ sampled from $F$, the sequence from $F^\delta$ is $L^\delta := \{ Y_1^\delta, \ldots, Y_T^\delta \}$, and the sequence from $F_\delta$ is $L_\delta = \{{Y_1}_\delta, \ldots , {Y_T}_\delta\}$. Now a packing of $L^\delta$ gives a feasible packing for $L$. Similarly, a packing of $L$ gives a feasible packing for $L_\delta$.

To prove the second statement we will modify $F_\delta$ to construct a distribution that stochastically dominates $F^\delta$. Let $p_\delta(x)$ denote the mass on $x$ according to distribution ${F_{cont}}_\delta$ (that is, mass on $x$ only due to the continuous part of $F$). Let 
\[ \gamma := \max_{x \in \{ 0, \delta, 2\delta, \ldots, B-\delta\} } p_\delta(x) .\]
Note that $\gamma \leq \delta D$. Define a new mass function $p_\delta^\gamma$ as follows:
\begin{align}
p_\delta^\gamma(B) &= p_\delta(B) + \gamma \ ,\\
p_{\delta}^\gamma(x) &= \left[ p_{\delta}(x) - \left( \gamma - \sum_{y < x} p_\delta(y) \right) \right]^+ \ , \quad x \leq B-\delta.
\end{align}
In simple words, we move $\gamma$ mass from the lowest sized items in $p_\delta$ to size $B$. Let ${F_{cont}}_\delta^\gamma$ be the discrete distribution corresponding to the mass function $p_\delta^\gamma$, and let $F^{\gamma}_\delta = F_{disc}+{F_{cont}}_\delta^\gamma$. Now it is easy to see that $ F^{\gamma}_\delta(x) \leq F^\delta(x)  $, or in other words $F^\gamma_\delta$ is stochastically larger than $F^\delta$. Therefore, we also have:
\[ b(F^\delta) \leq b(F^\gamma_\delta). \]
It is also easy to see that $b(F^\gamma_\delta) \leq b(F_\delta) + \gamma$: the items corresponding to $\gamma \leq \delta D$ mass added to $B$ can be packed in bins by themselves (1 per bin), and the remaining distribution (which is not proper) is dominated by $F_\delta$. This proves the second claim of the Lemma.
\endproof \Halmos

We are now ready to complete the proof of Theorem~\ref{thm:PDexp_continuous}. Consider phase $i$ of the algorithm such that $I_{i+1} \geq T$. That is, phase $i$ ends before the time horizon $T$. The item size distribution in this phase is $F^{\delta_i}$, for which the lemma above proves:
\[ b(F^{\delta_i}) \leq b(F) + \delta_i D.  \]
The regret analysis of Theorem~\ref{thm:PDexp_fixed_epsilon} (inequality \eqref{eqn:final_general_bound})
proves that the expected number of bins opened during phase $i$ is bounded by 
\begin{align*}
M_i & \leq \tau_i \cdot b(F^{\delta_i}) + \tau_i \epsilon_i + \frac{B_i}{\epsilon_i}  \\
& \leq \tau_i b(F) + \tau_i \delta_i D + \tau_i \epsilon_i + \frac{B_i}{\epsilon_i} \\
&=  \tau_i \cdot b(F) + 4^i \cdot (D +  2 \sqrt{B}  )
\end{align*}
For the last phase $ k$ with $I_k + \tau_k > T$:
\begin{align*}
M_k &\leq (T - I_k)  \cdot b(F^{\delta_k}) + (T - I_k) \epsilon_k + \frac{B_k}{\epsilon_k}  \\
& \leq (T - I_k) b(F) + \tau_k \delta_k D + \tau_k \epsilon_k + \frac{B_k}{\epsilon_k} \\
&=  (T-I_k) \cdot b(F) + 4^k \cdot (D +  2 \sqrt{B}  )
\end{align*}
Finally, total expected bins used:
\begin{align*}
\sum_{i=1}^k M_i & \leq T \cdot b(F) + (D+2 \sqrt{B}) \sum_{i=1}^k 4^i  \\
& \leq T \cdot b(F) + (D+2\sqrt{B}) \frac{16}{3}  T ^{2/3}.
\end{align*}
The last step follows since $ 8^{k-1} \leq T$ gives $8^k \leq 8T $, and $\sum_{i=1}^{k} 4^i \leq \frac{4}{3} 4^k = \frac{4}{3} (8^k)^{2/3} \leq \frac{4}{3} (8T)^{2/3} = \frac{16}{3} T^{2/3}$.


\subsection{Proof of Theorem~\ref{thm:adversarial_smoothed}}


We will show that given any $L$-bounded fractional partition $\pi = \{q_1, \ldots, q_T\}$,
\begin{align}
\label{eqn:nonstat_telescope}
\expct{\mL^{exp}(\bN^{\PDexp}_{T})} - \mL^{exp}(\bN^{\PDexp}_0) & \leq \sum_{t=1}^T \left( \hat{b}_{\pi,t} + \kappa \epsilon \left( L + \frac{3}{2} \right)  \right)
\end{align}
which combined with $ \mL^{exp}(\bN^{\PDexp}_0) \leq \frac{B\kappa}{\epsilon}$ gives the result in the theorem.

Given $\pi$, let $\{S_1, S_2, \ldots, S_T\}$ be a random vector such that the marginal distribution of $S_t$ is $q_t$, and $\{S_1, \ldots, S_T\}$ is a permutation of $\{1,\ldots, T\}$ with probability 1. Since we only require the notion of $L$-bounded fractional partition for analysis, we assume that $\{S_1, \ldots, S_T \}$ is independent of $\{Y_1, \ldots, Y_T\}$. That such a distribution exists is a corollary of Birkhoff-von Neumann Theorem which says that the matrix $Q_{ij} = q_i(j)$ which is doubly stochastic has a decomposition as a convex combination of permutation matrices.

Then,
\begin{align}
\mL^{exp}(\bN^{\PDexp}_T) - \mL^{exp}(\bN^{\PDexp}_0) &= \sum_{t=1}^{T} \mL^{exp}(\bN^{\PDexp}_{S_t}) - \mL^{exp}(\bN^{\PDexp}_{S_t - 1}).
\end{align}
Taking expectation with respect to the sequence $\{Y_1, \ldots, Y_T \}$ and $\{S_1, \ldots, S_T\}$:
\begin{align}
\expct{\mL^{exp}(\bN^{\PDexp}_T)}
- \mL^{exp}(\bN^{\PDexp}_0) 
&= \sum_{t=1}^{T}
\expct{\mL^{exp}(\bN^{\PDexp}_{S_t}) - \mL^{exp}(\bN^{\PDexp}_{S_t-1})}
\end{align}

We now focus on the term $\expct{\mL^{exp}(\bN^{\PDexp}_{S_t}) - \mL^{exp}(\bN^{\PDexp}_{S_t-1})}$ for some $1 \leq t \leq T$. Since $S_t$ has a marginal distribution $q_t$, the arriving job at time $S_t$ is a random sample from the smoothed distribution $\hat{F}_{\pi,t}$. However we can not apply Lemma~\ref{lem:AFbound} with $F$ replaced by $\hat{F}_{\pi, t}$ to upper bound the change in the Lagrangian because the description of $A_F$ policy in Lemma~\ref{lem:AFbound} had assumed that the packing seen by the arriving item $Y \sim F$ item is independent of $Y$. This is not true anymore. 

Instead, we will exploit the fact that support of $S_t$ is on an interval $[\ell_t, \ell_t+1,\ldots , \ell_t + L -1]$. Therefore, while the packing seen by an arrival at time $S_t$ is random and possibly correlated with the item size, all packing states in the support of this distribution differ by at most $L$ arrivals, and therefore are close to each other.

We now formalize the argument. Conditioning on the prefix $\{Y_1, \ldots, Y_{\ell_t-1}\}$, the law of total expectation gives:
\begin{align}
\expct{
	\mL^{exp}(\bN^{\PDexp}_{S_t}) 
	- \mL^{exp}(\bN^{\PDexp}_{S_t-1}) 
}
&= \expctsub{ \{Y_1 , \ldots, Y_{\ell_{t}-1}\} }{   
	\expct{ 
		\mL^{exp}(\bN^{\PDexp}_{S_t}) 
		- \mL^{exp}(\bN^{\PDexp}_{S_t-1})
		| \{Y_1, \ldots, Y_{\ell_t -1 }\}
		}
} 
\end{align}
We now make two observations:
\begin{enumerate}
\item $Y_{S_t}$ is independent of $\{ Y_1, \ldots, Y_{\ell_t - 1}\}$. \\
This follows since $S_t \geq \ell_t$ with probability 1 and independent of $\{Y_1,\ldots, Y_{T}\}$, and $\bY = \{Y_1, \ldots, Y_T\}$ is a sequence of mutually independent random variables. As a corollary, conditioned on $\{ Y_1, \ldots, Y_{\ell_t -1}\}$, the distribution of $Y_{S_t}$ is $\hat{F}_{\pi,t}$.
\item  For any $(y_1,\ldots, y_{\ell_t-1})$ and two packing states $\bN , \bN'$ such that,
\begin{align*}
\prob{ \bN^{\PDexp}_{S_t - 1} = \bN | \{Y_1,\ldots, Y_{\ell_t -1}\} = (y_1,\ldots, y_{\ell_t -1})} > 0 \\
\prob{ \bN^{\PDexp}_{S_t - 1} = \bN' | \{Y_1,\ldots, Y_{\ell_t -1}\} = (y_1,\ldots, y_{\ell_t -1})} > 0
\end{align*}
we must have ,
\[  \max_{ 1 \leq h \leq B } |N(h) - N'(h)| \leq  L. \]

In words, conditioned on $\{Y_1,\ldots, Y_{\ell_t -1}\}$, the random packing $\bN^{\PDexp}_{S_t-1}$ that an arrival at time $S_t$ sees is supported on an $L$-narrow set. \\
This observation follows because $\bN^{\PDexp}_{\ell_t - 1}$ is a deterministic function of $\{Y_1, \ldots, Y_{\ell_t - 1}\}$. Since $S_t - 1 \leq \ell_t + L - 2 = (\ell_t - 1) + L - 1$ with probability 1, $\bN^{\PDexp}_{S_t - 1}$ and $\bN^{\PDexp}_{\ell_t - 1}$ differ by at most $L-1$ arrivals, each of which changes $N^{\PDexp}(h)$ by at most $\pm 1$ for any $h$.
\end{enumerate}
Now invoking Lemma~\ref{lemma:one_step_drift}:
\begin{align}
\expct{ 
	\mL^{exp}(\bN^{\PDexp}_{S_t}) 
	- \mL^{exp}(\bN^{\PDexp}_{S_t-1})
	| \{Y_1, \ldots, Y_{\ell_t -1 }\}
	}
\leq \hat{b}_{\pi,t} + \kappa \epsilon \left( L + \frac{3}{2} \right).
\end{align}

Therefore,
\begin{align*}
\expct{\mL^{exp}(\bN^{\PDexp}_{T})} - \mL^{exp}(\bN^{\PDexp}_0) 
 &= \sum_{t=1}^{T} \expct{\mL^{exp}(\bN^{\PDexp}_{S_t}) - \mL^{exp}(\bN^{\PDexp}_{S_t - 1})} \\
 &= \sum_{t=1}^{T} \expctsub{ \{Y_1 , \ldots, Y_{\ell_{t}-1}\} }{   
	\expct{ 
		\mL^{exp}(\bN^{\PDexp}_{S_t}) 
		- \mL^{exp}(\bN^{\PDexp}_{S_t-1})
		| \{Y_1, \ldots, Y_{\ell_t -1 }\}
		} } \\
& \leq \sum_{t=1}^T \left( \hat{b}_{\pi,t} + \kappa \epsilon \left( L + \frac{3}{2} \right)  \right)
\end{align*}

giving the desired \eqref{eqn:nonstat_telescope}.

\begin{lemma}
\label{lemma:one_step_drift}
Consider the following one-step bin packing problem: An item type $Y$ is sampled from distribution $F$, with $b(F)$ denoting the optimal bin-rate under $F$ returned by the LP $\mathbf{P_{1d-level}}$. Conditioned on the event $(Y=j)$, an initial packing is sampled according to a measure $\mu_j$ on a set of packings $\mathcal{P}$. The universe of initial packings $\mathcal{P}$ is the same for all types $j$, and further satisfies the condition that it is $L$-\emph{narrow}, meaning:
\begin{align*}
\forall \ P_1, P_2 \in \mathcal{P}\ :\ \max_{ 1 \leq h \leq B-1 } \left| N^{P_1}(h) - N^{P_2}(h) \right|  \leq L
\end{align*}

On seeing the item type $Y \sim F$, and the initial packing $P \sim \mu_Y$ with $\bN := \bN^{P}$, the item is packed using action $\bU$ that minimizes the penalized Lagrangian function:
\begin{align}
\mL^{exp}(\bN') &= \sum_{h=1}^B N'(h) + \frac{\kappa}{\epsilon} \sum_{h=1}^{B-1}e^{-\epsilon N'(h)}, \\
\bN' &= \bN + \bC_{Y} \cdot \bU
\end{align}
where $L < \frac{1}{2\epsilon} \log  \frac{\kappa B}{B-1}$ (equivalently $\kappa e^{-2\epsilon L} \geq 1- \frac{1}{B}$) and $\epsilon L < 1$. Then:
\begin{align}
\label{eqn:drift_delayed_L}
\expctsub{Y \sim F}{\expctsub{P \sim \mu_Y}{\mL(\bN') - \mL(\bN)}} \leq b(F) + \kappa \epsilon \left( L + \frac{3}{2} \right).
\end{align}
\end{lemma}

\proof{Proof:} 
Following the outline of the previous results, we will rely on the fact that the Primal-Dual algorithm minimizes the Left Hand Side of \eqref{eqn:drift_delayed_L}, and bound it by the drift of a randomized policy as in Lemma~\ref{lem:AFbound}. We will pick a randomized policy very similar to the $A_F$ policy that $(i)$ starts with an optimal solution $\balpha^*$ to $\mathbf{P_{conf}}$ corresponding to $F$, $(ii)$ samples a configuration $X$ according to \eqref{eqn:Xt_given_Yt}  and orders the items in this configuration $\mJ_X = (j_1,\ldots, j_{|X|})$, and $(iii)$ then proceeds to pack the item $Y$ in a bin in $\bN$ by mapping it to an item in ordering $\mJ_X$ of the same type uniformly at random. However, in Lemma~\ref{lem:AFbound}, the current packing state $\bN$ was deterministic, and this was used in the alternate equivalent view where we generated $(X,\mJ_X,Y)$ by first sampling configuration $X$ according to \eqref{eqn:Xt}, and then generating the ordering $\mJ_X$ using the current packing $\bN$ which was independent of the item type $Y$. In the setup of Lemma~\ref{lemma:one_step_drift} the independence of $\bN$ and $Y$ is not true anymore.

To remedy this, we define the following \emph{meet} packing corresponding to $\mathcal{P}$ (the term meet comes from orders on lattices, where meet denotes the greatest lower bound of a partially ordered set of elements):
\begin{align}
\underline{P} & \doteq \bigwedge \mathcal{P} \doteq P_{1} \wedge P_{2} \wedge \cdots \wedge P_{|\mathcal{P}|}
\end{align}
That is:
\[ N^{\underline{P}}(h) = \min_{P \in \mathcal{P}} N^P(h).  \]
The $A_F$ policy is now almost the same as before, except we use the packing $\bN^{\underline{P}}$ to decide the ordering $\mJ_X$ and the placement of the item, and require a stricter lower bound on when we consider a level to be non-existent. Two crucial observations are:
\begin{enumerate}
\item $N^P(h) \geq N^{\underline{P}}(h)$ (almost surely): Therefore if it is feasible for $A_F$ to send an item to level $h$, then it is also feasible with probability 1 for our Primal-Dual algorithm to do so.
\item $N^P(h)  \leq N^{\underline{P}}(h) + L$ (almost surely): For the packing $\underline{P}$, we expect the drift in the Lagrangian to be $b(F) + \Theta(\epsilon)$. Since the initial packing is $L$-close to $\underline{P}$, we will prove that the drift only worsens by a small amount.
\end{enumerate}

We will use the following tweaked policy $A_F$:
\begin{quote}
Given a configuration $X$ and the packing $\underline{P}$, we first find an ordering $\mJ_X = (j_1,j_2,\ldots, j_{|X|})$ of the items in $X$, and a threshold index $last(X), \ 0 \leq last(X) \leq |X|$, such that if we set $h_k \equiv \sum_{k=1}^{i} j_k $, then:
\begin{itemize}
\item $\underline{P}$ has at least $L$ partially filled bins with each level $h_k$, $0 \leq k \leq last(X)$
\item $\underline{P}$ has at most $L-1$ partially filled bins in level $h_{last(X)} + j_k$ for any $k > last(X)$
\end{itemize}
\end{quote}

Below we show the change in Lagrangian for various cases, conditioned on the configuration $X$ and the ordering $\mJ_X = (j_1,\ldots, j_{|X|})$. The remaining randomness is from  $K \sim \mathsf{Unif}[|X|]$ given which the item $Y = j_{K}$. Finally, given $Y$, the packing is sampled as $P \sim \mu_Y$. Our analysis will mirror the case analysis in Lemma~\ref{lem:AFbound}, and we will use the same notation as in the proof of the Lemma ($\Delta \mL, \Delta N, \Delta V_h$). 
\begin{enumerate}
\item Case $last(X) =0$ : In this case for all $1\leq K \leq |X|$, a new bin is opened, but in $\underline{P}$ there are at most $L-1$ bins of the levels created thus, and therefore in the original packing $P$ there may be up to $2L$ bins of the new level created. Therefore the change in Lagrangian is:
\begin{align}
\nonumber \expct{\Delta \mL | X, \mJ_X} &\leq 1 + \frac{\kappa}{\epsilon} e^{-\epsilon 2L} (e^{-\epsilon} - 1) \\
\nonumber & = 1 - \kappa e^{-\epsilon 2L} \left(1 - \frac{\epsilon}{2!} + \frac{\epsilon^2}{3!} - \ldots \right) \\
\nonumber & \leq  1 - \kappa e^{-2\epsilon L} \left( 1 - \frac{\epsilon}{2}  \right) \\
\nonumber & \leq \frac{1}{B} + \kappa \frac{\epsilon}{2} \\
& \leq \frac{1}{|X|} + \kappa \frac{\epsilon}{2}
\label{eqn:cor_case1}
\end{align}
In the second inequality, we have used the assumption $\epsilon < 1$, and in the third inequality the assumption $\kappa e^{-2\epsilon L} \geq 1 - \frac{1}{B}$.
\item Case $last(X) > 0$
\begin{enumerate}
\item Subcase $ 1 \leq K \leq last(X)$ : Exactly one of these items causes a new bin to open and increase the objective function term by $1$, the others cause the objective function term to change by $0$. 
\begin{align}
\expct{\Delta N | X, \mJ_X, 1 \leq K \leq last(X)} & = \frac{1}{last(X)}.
\end{align}

We now focus on the change in potential function terms. As before,
\begin{align}
\expct{\Delta V_h | X, \mJ_X, 1 \leq K \leq last(X)} & = 0, \qquad (h \neq h_k, 1\leq k \leq last(X)); \\
\expct{\Delta V_{h_{last(X)}} | X, \mJ_X, 1 \leq K \leq last(X)} & \leq  0.
\end{align}

For $1\leq k \leq last(X)-1$, $N'(h_k) - N(h_k) = \pm 1$ with equal probability. However, the change is not independent of $N(h_k)$, and the actual value of $N_{h_k}$ when the change is $+1$ and when it is $-1$ can be different by up to $L$ (example, change is $-1$ on arrival of item of size 2 for which case $N(h_k)=L$, and $+1$ on arrival of size 3 for which case $N(h_k)=2L+1$). \\
It is easy to see that the above case is indeed the worst for the change in the Lagrangian, and therefore the expected change due to level $h_k$ is almost surely bounded by :
\begin{align*}
\expct{ \Delta V_{h_k} | X, \mJ_X , 1\leq K \leq last(X)}  &  \leq 
\frac{\kappa}{last(X) \cdot \epsilon} \left[ \left( e^{-(2L+1)\epsilon} - e^{- 2L \epsilon} \right)  + \left( e^{-\epsilon(L-1)} - e^{- \epsilon L} \right) \right] \\
& = \frac{\kappa e^{-\epsilon L}}{last(X) \cdot \epsilon} \left[ e^{-\epsilon L} (e^{-\epsilon} - 1) + (e^{\epsilon} - 1)  \right] \\
& \leq \frac{\kappa e^{-\epsilon L}}{ last(X) \epsilon} \left[ (1-\epsilon L )\left(-\epsilon + \frac{\epsilon^2}{2} \right) + (\epsilon + \epsilon^2 )  \right] \\
& \leq  \frac{\kappa}{last(X)} \left[ (1-\epsilon L) (-1 + \frac{\epsilon}{2}) + 1 + \epsilon \right] \\
& = \frac{\kappa}{last(X) } \epsilon \left( L + \frac{3}{2} \right)
\end{align*}
In the above, for the second inequality, we have used the assumptions $\epsilon < 1$ and  $\epsilon L < 1$.
\item Subcase $K > last(X)$ : The item is sent to a bin of level $h_{last(X)}$ to create a bin of level $h_{last(X)}+j_K$. By definition of $last(X)$, we must have at least $N(h_{last(X)}) \geq L$, and $N(h_{last(X)}+j_K) \leq 2L$ with probability 1. There is no change in the objective function term, and a similar calculations as the previous subcase gives for $K > last(X) > 0$:
\begin{align*}
\expct{ \Delta \mL | X, \mJ_X , K} & = N' - N + 
\frac{\kappa}{\epsilon} \left( 
e^{-(N(h_{last(X)})-1)\epsilon} - e^{-(N(h_{last(X)}))\epsilon}  \right.  \\
& \qquad 
 \left. + e^{-( N(h_{last(X)+j_K}+1 )\epsilon} - e^{- N(h_{last(X)}+j_K) \epsilon} 
\right) \\
& \leq \frac{\kappa}{last(X) \cdot \epsilon} \left[ \left( e^{-(2L+1)\epsilon} - e^{- 2L \epsilon} \right)  + \left( e^{-\epsilon(L-1)} - e^{- \epsilon L} \right) \right] \\
& \leq   \kappa  \epsilon \left( L + \frac{3}{2}\right) 
\end{align*}
\end{enumerate}
Combining the two subcases, when $last(X) > 0$:
\begin{align}
\expct{\Delta \mL | X, \mJ_X } & \leq \frac{1}{|X|} + \kappa \epsilon \left( L + \frac{3}{2}  \right).
\label{eqn:cor_case2}
\end{align}
\end{enumerate}
Combining the two cases, \eqref{eqn:cor_case1} and \eqref{eqn:cor_case2}, conditioning on the configuration $X$ and the ordering $\mJ_X = (j_1, \ldots, j_{|X|})$:
\begin{align}
\expct{\Delta \mL | X, \mJ_X} & \leq \frac{1}{|X|} + \kappa \epsilon \left( L + \frac{3}{2}\right) .
\end{align}
Finally, unconditioning on the configuration $X$ as in the proof of Lemma~\ref{lem:AFbound} leads to the bound in the Lemma statement.
\endproof \Halmos

%% file: tex/alternate.tex
\section{Alternate interpretations of Primal-Dual algorithm}
\label{sec:interp}
In this section we provide some intuitive interpretations of the Primal-Dual algorithms proposed in Section~\ref{sec:PDlevel}. This section serves expository purpose and does not present new results. In Section~\ref{sec:SSrevisit} we revisit the Sum-of-squares heuristic and the reason for its failure on Linear Waste distributions. We then show how patching the SS heuristic leads us to the Primal-Dual algorithms with quadratic and translated-quadratic penalty functions. In Section~\ref{sec:OMDinterp} we interpret Primal-Dual algorithms as solving the maximization problem dual to $\mathbf{P_{1d-level}}$ using stochastic gradient ascent, which is why we name our algorithms Primal-Dual.

\subsection{Primal-Dual algorithm as `patching' the Sum-of-Squares (SS) rule for LW distributions} 
\label{sec:SSrevisit}

Recall that SS places arrivals so as to greedily minimize the penalty function:
\[ ss(\bN) = \sum_{h=1}^{B-1} N(h)^2 \]
where $N(h)$ denotes the number of level $h$ bins in the packing obtained after packing the arriving item. Therefore, on the arrival of an item of size $j$ at time $t$, SS sends the item to a bin of level $h^*$ where:
\[ h^* = \argmin_{h:N(h) > 0} \left[ N_{t-1}(h+j) - N_{t-1}(h) \right] \]
with the convention $N_{t}(0) = N_{t}(B) = 0$ for all $t$. In other words, SS tries to equalize the number of bins at different levels (except level $B$ bins, which are full). This heuristic works for perfectly packable distributions where all items can be packed without any $N(h)$ growing large as $t$ increases.
Thus if any $N(h)$ with $h<B$ grows too much, SS stops creating new bins of level $h$.
For example, when $B=5$ and the item size distribution is $p_2=p_3=\frac{1}{2}$, we expect size 2 and size 3 items to be paired together. Now $N(4)$ can not grow a lot because this requires $N(2)$ must also grow (since SS tries to equalize $N(2)$ and $N(4)$ when packing size 2 items). However, $N(2)$ can not grow too much because size 3 items annihilate them, and therefore no $N(h)$ grows as $\Theta(T)$.

 However, for Linear Waste distributions some $N(h)$ must grow as $\Theta(T)$, and SS heuristic `pulls along' the number of bins which have room for more items. E.g., for $B=5$ and only size 2 items, SS creates $\Theta(T)$ bins of level 4 which cause the number of level 2 bins to grow as $\Theta(T)$ as well. Given this intuition, can we modify the SS heuristic to avoid this mode of failure?

One proposal is to stop counting the number of bins once they reach a threshold. In other words, we use the following rule:
\begin{align}
\label{eqn:proptest}
 h^* = \argmin_{h:N(h) > 0} \left[ \left( \eta \wedge N_{t-1}(h+j) \right) - \left( \eta \wedge N_{t-1}(h) \right) \right] 
\end{align}
for a suitable threshold $\eta$, and the convention $N_t(0)=N_t(B)=0$ for all $t$. It turns out this simple change is not sufficient. Consider the example $B=8$ with $p_2 = p_5 = \frac{1}{2}$. The optimal packing is to pair size 2 items with a size 5 items. However under the placement rule given above, size 2 items prefer to be packed together to create level $B=8$ bins since level $8$ bins are ``free'' because of our convention that $N(B)=0$. This results in three times as much waste as the offline optimal packing.

{\bf Proposal 1:} We can fix the above problem by modifying the rule in equation \eqref{eqn:proptest}  
by keeping the convention $N_t(0)=0$, but $N_t(B) := \eta$ for all $t$. This makes level $B$ not free. Surprisingly, this heuristic works, and in fact, is identical to the Primal-Dual heuristic with translated quadratic penalty function as we show next.

By noting that the dynamics under $\PDtquad$ satisfies $\widetilde{N}_t(h) = (\eta \wedge N_t(h))$ for all $t$, we can write the penalized-Lagrangian for the translated quadratic penalty function by dropping the $\widetilde{\bN}$ argument as:
\[ \mL^{tquad}_\eta( \bN) = \sum_{h=1}^B N(h) + \frac{1}{2\eta}\sum_{h=1}^{B-1} \left((\eta - N(h))^+\right)^2 ,\]
Defining $V(h) = \frac{\partial \mL^{tquad}_\eta(\bN)}{\partial N(h)}$, we get:
\begin{align*}
V(h) &= \frac{\eta \wedge N(h)}{\eta},  \qquad  (1\leq h \leq B-1); \\
V(0) &:= 0 \  ; \quad V(B) := 1.
\end{align*}
giving the placement rule: $h^* = \argmin_{h:N_{t-1}(h)>0} V(h+j)-V(h) =  \argmin_{h : N_{t-1}(h) > 0} \left[ (\eta \wedge N_{t-1}(h+j)) - (\eta \wedge N_{t-1}(h)) \right] $ with the convention $N_{t-1}(0)=0, N_{t-1}(B) = \eta$. This is precisely Proposal 1.

{\bf Proposal 2:} Another way to use a sum-of-square potential term while not letting any $N(h)$ grow linearly in $t$ is to allow bins to be started at any level $h$. For example, in the $B=5$ example with only items of size $2$, if we allow bins to start at level $1$, then they will reach level $5$ and we do not accumulate level $4$ bins. This modification by itself is insufficient -- nothing prevents us from opening a fresh bin for each arriving item, and packing them at level $B-j$ to create a level $B$ bin (recall there is no $N(B)^2$ term in $ss(\bN)$). To fix this, we penalize items which cause the level of a bin to reach $B$. Therefore, on the arrival of $t$th item of size $j$, we send it to level $h^*$ where
\begin{align}
\label{eqn:prop2}
h^* &= \argmin_{h}  \left[ \lambda \cdot \mathbf{1}_{h=B-j} +  N_{t-1}(h+j) - N_{t-1}(h) \right] 
\end{align}
where $\lambda$ is the penalty for closing a bin. Note that in \eqref{eqn:prop2} we do not constrain $h^*$ to the set $\{h: N_{t-1}(h) > 0\}$. We can equivalently view this heuristic as minimizing the cost function:
\begin{align*}
\widehat{ss}(\bN) &= \lambda \cdot N(B) + \frac{1}{2} \sum_{h=1}^{B-1} N(h)^2
\end{align*}
which is a combination of the Best Fit and SS cost functions (but still rather odd due to penalty for `full bins'). This algorithm ``works'' and is equivalent to the Primal-Dual heuristic method with quadratic penalty function as we show next, but the output of the Primal-Dual packing must now be viewed upside-down!

For the quadratic penalty function, recall that we must violate all the constraints to get non-zero duals. However, the {\it (no floating items)} constraints of $\mathbf{P_{1d-level}}$ dictate that for all levels $h$, there are at least as many items which start at level $h$ than items which end at level $h$. Violation of these constraints mean that for all $h$, we must have more items that start at level $h$ than ending at level $h$. That is, we start packing our bins from top going downwards. Further, since the objective function of $\mathbf{P_{1d-level}}$ is $\sum_{j} v(j,0)$, we must pay the cost of a bin when an item touches level 0 during this top-down packing. 

Denote the violation of level $h$ constraint by $Q(h)$, and let $Q(0)$ denote the number of bins which reach level 0 (we have suppressed the dependence on time $t$). The penalized-Lagrangian with quadratic penalty function becomes:
\begin{align*}
 \mL^{quad}(\bQ) &= Q(0) + \frac{\epsilon}{2} \sum_{h=1}^{B-1} Q(h)^2
\intertext{and applying the transformation: $N(h) = Q(B-h) $, and $\lambda = \frac{1}{\epsilon}$:}
 & = N(B) + \frac{1}{2\lambda} \sum_{h=1}^{B-1} N(h)^2  = \frac{1}{\lambda} \cdot \widehat{ss}(\bN).
\end{align*}
Therefore the actions taken by the Primal-Dual algorithm with quadratic penalty function, and the heuristic in Proposal 2 will be identical.

\subsection{Online Mirror Descent view of PD-exp}
\label{sec:OMDinterp}
As the naming of our algorithms suggest, they are in fact Primal-Dual iterations whereby the maximization problem that is the dual to $\mathbf{P_{1d-level}}$ is solved using stochastic gradient ascent. We now briefly describe this interpretation of the $\PDexp$ algorithm. We start with the LP $\mathbf{P_{1d-level}}$, and perform a change of variables:
\[ u(j,h) := \frac{v(j,h)}{p_j}. \]
Note that $ u(j) = \left( u(j,0), \ldots, u(j,B-1) \right)$ now represents a probability vector  -- a randomized strategy for packing item $j$. Denote the set of feasible action for $u(j)$ by $\widehat{\mU}(j) := \widehat{\mU}(\mathbf{1},j)$ (see \eqref{eqn:feasible_simplex}). The objective function becomes:
\begin{align*}
\sum_{j} v(j,0) = \sum_{h=1}^B n(h) &= \sum_{h=1}^B \sum_j \left[ v(j,h-j)-v(j,h) \right] \\
 &= \sum_{h=1}^B \sum_j p_j \left[ u(j,h-j)-u(j,h) \right] 
\end{align*}

Introducing duals $\{\gamma(1), \ldots, \gamma(B-1)\}$ for the {\it (no floating items)} constraints, we write the Lagrangian:
\begin{align*}
L(\mathbf{u}, \boldsymbol{\gamma})&=  \sum_{h=1}^{B} \sum_j p_j \left[ u(j,h-j)-u(j,h) \right] 
-\sum_{h=1}^{B-1} \gamma(h) \sum_j p_j \left[ u(j,h-j)-u(j,h) \right] \\
&= \sum_j p_j \sum_{h=0}^{B-j} u(j,h) \left[ (1-\gamma({h+j})) - (1-\gamma(h)) \right]
\end{align*}
where we define $\gamma(0) := 1$, $\gamma(B) := 0$. This leads to the \emph{dual function}:
\begin{align}
\label{eqn:dualfn1}
q(\boldsymbol{\gamma}) & := \min_{ \{ u(j) \in \widehat{\mU}(j)\} } \sum_{j} p_j \sum_{h=0}^{B-j} u(j,h) [(1-\gamma({h+j})) - (1-\gamma(h))]  \\
& =  \sum_{j} p_j \min_{ 0 \leq h \leq B-j} [(1-\gamma({h+j})) - (1-\gamma(h))]   \\
& =  \sum_{j} p_j q_j(\boldsymbol{\gamma}).  
\label{eqn:dualfn2}
\end{align}

The dual optimization problem is to solve 
\begin{align}
\label{eqn:dual_level}
\boldsymbol{\gamma}^* = \argmax_{\boldsymbol{\gamma}\geq 0 : \gamma(0)=1,\gamma(B)=0 } q(\boldsymbol{\gamma}).  
\end{align}

The Primal-Dual algorithm $\PDexp$ can now be viewed as an Online Mirror Descent (OMD) algorithm (\cite{nemirovski1979efficient},\cite{beck2003mirror}) for problem \eqref{eqn:dual_level} :  In a typical application of OMD, a minimization problem  $\min_{x \in \Omega} f(x)$ is solved iteratively with access to subgradients $f'(x)$. One first defines a strongly-convex distance function $\omega(x)$ on $\Omega$, and the output $x(t)$ at the $t$th iteration is set as:
\begin{align}
\label{eqn:OMD_update}
x(t) = \argmin_{x \in \Omega} \left[ \epsilon \sum_{s=1}^{t-1} \inn{f'(x(s)) }{x} + \omega(x)  \right]. 
\end{align}
The parameter $\epsilon$ is called the learning rate. 


Applied to our setting, we are solving a maximization problem. Let $Y_t$ be the size/type of the $t$th arrival, and for now assume that the action $\bU_t \in \widehat{\mU}(Y_t)$ is chosen so that $U_t(h^*) = 1$ for
\begin{align}
\label{eqn:OMD_action}
h^* \in \argmin_{N_t(h-1) \geq 1}  \left[ (1-\gamma_{t-1}({h+j})) - (1-\gamma_{t-1}({h})).   \right]
\end{align}
(We will explain shortly, how $\PDexp$ is indeed choosing $\bU_t$ as above).
Observing the relationship,
\[ N_t(h) = \sum_{s=1}^t U_s(h-Y_s) - U_s(h) . \]
denote $\Delta N_t(h) = U_t(h-Y_t) - U_t(h)$, and $\Delta \bN_t = (\Delta N_t(1), \ldots, \Delta N_t(B-1))$. From \eqref{eqn:dualfn1}, and assuming \eqref{eqn:OMD_action}, 
we get that given $Y_t=j$, $\Delta \boldsymbol{N}_t \in \delta q_j(\boldsymbol{\gamma}_{t-1})$, the subgradient set of $q_j$. Therefore,  $\expctsub{Y_t}{ \Delta \boldsymbol{N}_t } \in \delta q(\boldsymbol{\gamma}_{t-1})$. That is, the primal packing decisions are unbiased stochastic subgradients of the dual function $q(\boldsymbol{\gamma})$ at the current solution $\boldsymbol{\gamma}_{t-1}$.

By choosing the distance generating function $\omega(\boldsymbol{\gamma}) = \sum_{h=1}^{B-1} \gamma(h) (\log \frac{\gamma(h)}{\kappa} -1 )$ (strongly convex with respect to $\ell_2$ norm), the duals $\boldsymbol{\gamma}$ at iteration $t$ become (using \eqref{eqn:OMD_update}):
\begin{align*}
\gamma_t(h) &= \argmax_{\gamma \in [0,1]} \left[ \epsilon_t \sum_{s=1}^{t} \inn{U_s(h)-U_s(h-Y_s) }{\gamma} - \gamma \left( \log \frac{\gamma}{\kappa} - 1\right) \right]\\
 &= \argmax_{\gamma \in [0,1]} \left[ \epsilon_t \inn{- N_t(h)}{\gamma } - \gamma \left( \log \frac{\gamma}{\kappa} - 1 \right) \right]  \\
& =  \kappa e^{-\epsilon_t N_t(h)}.
\end{align*}
One can now verify that $\PDexp$ algorithm where we greedily minimize the penalized-Lagrangian $\mL^{exp}_t( \bN_{t-1} + \bC_{Y_t} \cdot \bU_t )$ is in fact taking action as per \eqref{eqn:OMD_action}, and hence is a Primal-Dual iteration folded into a single step. 

More generally, to implement the OMD algorithm with a given separable distance generating function $\omega(\boldsymbol{\gamma}) = \sum_{h=1}^{B-1} \omega(\gamma(h))$ and learning rates $\{\epsilon_t\}$, we can use a greedy penalized Lagrangian function $\mL^{\Phi}_t(\bN) = N + \sum_{h=1}^{B-1} \Phi_t(N(h))$ where $\Phi_t, \omega$ satisfy:
\begin{align*}
1- \gamma_t(h) &= \frac{\partial \mL^{\omega}_t(\bN) }{ \partial N(h)} = 1 + \nabla \Phi_t(N(h)) , \\
\gamma_t(h) &= \argmax_{\gamma \in [0,1]} \left[ - \epsilon_t N(h) \gamma   - \omega(\gamma)  \right] = (\nabla \omega)^{-1}(-\epsilon_t \cdot N(h)) = \nabla \omega^*( -\epsilon \cdot N(h) )
\end{align*}
where $\omega^*(y) = \max_{x} x\cdot y - \omega(x)$ is the convex conjugate of $\omega$. Therefore,
\[ \nabla \Phi_t(N) = - \nabla \omega^*( -\epsilon_t \cdot N),  \]
or equivalently,
\[ \Phi_t(N) = \frac{1}{\epsilon_t} \omega^*( -\epsilon_t \cdot N )  . \]

%% file: tex/appendix2.tex
\section{Auxiliary Results}

\subsection{Generalization of Theorem~\ref{thm:PDexp_fixed_epsilon}}
\label{sec:generalizedPD}
As we mention in the paper, our interior point based algorithm is in fact much more general than the $\PDexp$ algorithm that uses an exponential penalty function. To demonstrate this, we consider the following general penalized-Lagrangian:

\begin{align}
\mathcal{L}(\mathbf{N}_t) &= \sum_{h=1}^{B} N_t(h)  + \sum_{h=1}^{B-1} \Phi(N_t(h)) 
\end{align}

where $\Phi()$ is a convex decreasing penalty function. We assume it is smooth, and has a second derivative almost everywhere. To understand how the regret of a Primal-Dual algorithm depends on the penalty function $\Phi$, we now highlight the places where the proof of Theorem~\ref{thm:PDexp_fixed_epsilon} changes.

\noindent \textbf{A generalized $A_F$ policy} \\
We consider the following generalization of the $A_F$ policy, parameterized by a threshold $\tau$. Once a configuration $X$ is sampled using the probability measure $\balpha^*$, the ordering $\mJ_X = (j_1,j_2,\ldots, j_{|X|})$ of the items in $X$ and $last(X), \ 0 \leq last(X) \leq |X|$ are chosen such that given the current packing $\bN$:
\begin{itemize}
\item $\bN$ has at least $\tau+1$ bins with level $h_k$ if $h_k < B$, $0 \leq k \leq last(X)$ where:
 \[ h_0 = 0 ; \quad h_k = h_{k-1} + j_k \]
\item $\bN$ has at most $\tau$ bins of levels $h'_k$ for any $k > last(X)$ where:
 \[  h'_k = h_{last(X)} + j_k \]
\end{itemize}

\noindent \textbf{Drift analysis of $A_F$} \\
Let $\bN$ denote the state before packing, $Y \sim F$ the item to be packed, $\bU$ the action, and $\bN' = \bN + \bC_{Y} \cdot \bY$ the state after packing. The change in Lagrangian is:
\[ \Delta \mL = \underbrace{N' - N}_{=: \Delta N} + \sum_{h=1}^{B-1} \underbrace{\Phi(N'(h)) - \Phi(N(h))}_{ = :\Delta V_h}.\]
To analyze $\expctsub{Y \sim F, \bU \sim \pi^{A_F}}{\Delta \mL}$, we will condition on the configuration $X$, and the ordering $\mJ_X = (j_1, \ldots, j_{|X|})$.
\begin{enumerate}
\item Case $last(X) = 0$: For all $1\leq K \leq last(X)$ the objective function term $\Delta N$ is $1$, $N'(j_K)=N(j_K)+1$ and $N'(h) = N(h)$ for $h \neq j_K$. Since $j_K \neq B$, the change in Lagrangian is:
\begin{align}
\expct{\Delta \mathcal{L} | X, \mJ_X, K} &= 1 + \Phi(N_{j_K}+1) - \Phi(N_{j_K}) \\
\intertext{Since $N_{j_k} \leq \tau$ and $\Phi(\cdot)$ is convex,}
& \leq 1 + \Phi(\tau+1) - \Phi(\tau)
\end{align}
Unconditioning on $K$,
\begin{align}
\expct{\Delta \mathcal{L} | X, \mJ_X} & \leq 1 + \Phi(\tau+1) - \Phi(\tau)
\end{align}
\item Case $last(X) > 0$ : 
\begin{enumerate}
\item Subcase $K > last(X)$ : 
Since $\Phi()$ is decreasing. $\Delta N = 0$, $N'(h_{last(X)}+j_K) = N(h_{last(X)}+j_K)+1 = 0$, and $N'(h_{last(X)}) = N(h_{last(X)}) - 1 \geq \tau$. Using convexity of $\Phi(\cdot)$, when $K > last(X)$
\begin{align}
\expct{\Delta \mathcal{L} | X, \mJ_X, K}  & \leq 0
\end{align}
\item Case $1\leq K \leq last(X)$ : As before, the change in the objective function term is
\begin{align}
\expct{\Delta N | X, \mJ_X, 1 \leq K \leq last(X)} &= \frac{1}{last(X)}.
\end{align}
Further, the change in potential function term for level $h \neq h_k$ for any $1\leq k \leq last(X)$ is 0. The analysis of the change in potential function term for $h_k$ ($1\leq k \leq last(X)$) becomes (note, we have assumed below that $ N(h_k) = n \geq \tau+1$ before the new item is packed ):
\begin{align*}
& \expct{\Delta V_{h_k} | X, \mJ_X, 1\leq K \leq last(X)}  \\
&= \sum_{a=1}^{last(X)} \expct{\Delta V_{h_k} | X, \mJ_X , K = a } \cdot \prob{ K = a | 1\leq K \leq last(X)} \\
& = \frac{1}{last(X)} \left( \expct{\Delta V_{h_k} | X, \mJ_X , K=k}  + \expct{\Delta V_{h_k} | X, \mJ_X, K=k+1} \right) \\
  &  \leq \frac{1}{last(X) } \left[ \left( \Phi(n+1) - \Phi(n) \right)  + \left( \Phi(n-1) -  \Phi(n) \right) \right] \\
\intertext{Under the assumption that $\Phi'()$ is continuous almost everywhere differentiable (this is true for the translated quadratic penalty function also), for some $0\leq \xi_1, \xi_2 \leq 1$ such that $\Phi''(\xi_1) , \Phi''(\xi_2)$ is a point of differentiability of $\Phi'()$:}
  &  \leq \frac{1}{last(X) } \left[ \left( \Phi'(n) + \frac{1}{2} \Phi''(n+\xi_1) \right)  + \left( - \Phi'(n) +  \frac{1}{2}\Phi''(n+\xi_2) \right) \right] \\ 
  &  \leq \frac{1}{last(X) } \Phi''_{max,\tau} 
\end{align*}
where $\Phi''_{max,\tau}$ is an upper bound on the second derivative of $\Phi$ in the domain $[\tau,T]$.
\end{enumerate}
 Combining the two subcases, when $last(X) > 0$, we get
\begin{align}
\nonumber \expct{\Delta \mL | X, \mJ_X} &= 
\expct{\Delta \mL | X, \mJ_X , K > last(X) } \cdot \prob{K > last(X)} \\ 
& \qquad + \expct{\Delta \mL | X, \mJ_X, 1 \leq K \leq last(X)} \cdot \prob{1\leq K \leq last(X)} \\ 
& \leq 0 \cdot \frac{|X|-last(X)}{|X|}  
+ \left(\frac{1}{last(X)} + \Phi''_{max,\tau}  \right) \cdot \frac{last(X)}{|X|} \\ 
& \leq \frac{1}{|X|} + \Phi''_{max,\tau}.
\end{align}
\end{enumerate}
Combining cases $last(X) = 0$ and $last(X) > 0$:
\begin{align}
\expct{\Delta \mL | X, \mJ_X} & \leq \frac{1}{|X|} + \max\{ 1 +\Phi(\tau+1) - \Phi(\tau) , \Phi''_{max,\tau}  \}.
\end{align}

Finally, this yields the following bound on $\Delta \mathcal{L}$:
\begin{align}
\nonumber \expctsub{Y \sim F, \bU \sim \pi^{A_F}}{\mL(\bN') - \mL(\bN) | \bN} &= 
\expctsub{X, \mJ_X, K \sim \mathsf{Unif}[|X|],  \bU = \be_{ \min\{ h_{K-1} ,  h_{last(X)}\}}}{\mL(\bN') - \mL(\bN) | \bN} \\
& \leq b(F) + \max \{ 1 + \Phi(\tau+1) - \Phi(\tau) , \Phi''_{max,\tau} \} 
\end{align}

\noindent \textbf{Regret Bound for Primal-Dual with penalty function $\Phi$}\\
Upper bounding the change in Lagrangian of Primal-Dual by the change in Lagrangian of $A_F$ gives:
\begin{align*}
\expct{ \mL (\bN^{\PD}_T)} - \mL(\bN^{D}_0)
& \leq T \left( b(F) + \max\{1 + \Phi(\tau+1) - \Phi(\tau) , \Phi''_{max, \tau}\} \right) 
\end{align*}
Therefore, for any $\tau \geq 0$:
\begin{align*}
\expct{ N^{\PD}_T}  & \leq T b(F) + T \max\{1+\Phi(\tau+1) - \Phi(\tau) , \Phi''_{max,\tau} \} + \sum_{h=1}^{B-1} \left( \Phi(N^{\PD}_T(h) - N^{\PD}_0(h)  \right) \\
& \leq T b(F) + T \max\{1+\Phi(\tau+1) - \Phi(\tau) , \Phi''_{max,\tau} \} + B (\Phi(0) - \Phi(T)) 
\end{align*}

\noindent \textbf{Example: $\log$-barrier penalty }

Consider the following $\log$-barrier penalty function:
\[ \Phi(x) = \frac{1}{\epsilon} \log (1 + x). \]
We have:
\begin{enumerate}
\item $1+\Phi(\tau+1) -\Phi(\tau) = 1 - \frac{1}{\epsilon} \log (1 + \frac{1}{\tau}) \leq 1 - \frac{1}{2\epsilon \tau} $. This is bounded by $0$ if $\tau \leq \frac{1}{2\epsilon}$ 
\item $\Phi''_{\max, \tau} = \Phi''(\tau) \leq \frac{1}{\epsilon \tau^2}$
\item $\Phi(T) \geq - \frac{1}{\epsilon} \log (T+1)$ 
\end{enumerate}
Finally, for $\tau \leq \frac{1}{2\epsilon}$:
\begin{align*}
\expct{N^{\PD}_T } & \leq T b(F) + \frac{T}{\epsilon \tau^2}  + \frac{B}{\epsilon} \log (T+1) \\
\intertext{Choosing: $\epsilon = \sqrt{ \frac{B \log (T+1)}{4T} }$ and $\tau = \sqrt{\frac{T}{B \log (T+1)}}$,}
\expct{N^{\PD}_T}& \leq T b(F) + 4 \sqrt{ B T \log (T+1)}.  
\end{align*}

\noindent \textbf{Example: translated-Quadratic penalty}

Consider the following translated-quadratic penalty function:
\[ \Phi(x) = \begin{cases}
\frac{1}{A} \left( 1 - Ax + \frac{A^2}{4}x^2  \right) & 0 \leq x \leq \frac{2}{A} \\
0 & x \geq \frac{2}{A}
\end{cases} \]
where $A = \sqrt{\frac{2 B}{T}}$.

With $\tau=0$, we have:
\begin{enumerate}
\item $1+\Phi(1) -\Phi(0) =  \frac{A}{4}  $
\item $\Phi''_{\max, 0} = \Phi''(0)= \frac{A}{2} $
\item $\Phi(0) - \Phi(T)  = \Phi(0) = \frac{1}{A} $ 
\end{enumerate}

Finally, 
\begin{align*}
\expct{ N^{\PD}_T } & \leq T b(F) + T \cdot \frac{A}{2}  + B \cdot \frac{1}{A} = T b(F) + \sqrt{2 BT}.
\end{align*}